%% file: qcnn.tex
\documentclass[fleqn,usenatbib]{mnras}
\pdfoutput=1
\nonstopmode

\usepackage{newtxtext,newtxmath}
\usepackage[T1]{fontenc}

\DeclareRobustCommand{\VAN}[3]{#2}

\let\VANthebibliography\thebibliography
\def\thebibliography{\DeclareRobustCommand{\VAN}[3]{##3}\VANthebibliography}

\newcommand{\lya}{\text{Ly-}\alpha }
\usepackage{caption}
\usepackage{graphicx}	% Including figure files
\usepackage{amsmath}	% Advanced maths commands
\usepackage{amssymb}	% Extra maths symbols

\title[iQNet]{A Deep Learning Approach to Quasar Continuum Prediction}

\author[Liu \& Bordoloi ]{
Bin Liu,$^{1}$\thanks{Email: bliu8@ncsu.edu}
Rongmon Bordoloi,$^{1}$\thanks{Email: rbordol@ncsu.edu}
\\
$^{1}$Department Of Physics, North Carolina State University, Raleigh, North Carolina, 27695\\
}

\date{Accepted ---. Received ---; in original form ---}

\pubyear{2021}

\begin{document}
\label{firstpage}
\pagerange{\pageref{firstpage}--\pageref{lastpage}}
\maketitle

\begin{abstract}
We present a novel intelligent quasar continuum neural network (iQNet), predicting the intrinsic continuum of any quasar in the rest-frame wavelength range $1020 \text{\AA} \leq \lambda_{\text{rest}} \leq 1600 \text{\AA}$. We train this network using high-resolution Hubble Space Telescope/Cosmic Origin Spectrograph ultraviolet quasar spectra at low redshift ($z \sim 0.2$) from the Hubble Spectroscopic Legacy Archive, and apply it to predict quasar continua in different astronomical surveys. We utilize the HSLA quasar spectra that are well-defined in the rest-frame wavelength range [1020, 1600]\text{\AA} with an overall median signal-to-noise ratio of at least five. The iQNet model achieves a median AFFE of 2.24\% on the training quasar spectra, and 4.17\% on the testing quasar spectra. We apply iQNet and predict the continua of $\sim$ 3200 SDSS-DR16 quasar spectra at higher redshift ($2<z\leq5$) and measure the redshift evolution of mean transmitted flux ($\langle F\rangle$) in the $\lya$ forest region. We measure a gradual evolution of $\langle F \rangle$ with redshift, which we characterize as a power-law fit to the effective optical depth of the $\lya$ forest. Our measurements are broadly consistent with other estimates of $\langle F \rangle$ in the literature but provide a more accurate measurement as we are directly measuring the quasar continuum where there is minimum contamination from the $\lya$ forest. This work proves that the deep learning iQNet model can predict the quasar continuum with high accuracy and shows the viability of such methods for quasar continuum prediction.
\end{abstract}


%%%%%%%%%%%%%%%%%%%%%%%%%%%%%%%%%%%%%%%%%%%%%%%%%%

%%%%%%%%%%%%%%%%% BODY OF PAPER %%%%%%%%%%%%%%%%%%

\section{Introduction}
The $\lya$ forest absorption in the spectra of high redshift objects is imprinted by neutral hydrogen in the intergalactic medium (IGM), which makes up most of the baryons in the early Universe \citep{Shull_2012}. This is unlike lower redshifts, where a significant fraction of the baryons reside in the circumgalactic medium of galaxies \citep{Tumlinson_2013, Werk_2014, Bordoloi_2014, Chen_2018}. Our understanding of the $\lya$ forest changed dramatically in the late 1990s when the pioneering hydrodynamical simulations showed that the $\lya$ forest statistics could be understood if the neutral gas followed the underlying dark matter distribution in the cosmic web of filaments, voids, and clusters \citep{Cen1994,Hernquist__1996, Croft_1998}. The properties of the $\lya$ forest inform us about the density, temperature, and ionization state of the IGM. Subsequently, both the 1-D power spectrum of HI column density, and the evolution of the mean transmitted flux in the $\lya$ forest along the lines of sight to distance quasars are studied for a small sample of bright high-resolution quasar spectra \citep{Rauch_1997, Schaye_2003, Songaila_2004, Kirkman_2005, Becker_2010, Faucher_Gigu_re_2008}, and large samples of moderate resolution SDSS spectra \citep{Bernadi_2003, McDonald_2005, Paris_2011, Lee_2012, Becker_2013, Davies_2018}.

One of the basic observable used to quantify flux distribution in the $\lya$ forest is the evolution of mean transmitted flux, $\langle F\rangle$, with redshift \citep[e.g.][]{Faucher_Gigu_re_2008, Becker_2010}. The effective optical depth, $\tau_{\text{eff}}$, of the $\lya$ forest ($\tau_{\text{eff}}=-\ln \langle F\rangle$) can be used to constrain the metagalactic ionizing background intensity \citep{Rauch_1997, Bolton_2005, McDonald_2001, McDonald_2005, Becker_2013}. The transmitted flux $F$ for any arbitrary quasar spectrum is defined as 

\begin{equation}
    F =\frac{F_{\text{spec}}}{F_{\text{cont}}},
    \label{eq:transmitted_flux}
\end{equation}
where $F_{\text{spec}}$ is the observed quasar flux in its rest frame, and $F_{\text{cont}}$ is the true quasar continuum without any absorption.

To study the properties of the $\lya$ forest, we need to know the true quasar continuum ($F_{\text{cont}}$, equation \ref{eq:transmitted_flux}) to make correct estimates of $\langle F\rangle$ and therefore $\tau_{\text{eff}}$. The uncertainty in predicting the true continuum is often one of the largest uncertainties in studying the IGM \citep[e.g.][]{Croft_2002}. Thus, an accurate model to predict the quasar continuum in the presence of $\lya$ forest absorption is necessary for scientists to study neural hydrogen in the IGM.

Previous studies have defined the transmission peaks in the $\lya$ forest region as the continuum in high redshift quasars using high resolution-echelle spectra \citep[e.g.][]{Rauch_1997, Schaye_2003}. However, this method may note be accurate at higher-$z$, particularly as the IGM becomes increasingly more neutral (i.e. $\langle F\rangle$ is very small). Additional small biases may arise owing to presence of cosmic voids, large regions of mild over-density and high temperature, or metal absorption lines along the lines of sight. $F_{\text{cont}}$ can be statistically corrected  by using synthetic spectra from hydrodynamical simulations \citep{Faucher_Gigu_re_2008}. Unfortunately, such correction also relies on the veracity of the simulations/models themselves and is not, in fact, a direct measurement of $\tau_{\text{eff}}$.  

Other studies used a large number of moderate-resolution spectra to either statistically correct for the continuum \citep{Bernadi_2003}, used composite spectra of $z > 2$ quasars to statistically study the evolution of $\tau_{\text{eff}}$ \citep{Becker_2013}, or used principal component analysis (PCA) of $z \sim$ 3 quasars to predict the true continuum \citep{Paris_2011}. Composite spectra can be used to estimate very robust "differential" $\tau_{\text{eff}}$ evolution from the stacked spectra in different redshift bins. However, this method relies on two assumptions: (1) all stacked quasar spectra have the same "shape", and (2) the true quasar continuum can be represented by the composite spectra at $z \sim 2.15$. Both these assumptions may not be always very accurate.

PCA-based prediction models are popular due to the rise of machine learning applications, \citep[e.g.][]{Suzuki_2005, Paris_2011, Lee_2012, Davies_2018}.\footnote{We follow the same steps in \citet{Suzuki_2005} and \citet{Paris_2011} to construct our PCA prediction models by constructing the square-definite transformation matrix. Other PCA methods may not require the transformation matrix to be square definite.} However, the main drawback of this type of model is its limited generalizability. A PCA prediction model usually consists of two sets of principal components, which recover the full wavelength coverage of the quasar continuum and the part of the quasar continuum redward of quasar $\lya$ emission, respectively. Further, a transformation matrix is used to transform the PCA weights of continuum redward of $\lya$ emission into the weights of the whole continuum. Most PCA reconstruction models are usually required to recover at least 95\% of the explained variance of the flux within the wavelength range of interest. \citep{Suzuki_2005, Paris_2011, Davies_2018, Durov_kov__2020}.

Another important step in evaluating the performance of a PCA model is to use an independent testing data set that should never be used for training and only reserved to test and evaluate the researched model. Evaluating the model accuracy based on a training data set only reflects how well it performs through the training process but does not reflect how well it will work on a testing data set. Because the model has been constructed based on the information extracted from the training data set, the model will, unsurprisingly, perform well on the training data. Several published PCA models used for quasar continuum estimation use the same data set to both train the weights of the PCA models and test the performance of those models. Therefore, the robustness of PCA models on a testing data set is never known if the performance of researched models is evaluated on a training data set. \citet{Paris_2011} applies the leave-one-out strategy to keep one continuum for testing purpose. Recent studies have started to address this issue, e.g.   \citet{Durov_kov__2020} and \citet{bosman2020comparison} did split the data properly into a training set and a testing set as we do. A robust model needs to be tested so that it performs well not only on the training data but also on the testing data that have never been exposed to the model. The testing data set will be the model benchmark to make sure that the model generalizes well and robustly on the unseen data.

To predict the true quasar continua, several model independent methods are widely used for IGM studies (e.g. power-law extrapolation method \citep{Oke82, Songaila_2004}, variational method \citep{Press_1993}). The power-law extrapolation method was applied in \citet{Fan_2006, Becker_2015, Bosman_2018} with the assumption that the quasar spectra in the ultraviolet or optical range behave like black-bodies. The variational method was applied in \citet{Kim+07, Kamble_2020} with assumption that the IGM absorption has a faster variation with wavelength than the quasar continuum.  Several recent studies estimate/correct for high-$z$ $\tau_{\text{eff}}$ values using simulations or models, or use $z \sim$ 2 quasar spectra as the ground truth continua (e.g. \citealt{Faucher_Gigu_re_2008,Paris_2011,Becker_2013}). In this paper, we present an alternative approach, to predict the true quasar continua (uncontaminated by $\lya$ forest) with accuracy at high redshifts; we employ $z < 1$ quasar spectra observed with the Hubble Space Telescope's Cosmic Origin Spectrograph (HST/COS). We develop a novel deep learning approach using quasar spectra at low redshift ($z<1$) where the spectra are least affected by the $\lya$ forest and predict quasar continua at higher redshift ($2<z\leq 5$). This method allows us to compute $\tau_{\text{eff}}$ by predicting the true quasar continuum at high redshift over the range $2<z_{\text{QSO}}\leq 5$ directly, and to test the performance of our model by using a testing data set that is never used during the training phase on the quasar continuum model.

This paper is organized as follows: in Section \ref{sec:description}, we present the quasar spectra used in this work from Hubble Spectroscopic Legacy Archive and Sloan Digital Sky Survey, respectively. In Section \ref{sec:methods}, we construct our deep neural network, iQNet, and apply a standardization process and present the use of principal component analysis. In Section \ref{sec:results}, we compute the absolute fractional flux error and use it as a goodness of fit metric to evaluate the model performance among iQNet and PCA prediction models. We measure the evolution of the mean transmitted flux in the $\lya$ forest with redshift and compare our findings with literature measurements in Section \ref{subsec:comparison_literature}. Throughout this work, we adopted a $\Lambda$CDM cosmology  ($\rm{\Omega_{m}}$ = 0.286, $\rm{\Omega_{\Lambda}}$ = 0.71, $\rm{H_{0}}$ = 69.32 $\rm{kms^{-1}\; Mpc^{-1}}$, $\rm{\Omega_{b}}$ = 0.04628),  from the Wilkinson Microwave Anisotropy Probe nine-year data \citep{Hinshaw_2013}.

\section{Description of Observations}
\label{sec:description}
\subsection{Quasar Spectra from the Hubble Spectroscopic Legacy Archive (HSLA)}
We use the UV quasar spectra from the Hubble Spectroscopic Legacy Archive Data Release 2 \citep{2017cos..rept....4P} in this research. This data set contains 799 unique quasar spectra, with  mean redshift $\langle z_{\text{QSO}}\rangle=0.689$. The data set comprises of  542, 326, 197 HST/COS \citep{Green_2011} spectra observed with the G130M, G160M, G140L far-UV gratings, 43, 23, and 15 HST/COS spectra observed with the G230L, G185M, and G225M near-UV gratings, respectively. We cross-match HST/COS quasars with the Million Quasars Catalog \citep{flesch2019million} to obtain the quasar redshifts. We select the HST-COS quasars that fully cover the rest-frame wavelength range of $1020\text{\AA}\leq \lambda_{\text{rest}}\leq 1600\text{\AA}$ and have a median signal to noise ratio (S/N)$>$5 per resolution element. These selection criteria yield 63 quasars with $z_{\text{QSO}}<1$. The left panel in Figure \ref{fig:histo_z_hsla_sdss} shows the redshift distribution of these selected quasars from the HSLA. We bin each HST-COS spectrum to Nyquist sampling with three pixels per COS resolution element (FWHM $\sim 18$ km/s ). We shift each quasar spectrum to its rest frame, and re-sample it on a uniform wavelength grid of 0.05\text{\AA} per pixel, which is approximately the mean resolution of those 63 selected quasar spectra. We compute the relative flux of each quasar spectrum by dividing the flux by the average flux of 40 pixels centered at 1280\text{\AA}. Throughout the paper, we will use this relative flux to predict the shape of the quasar continuum.

To create the ground truth for a deep learning interface, we first need to identify the continuum for each of these $z<1$ quasars. We perform this task using the interactive continuum fitting tool, Lintools API\footnote{https://linetools.readthedocs.io}. We fit a quasar continuum for each quasar spectrum by masking out all the absorption lines and locally fitting a spline curve to trace the strong emission lines of OVI, $\lya$, SiIV, NV, and CIV. We also mask out the Damped $\lya$ absorption (DLA) associated with the Milky-Way galaxy and the geocoronal $\lya$ emission in the selected spectra when performing a manual fitting. We use these quasar continua as the ground truth and they are used to test the robustness of our fitting procedures. We randomly select 85\% of the 63 fitted continua and their corresponding quasar spectra to form the training data set. We treat the remaining 15\% of the sample as the testing data, which are not used in the training phase of neural network and only reserved for performance evaluation of the researched models.

\subsection{Quasar Spectra from Sloan Digital Sky Survey (SDSS)}
The Sloan Digital Sky Survey Data Release 16 contains optical spectroscopy of more than 750,000 unique quasars \citep{ahumada2019sixteenth,sdss}. We proceed to use these observations to predict the quasar continua at different redshifts based on the continua obtained from the low redshift HSLA data. To that end, we select all quasar spectra, approximately 5,000 spectra in total, through the SDSS SkyServer\footnote{http://skyserver.sdss.org/dr16/en/tools/search/sql.aspx} that satisfy the following criteria:
\begin{itemize}
    \item Median signal-to-noise ratio S/N$>$5 per pixel
    \item QSO redshifts $2.0 \leq z_{\text{QSO}}\leq 5.0$
    \item Minimum rest-frame wavelength $\lambda_{\min} \leq  1080\text{\AA}$
\end{itemize}

\begin{figure}
	\includegraphics[width=\columnwidth]{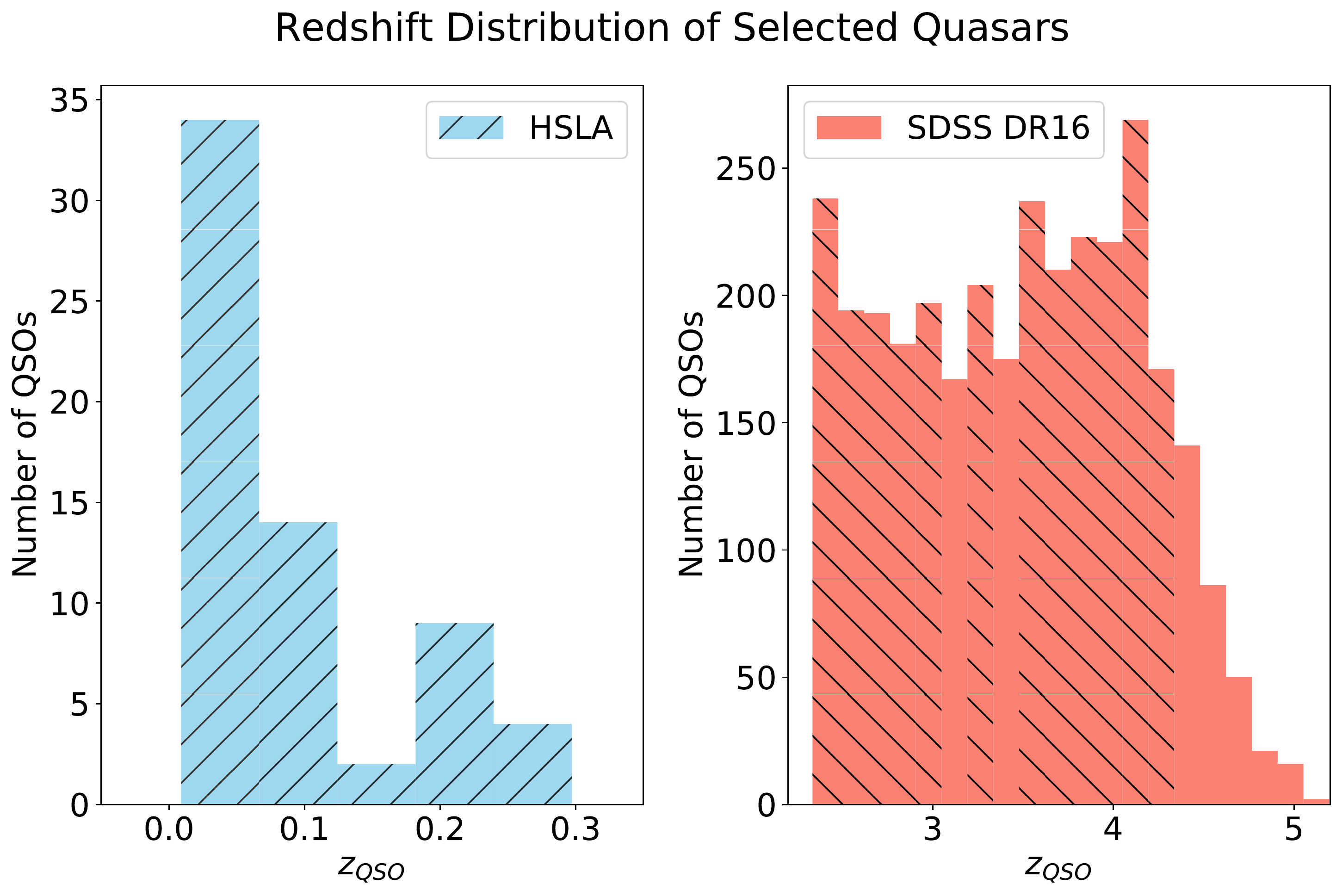}
    \caption{Redshift distributions of the selected quasars from HSLA (blue) and SDSS DR16 (red), respectively. We utilze 63 quasars from HSLA and 3,196 quasars from SDSS DR16 to train and test the performance of iQNet.}
    \label{fig:histo_z_hsla_sdss}
\end{figure}

We visually verify the downloaded quasar spectra and further discard spectra that are misclassified as quasars or have a bad $z_{\text{QSO}}$. This leaves us a final sample of 3196 spectra. The right panel of Figure \ref{fig:histo_z_hsla_sdss} shows the redshift distribution of these quasars.The quasars continuously cover the redshift range between $z_{\text{QSO}}=2.3$ and $z_{\text{QSO}}=5.1$ with the mean value of $\langle z_{\text{QSO}}\rangle=3.5$. We will use these SDSS quasar spectra to study the evolution of the mean effective optical depth in Section \ref{subsec:mean_flux}.

We transform the quasar spectra into their rest frame and re-sample them onto a uniform wavelength grid. Each spectrum flux is then re-normalized with the mean flux at $\sim$ 1280 \text{\AA}.  We are interested in predicting the rest frame quasar continuum within the wavelength range [1020, 1600]\text{\AA}, therefore for all the SDSS spectra, only the flux in that wavelength range is retained for further analysis.

\section{Methods}

\label{sec:methods}
\subsection{intelligent Quasar Continuum Neural Network (iQNet)}
Deep neural networks have a proven track record for image restoration and image synthesis \citep{Zhou_1988}. In this work, we show how deep neural networks can be applied to successfully predict quasar continuum at any redshift. We treat a 1-D rest-frame UV quasar spectrum as a 1-D image array. Each input quasar spectrum to this  network is a superposition of the true quasar continuum, random noise owing to finite signal-to-noise ratio, and intervening HI and metal absorption line systems that are imprinted on the quasar spectra along the line of sight. Instead of building a more advanced architecture (i.e., the Variational Autoencoder (VAE) and the Generative Adversarial Network (GAN)), we start with the simpler autoencoder (AE) architecture. Occam's razor in deep learning suggests that it is not necessary to apply a complicated network structure if the simpler network is able to solve the problem. The inspiration for our deep neural network architecture is analogous to the idea of the stacked denoising autoencoder \citep[e.g.][]{Vincent_2010}. However, the structure shapes of the encoder and decoder are not symmetric to the coding layer. Our deep neural network, iQNet, will take the part of the quasar spectrum redward of the  $\lya$ line ([1216, 1600]\text{\AA}) as input and generate the whole ([1020, 1600]\text{\AA}) quasar continuum as output.

The architecture of our intelligent quasar Continuum Neural Network, (iQNet) is shown in Table~\ref{tab:iQNet}. The iQNet is compiled and trained using the Keras\footnote{https://keras.io/} with Tensorflow\footnote{https://www.tensorflow.org/} backend, \citep[e.g.][]{keras,tensorflow}. To utilize the power of denoising autoencoder, we treat the quasar spectra redward of $\lya$ emission as a superposition of ground truth continua, random noise, and intervening absorption line systems imprinted on the spectra. The objective of the neural network is to discover the underlying ground truth continua, despite the presence of noise and absorption lines. The iQNet network takes a 1-D quasar spectrum in the rest-frame redward wavelength range $1216 \text{\AA} \leq \lambda_{rest} \leq 1600 \text{\AA}$ as input and generates the corresponding quasar continuum in the full wavelength range of interest ($1020 \text{\AA} \leq \lambda_{rest} \leq 1600 \text{\AA}$). The loss function to optimize our neural network is binary cross-entropy with Adam optimization \citep{kingma2014adam} and a training batch size of 512 quasar spectra redward of $\lya$ emission from the training set. The order of training spectra is also randomly shuffled every epoch (training cycle) before they are fed into our neural network to reduce the training bias. To quantify the performance of our neural network model and the goodness of fit, we use the absolute fractional flux error (AFFE) to evaluate our model to make sure that the weights of the iQNet are updated to achieve a lower AFFE and reach to the minimum of the loss function at the same time. We define the absolute fractional flux error, $|\delta F|$, as
\begin{equation}
    |\delta F|=\int^{\lambda_2}_{\lambda_1}\left|\frac{F_{\text{pred}}(\lambda)-F_{\text{true}}(\lambda)}{F_{\text{true}}(\lambda)}\right| d\lambda \left/\int^{\lambda_2}_{\lambda_1}d\lambda\right.
    \label{eqn:affe}
\end{equation}
where $F_{\text{pred}}$ is the predicted continuum and $F_{\text{true}}$ is the true continuum. An Early Stopping \citep{Prechelt1996EarlySW} strategy is also used to terminate the training process if there is no weight update available. We also add a median filter of the size of 50 pixels to smooth out the output continuum. The size of the median filter is a hyper-parameter and we have tested that a median filter of 50 pixels provides the lowest AFFE among filter size range from 0 to 100 pixels with a 10-pixel increment.

\begin{table}
	\centering
	\caption{intelligent Quasar Continuum Neural Network (iQNet) Architecture. The FC layers represent neural layers with all fully-connected neurons. The ELU activation function represents the exponential linear unit function.}
	\label{tab:iQNet}
	\begin{tabular}{ccc} % four columns, alignment for each
		\hline
		Layer Type & Number of Neurons & Activation Function\\
		\hline
		Input Layer & 7680 & N/A\\
		FC Layer & 1024 & ELU\\
		FC Layer & 512 & ELU\\
		FC Layer & 256 & ELU\\
		FC Layer & 256 & ELU\\
		FC Layer & 512 & ELU\\
		FC Layer & 1024 & ELU\\
		FC Layer & 2048 & ELU\\
		Output Layer & 11600 & Sigmoid\\
		\hline
	\end{tabular}
\end{table}

\subsection{Standardization Process}
\label{subsec:standardization}
To avoid model over-fitting and reduce model bias, we add a pre-processing step prior to model training called standardization. The standardization process is defined as

\begin{equation}
    f_{\text{input}}=\frac{f - f_{\min}}{f_{\max}-f_{\min}}
\end{equation}
where $f_{\text{input}}$ is the scaled flux used as model input after the standardization process, $f$ is the original relative flux, $f_{\max}$ is the maximum value of quasar flux in the training spectra, and $f_{\min}$ is the minimum flux in the training spectra. This process applies to the individual relative flux of each quasar spectrum or continuum in both the training and testing data sets. After this transformation, all quasar scaled fluxes should be within the range [0, 1]. We apply this transformation to scale down the dominant $\lya$ and CIV emissions and to un-correlate the emission features in quasar spectra.

In addition, we define the corresponding inverse transformation of standardization as
\begin{equation}
    f_{\text{pred}} = f_{\text{output}}\cdot (f_{\max}-f_{\min}) + f_{\min}
\end{equation}
where $f_{\text{output}}$ is the scaled flux output from the prediction model and $f_{\text{pred}}$ is the predicted relative flux after the inverse transformation of this standardization process. Hence, the standardization process and its inverse transformation are internal processes added to the prediction model and this process removes the quasar intrinsic luminosity for quasar study.

\subsection{Principal Component Analysis (PCA)}
\label{subsec: pca}
We have only a limited number of training quasar spectra available from HSLA (53 quasars) that meet our wavelength coverage and S/N cutoff requirements. Therefore to increase the number of training spectra, we need to synthesize mock 1-D quasar spectra using the selected training continua.

The training spectra cover [1020, 1600]\text{\AA} with 0.05\text{\AA} per pixel; i.e., there are 11,600 pixels for each quasar spectrum). To reduce the computational complexity, we apply the Principal Component Analysis\footnote{https://scikit-learn.org/stable/modules/generated/sklearn.decomposition.\\PCA.html} by \citet{sklearn_api} (PCA) as a dimensionality reduction tool to reduce the 11,600 data points on each quasar spectrum to a small number of principal components that can reconstruct the spectral flux of the training HSLA quasar continua.

\begin{figure}
    \centering
    \includegraphics[width=\columnwidth]{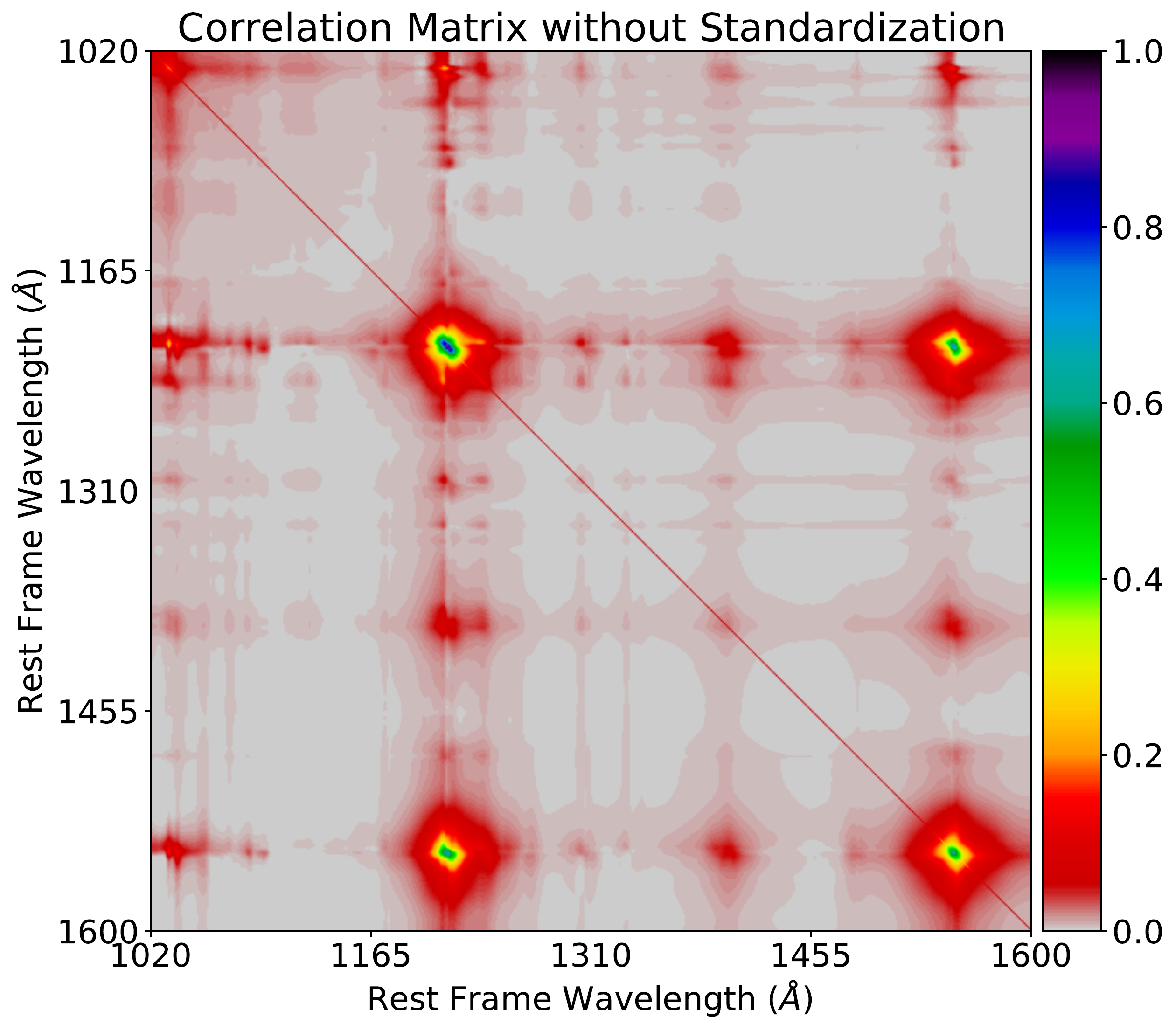}
	\includegraphics[width=\columnwidth]{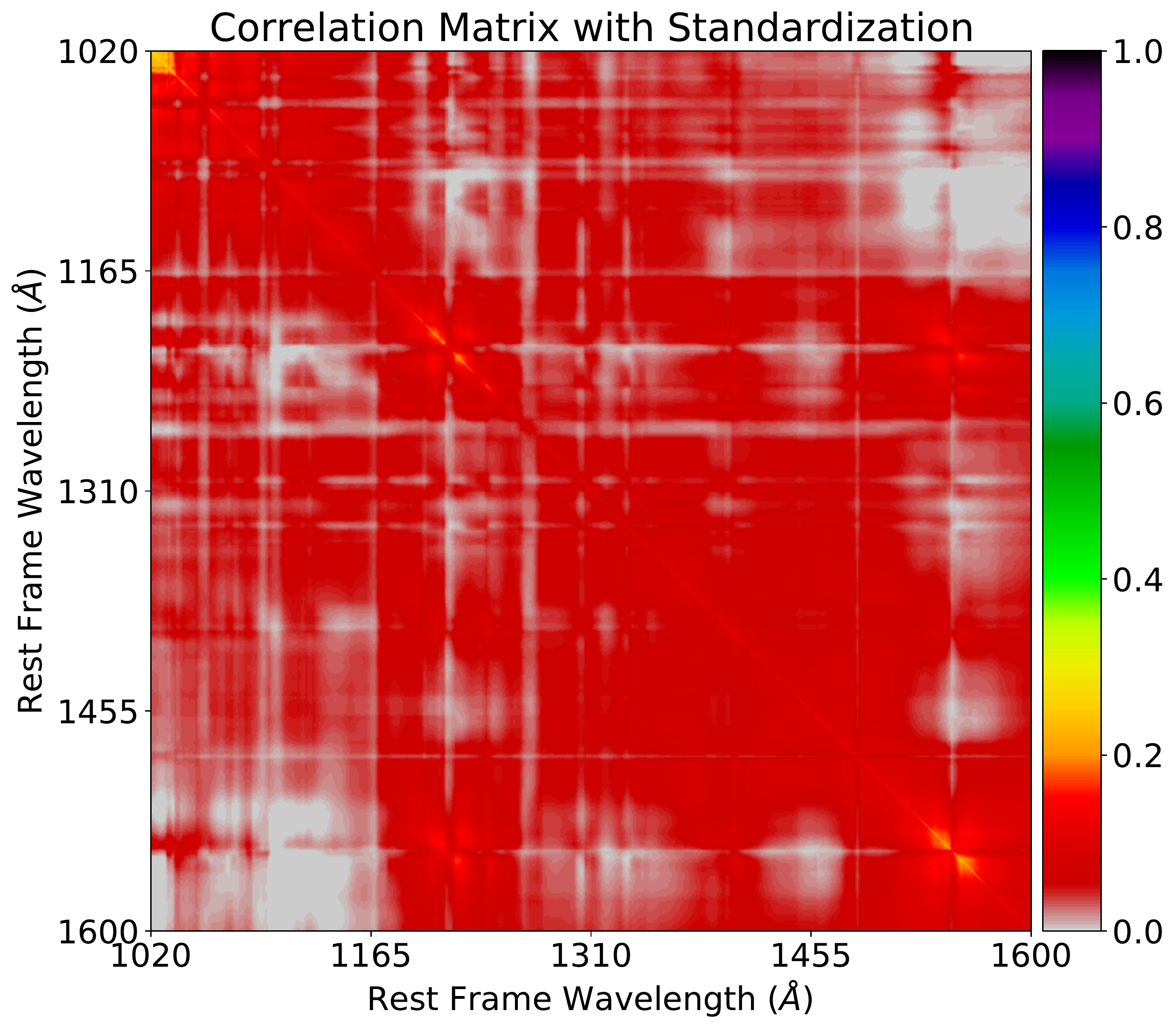}
    \caption{Difference between correlation matrices with and without standardization. Top panel shows strong correlations at $\lya$ and CIV transitions indicating that these two features are positively correlated and introduce bias to the PCA model without standardization. Bottom panel illustrates no strong correlation shows up with the addition of standardization process prior to PCA construction, indicating that the PCA model with standardization is not biased from the dominant $\lya$ and CIV features. Note that the diagonal elements of the correlation matrix with standardization are unity but not visible due to its large size ($11600\times 11600$.) The color bar in both panels illustrates the correlation coefficients within the rest-frame wavelength range $1020\text{\AA}\leq\lambda_{\text{rest}}\leq 1600\text{\AA}$}
    \label{fig:correlation}
\end{figure}

\begin{figure}
	\includegraphics[width=\columnwidth]{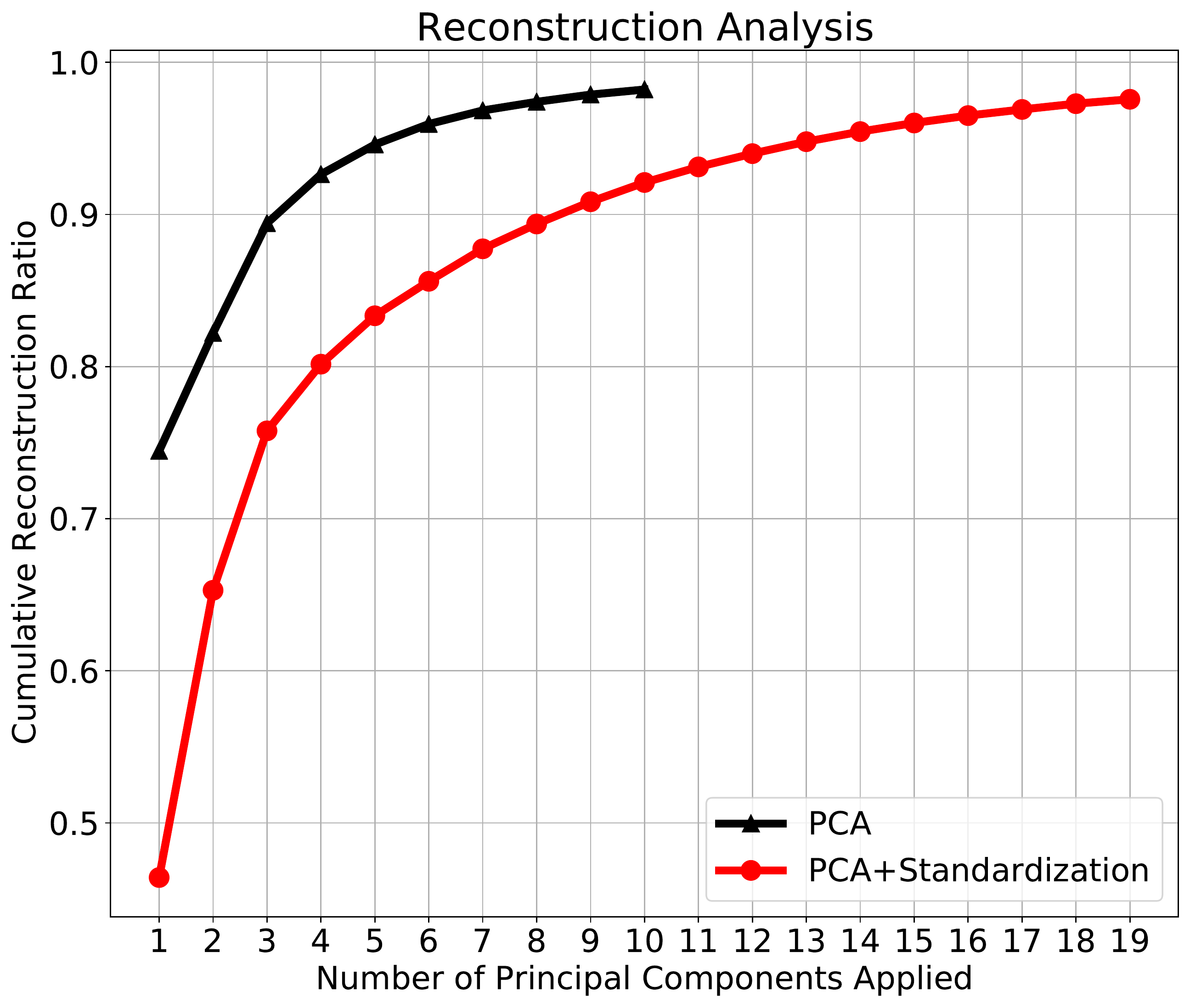}
    \caption{PCA reconstruction ratio is also known as the explained variance ratio. 
    The cumulative reconstruction ratio tells how much percentage the PCA recovers the original training continua with an increasing number of principal components up to 97.5\% of original training quasar continua. The PCA model without standardization requires 10 principal components because it is biased by the dominant $\lya$ and CIV correlated features. The PCA model with the standardization process needs 19 principal components because the standardization properly scales down and un-correlates the $\lya$ and CIV features.}
    \label{fig:pca_cum_evr}
\end{figure}

We perform two sets of PCA decomposition, both with and without the standardization process described in Section \ref{subsec:standardization}. After the PCA model construction, we are able to compute the correlation matrix using PCA models to demonstrate the data correlation. The top panel of Figure \ref{fig:correlation} shows that there are strong correlations at $\lya$ and CIV emissions. In other words, this PCA model is biased by the strong $\lya$ and CIV emissions. For the PCA model without the standardization process, Figure \ref{fig:pca_cum_evr} illustrates that it only requires 10 principal components to reconstruct 97.5\% of the training spectra because it captures the dominant emission features and ignores other weak features to represent a quasar continuum.

The PCA model with the added standardization process requires 19 principal components to reconstruct 97.5\% of the training quasar continua. The bottom panel in Figure \ref{fig:correlation} shows that there is no correlation coefficient greater than 0.4. This illustrates that all features in the training data set have been uncorrelated prior to PCA model construction by the standardization process. Figure \ref{fig:pca_cum_evr} demonstrates the cumulative reconstruction ratio with the number of principal components for the PCA model with the added standardization process. The first 10 principal components of this PCA model can reconstruct $\sim$92\% of the original quasar continua. To achieve our goal of 97.5\% reconstruction of the original spectra, we use 19 PCA components.

A rule of thumb of using PCA is to set explained variance ratio is at least 97.5\%. However, different authors have used different criteria to set this threshold. Recently \cite{Durov_kov__2020} used 99\% explained variance for their PCA analysis and found an increase in accuracy of their model when more than 60 principal components were used. We test the impact of adopting different explained variance ratios in our PCA models as follows. We first test two PCA models without standardization, one having 97.5\% explained variance and requiring 9 principal components and the other having 99\% explained variance and requiring 14 principal components. We further test two PCA models with standardization, one having 97.5\% explained variance and requiring 19 principal components and the other having 99\% explained variance and requiring 27 principal components. In all cases, even though the mean AFFEs of the PCA reconstruction models improve by $\sim$ 2\% when we switch the explained variance ratio from 97.5\% to 99\%, the mean AFFEs of the training set and the testing set of the PCA prediction models only differ by approximately 0.5\% and 0.2\%, respectively. Because the additional eight principal components for PCA with standardization and the extra five principal components for PCA without standardization only improve the mean testing AFFE by 0.2\%, we decide to set the explained variance ratio to 97.5\% for our PCA models. In this paper, we only compare quasar continuum predictions with results from \citet{Suzuki_2005} and \citet{Paris_2011} because of identical rest frame wavelength range of interest. However, neither \citet{Suzuki_2005} nor \citet{Paris_2011} further solved and scaled down the strong $\lya$ and CIV correlations between the reconstruction ratios of the weights of those two sets of principal components.

Our PCA models predict the rest-frame UV quasar continuum in the wavelength range $1020 \text{\AA} \leq \lambda_{\text{rest}} \leq 1600 \text{\AA}$. Even though this PCA+Standardization model needs the additional 9 components; we demonstrate in Figure \ref{fig:correlation} and Table \ref{tab:affe_pca} that the PCA model without standardization is biased by $\lya$ and CIV emissions, as those two emission lines are the dominant emission features in a quasar spectrum.
 In addition, Figure \ref{fig:bias_uncer_allmodels} in the Appendix \ref{appendix: bias_uncer} confirms that the standardization process helps the PCA prediction models reduce the fractional bias and uncertainty in wavelength region [1280, 1500]\text{\AA} where there are no strong emission features. Table \ref{tab:affe_blue_red} also shows that the standardization process helps the PCA prediction model to reduce the AFFE by half for the predicted continua redward of $\lya$ emission while keeping the AFFE of the blueward continua the same. We prefer the PCA model with the addition of the standardization process because the AFFE (see Section \ref{subsec:hst_prediction}) for a PCA model with standardization is much lower than that of the PCA model without standardization. A lower AFFE indicates that the PCA reconstruction is closer to the original quasar spectra. Therefore, the standardization process is necessary for PCA construction.

We note that, while we apply the Principal Component Analysis to reduce the computational complexity, others have proved that PCA can be used to predict quasar continua as a prediction model. \citet{Suzuki_2005}, \citet{Paris_2011} obtain such PCA prediction models by constructing two PCA models, one covering the rest-frame wavelength range ([1020, 1600]\text{\AA}) and the other covering the part of the spectra redward of the $\lya$ line ([1216, 1600]\text{\AA}). They further use a transformation matrix, which converts the weights of principal components in the eigenspace covering the continuum redward of $\lya$ into the weights in the eigenspace that covers the whole continuum. Both of these approaches do not utilize the standardization approach introduced here. We follow the same procedure discussed in \citet{Suzuki_2005} and compare the results in Section \ref{sec:results} to demonstrate the limited generalization ability of the PCA prediction model.

\subsection{Generation of Synthetic Quasar Spectra}
To augment our training set, we proceed to create a large number of synthetic quasar spectra as described below. To create a representative training set, we first have to identify how many different classes of quasar continua are present in the training set. To estimate this, we apply the Gaussian mixture model\footnote{https://scikit-learn.org/stable/modules/generated/sklearn.mixture.\\GaussianMixture.html} (GMM, \citealt{reynolds2009gaussian}) clustering algorithm to classify the training quasar continua into different classes. To reduce the computational complexity, we apply the PCA with standardization and use 19 principal components as the input to the GMM model instead of the original quasar spectra.

\begin{figure}
    \centering
    \includegraphics[width=\columnwidth]{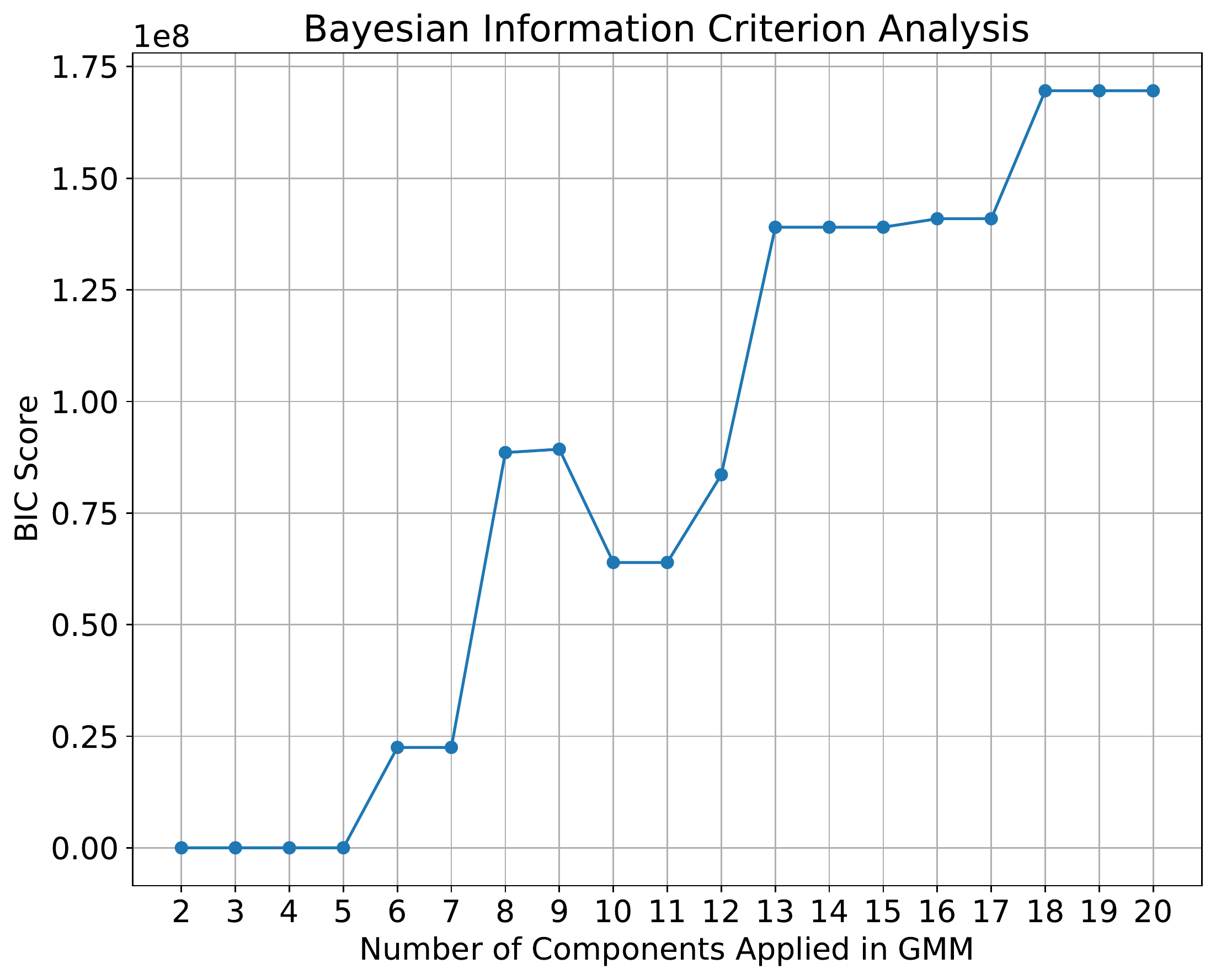}
	\includegraphics[width=\columnwidth]{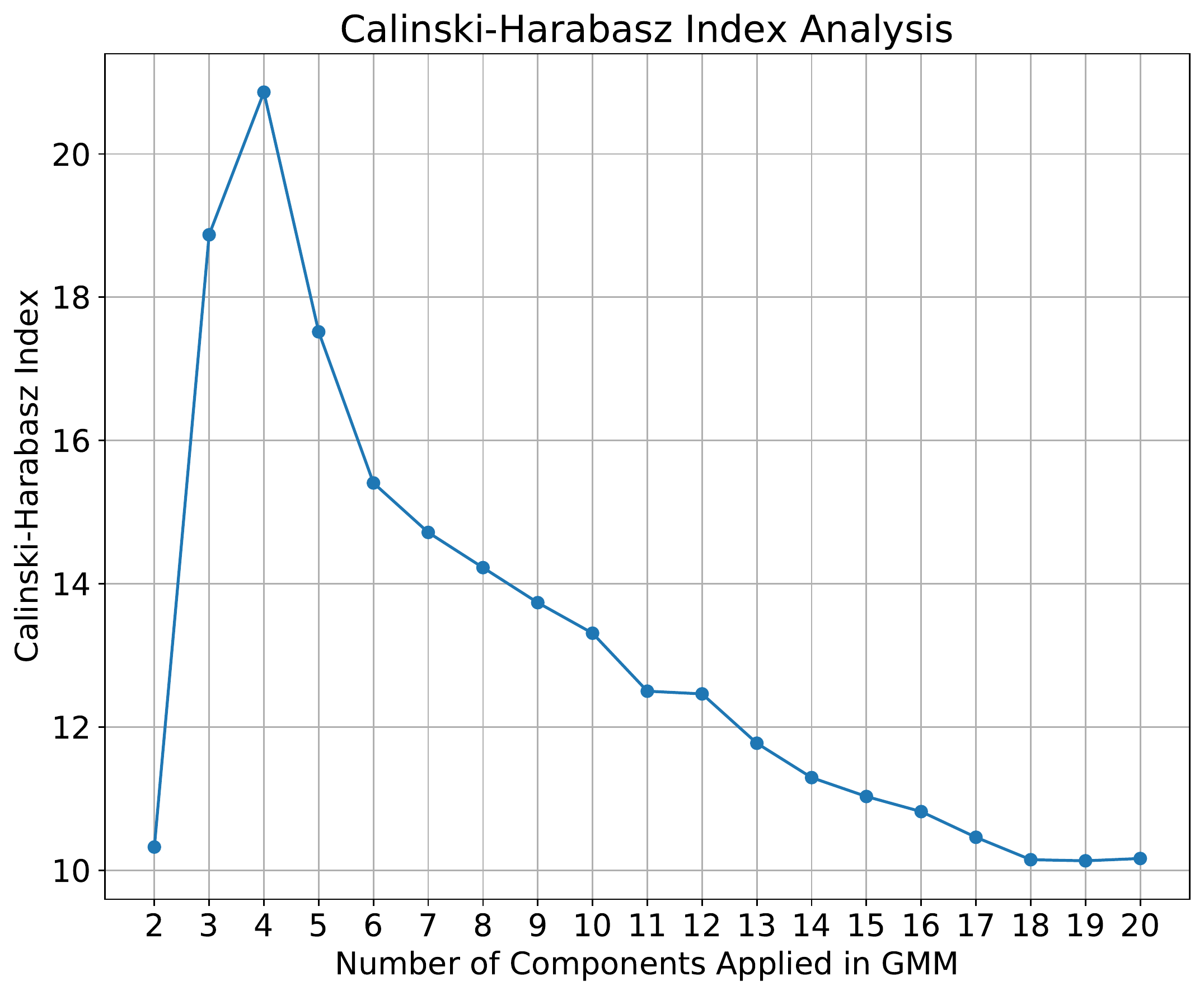}
    \caption{Bayesian Information Criterion (BIC) Scores and Calinski-Harabasz Index Scores from 2 Classes to 20 Classes. The numbers of quasar classes, 2, 3, 4, and 5, all provide the lowest BIC scores, indicating that those numbers of classes are representative numbers in the GMM clustering analysis. Four quasar classes in the Calinski-Harabasz Index Analysis illustrate the highest score, indicating that four classes are the best number of quasar classes pre-set to the GMM model.}
    \label{fig:bic_chindex}
\end{figure}

To determine the correct number of quasar classifications of the training data set and quantify the goodness of clustering, we construct 19 GMM models with the pre-defined class numbers ranging from 2 classes to 20 classes and compute the Bayesian Information Criterion  (BIC) \citep{schwarz197801} and the Calinski-Harabasz Index \citep{calinski1974}, as shown in Figure \ref{fig:bic_chindex}. Both Bayesian Information Criterion and Calinski-Harabasz Index are common metrics used for model selection in the clustering algorithm, like GMM, to determine the best model among the available candidates in machine learning. A lower BIC score indicates a better clustering model. Therefore, the GMM results with 2, 3, 4, and 5 classes provide the same clustering goodness given the training data set based on the BIC score, and we need an additional metric, the Calinski-Harabasz Index, to further determine the best clustering number among 2,3,4, and 5 classes. We are able to apply the Calinski-Harabasz Index to the training data because the correct number of quasar classes is unknown. The higher the Calinski-Harabasz Index score, the better the GMM model predicts the clusters given a pre-selected number of clusters. The bottom panel in Figure \ref{fig:bic_chindex} shows the highest Calinski-Harabasz Index is at 4-class clustering GMM model. Thus, 4 classes provide the best clustering result based on the GMM clustering algorithm. 

\begin{figure*}
	\includegraphics[width=0.8\textwidth, height = 9cm]{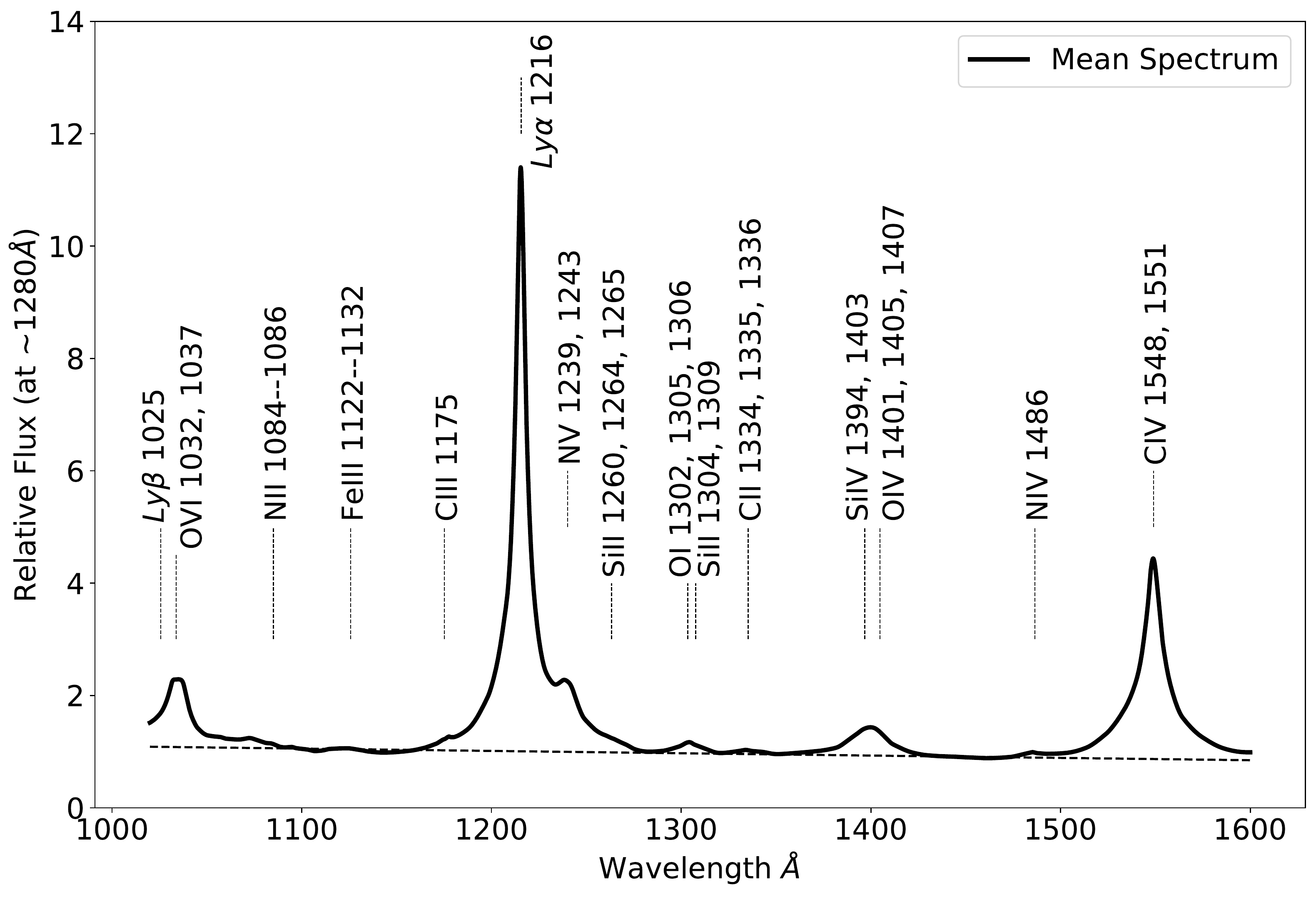}
	\includegraphics[width=0.8\textwidth, height =9cm]{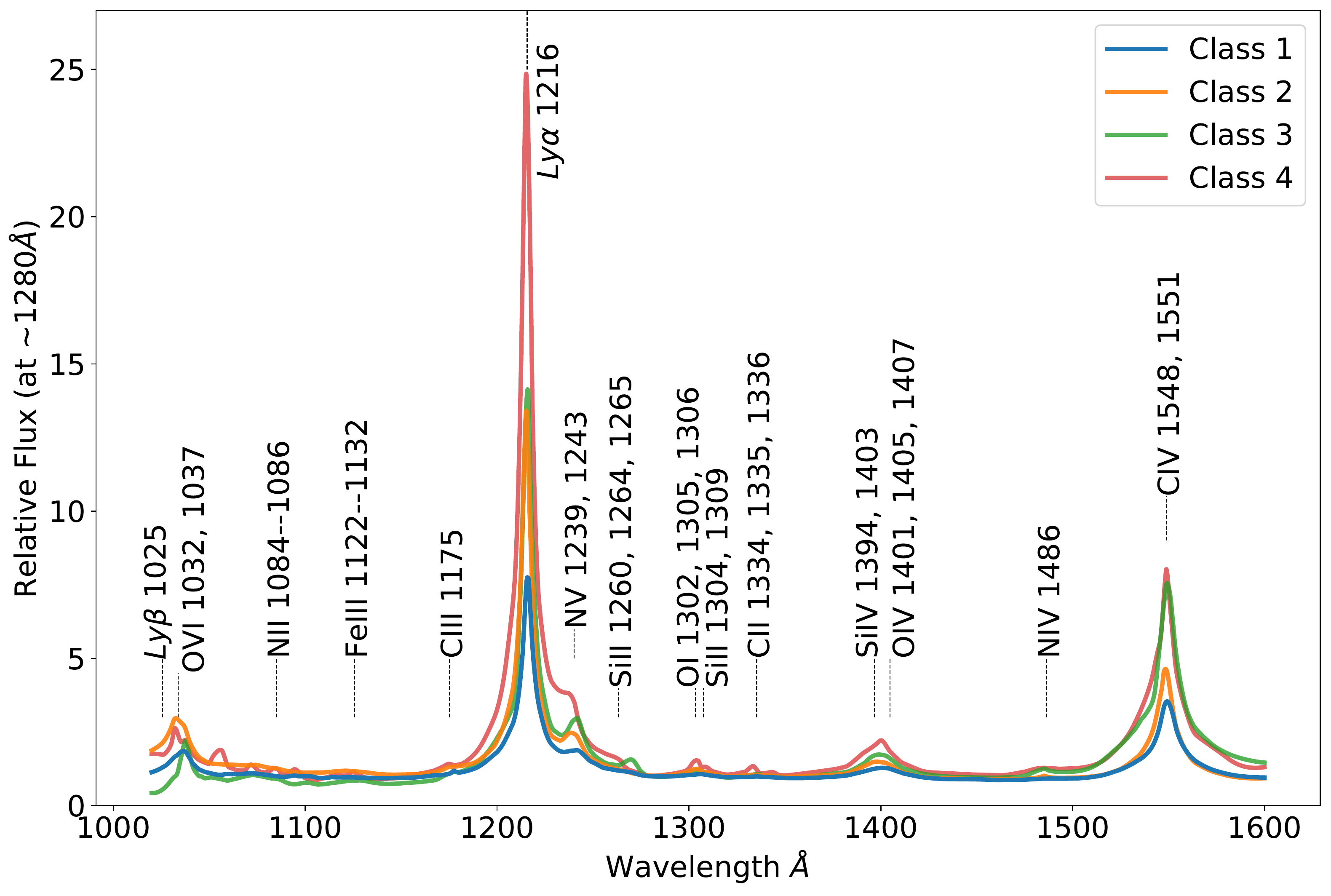}
    \caption{ \textit{Top Panel:} Mean quasar spectrum obtained from the 63 ($z \sim 0.2$)  quasars observed with high resolution HST/COS spectroscopy covering the full rest-frame wavelength range of $1020\text{\AA} \leq \lambda_{\rm{rest}} \leq 1600 \text{\AA}$. Prominent emission features are marked with vertical dashed lines. The spectrum is normalized relative to 1280 \text{\AA}. \textit{Bottom Panel:} Mean spectra of four quasar classes showing how the mean spectra from the top panel can be decomposed into four distinct classes. We generate these mean quasar spectra by classifying the PCA eigenvalues into four distinct classes using a Gaussian Mixture Model. The primary difference between different classes are the $\lya$ and CIV emission strengths. The slope of the continua also show small change blueward of $\lya$ emission between different classes. The vertical dotted lines mark the position of prominent emission features in the quasar continua. All the spectra are normalized relative to 1280\text{\AA}. The lines shown in both panel are obtained from \citet{Shull_2012_linelist}.}
    \label{fig:gemm_spec}
\end{figure*}

The top panel of Figure \ref{fig:gemm_spec} shows the mean quasar spectrum obtained by averaging the 63 HSLA quasar spectra selected for this work. All these spectra completely cover the wavelength range of $1020\textup{\AA} \leq \lambda_{\rm{rest}} \leq 1600 \textup{\AA}$. Prominent emission features, such as $\lya$, CIV, Ly$-\beta$, OVI, and SiIV, are observed and marked with vertical dotted lines. The horizontal dashed line shows the approximate power law continuum of this composite quasar spectrum. The four classes of spectra identified are essentially a decomposition of this mean quasar spectrum. Figure \ref{fig:gemm_spec}, bottom presents the mean quasar spectra from all four classes. The primary difference among different quasar classes is particularly seen in the variation of $\lya$ and CIV emission strengths. Additionally, we see secondary differences as small variations in the slope of the continua blueward of $\lya$ emission. In addition to those four types of quasar spectra, we add the fifth quasar class with a constant relative flux of 1 representing a BL-Lac like AGN spectra, which is not well represented in our training spectra.

We use these five classes of quasar spectra to generate mock spectra to augment our training set. The mock quasar spectra are synthesized with the following steps:
\begin{itemize}
    \item randomly selecting a quasar continuum from one of the five quasar classes,
    \item shifting the quasar to a higher redshift ($z_{QSO}$),
    \item injecting Voigt profiles of a random column density of strong ISM and HI absorption lines for the Milky Way and at higher random absorber redshift ($z_{abs} \leq z_{QSO}$),
    \item adding  Gaussian noise to the spectra with S/N = 5, 10, and 20.
\end{itemize}

We maintain the same number of mock quasar spectra in each class such that the neural network is not biased on a specific class of quasar spectra during the training phase. We finally have $\sim$ 12,000 training spectra, including real HSLA quasar spectra and synthesized mock spectra. We train the iQNet model on these 12,000 mock + real HSLA quasar spectra in the Google Colab\footnote{colab.research.google.com} environment with GPU acceleration enabled for the maximum time of 40 min or an EarlyStopping condition is satisfied.

\subsection{Mean Transmitted Flux Estimate}
\label{subsec:mean_flux}
We now use our iQNet predicted continua at higher redshift to estimate the evolution of mean transmitted flux in the $\lya$ forest with redshift. The mean transmitted flux in the $\lya$ forest can be defined as follows:

If $F_{\text{spec}}$ and $F_{\text{cont}}$ are the relative flux (normalized to average flux at 1280 {\AA}) and the corresponding continuum as a function of the observed wavelength \citep[e.g.][]{Faucher_Gigu_re_2008}, respectively, then the corresponding redshift of $\lya$ transition along the line of sight is  $z_{\lya}=\lambda/\lambda_{\lya}-1$, where $\lambda_{\lya}=1215.67$\text{\AA}. The transmitted flux $F(z_{\lya})$ as a function of $\lya$ redshift $z_{\lya}$ is:

\begin{equation}
    F(z_{\lya})=\frac{F_{\text{spec}}(z_{\lya})}{F_{\text{cont}}({z_{\lya}})}.
    \label{eq:transmitted_flux_detail}
\end{equation}

The mean transmitted flux $\langle F(z_{\lya})\rangle$ is the ensemble average of $F(z_{\lya})$ within a redshift bin, in the $\lya$ forest region.

We follow a method similar to the one outlined in \citet{Faucher_Gigu_re_2008} to compute $ F(z_{\lya})$, and we summarize below:
\begin{itemize}
    \item Extract the quasar spectrum and its corresponding predicted continuum in the $\lya$ forest region and calculate its transmitted flux $F(z_{\lya})$ using Equation \ref{eq:transmitted_flux_detail}.
    \item  Mask out the flux regions near the quasar proximity zone. In this work, we conservatively mask out any pixels that are within 25 proper Mpc of the quasar redshift. The typical size of HI-ionizing radiation of a quasar is $\lesssim$ 12 proper Mpc \citep{Davies_2019,Eilers2020}, thus there should be little contamination left after we mask out the 25 proper Mpc.
    \item Re-sample all pixels in the $\lya$ forest regions in 3 proper Mpc intervals.  We bin all pixels in 3 proper Mpc intervals to un-correlate the values of the optical depth in the $\lya$ forest region and make them independent of their neighbors. \citep{Faucher_Gigu_re_2008}.
    \item Divide the $\lya$ forest region into 30 redshift bins with $\Delta z_{\lya} = 0.1$.
    \item Within each redshift bin, we define the mean transmitted flux ($\langle F(z_{\lya})\rangle$) as the 3-sigma-clipped mean of the distribution, and convert into the effective optical depth ($\tau_{\text{eff}}=-\ln\langle F\rangle$). Sigma clipping is used to remove extreme absorption troughs like Dampled $\lya$ absorbers (DLAs), any bad pixels, ultra strong metal lines at low redshifts, or other contamination (sky residuals) that might be present in each bin. This process does not bias or over-correct for metal line contamination at higher redshifts as very few pixels are clipped in that range.
    \item We estimate uncertainty, $\sigma_{F(z_{\lya})}$, on $\langle F(z_{\lya})\rangle$ estimates using a bootstrap with replacement approach. In each redshift bin, we re-sample the $F(z_{\lya})$ values with replacement 200 times. In each iteration, we compute a sigma clipped mean ($\langle F(z_{\lya})\rangle$) as described in the previous step. We measure the final $\langle F(z_{\lya})\rangle$ as the mean of these 200 bootstrapped estimates. We estimate the uncertainty on $\langle F(z_{\lya})\rangle$ as the 16th and 84th percentile of the bootstrapped mean estimates, which should account for sample variance, bad pixel contamination, and metal line contamination uncertainties. In addition, since deterministic neural networks do not predict uncertainties, we follow the following method to estimate and propagate the continuum uncertainties. We use 2400 mock testing quasar spectra with different S/N, metal absorption lines, and redshifts and use the iQNet to predict their continua. For each spectra we compute the error in continuum estimation as a function of wavelength. These are rebinned in 3 proper Mpc bins similar to the previous steps. The mean error per 3 proper Mpc bin is adopted as the continuum uncertainty $\sigma_{\text{sys}}$ for that redshift bin. This is propagated to estimate the total uncertainty ($\sigma_{\text{tot}}$) on mean flux evolution as
    \begin{equation}
        \sigma_{\text{tot}}^2= \sigma_{F(z_{\lya})}^2 + \sigma_{\text{sys}}^2
    \end{equation}
    \item As the final step, we apply prescriptions to correct for metal line contamination as described below. We follow the methods of metal line contamination removal process discussed in \citet{Schaye_2003} and \citet{Kirkman_2005} and then compute the corrected effective optical depth, and compare results in the Section \ref{subsec:mean_flux}.
\end{itemize}

\section{Results}
\label{sec:results}
\subsection{Predicting Quasar Continua on $z<1$ HST-COS Spectra}
In this section we will describe the performance of both our PCA and iQNet deep learning neural network in predicting the UV continua of $z<1$ HST-COS quasars.

\label{subsec:hst_prediction}
\subsubsection{PCA Performance}

We first discuss the quasar continuum \textit{reconstruction} performance of our PCA models over $1020 \text{\AA} \leq \lambda_{\text{rest}} \leq 1600 \text{\AA}$ and compare them with the PCA reconstruction models in \citet{Suzuki_2005} and \citet{Paris_2011}, respectively. We apply the PCA reconstruction models on the quasar spectra with their full set of  corresponding principal components, in the wavelength range ($1020 \text{\AA} \leq \lambda_{\text{rest}} \leq 1600 \text{\AA}$). We stress that this method cannot be used to predict the true continuum blueward ($1020 \text{\AA} \leq \lambda_{\text{rest}} \leq 1216 \text{\AA}$) of $\lya$ emission, and discuss this in the context of PCA \textit{reconstruction} performance only. 

\begin{figure*}
	\includegraphics[width=0.9\textwidth]{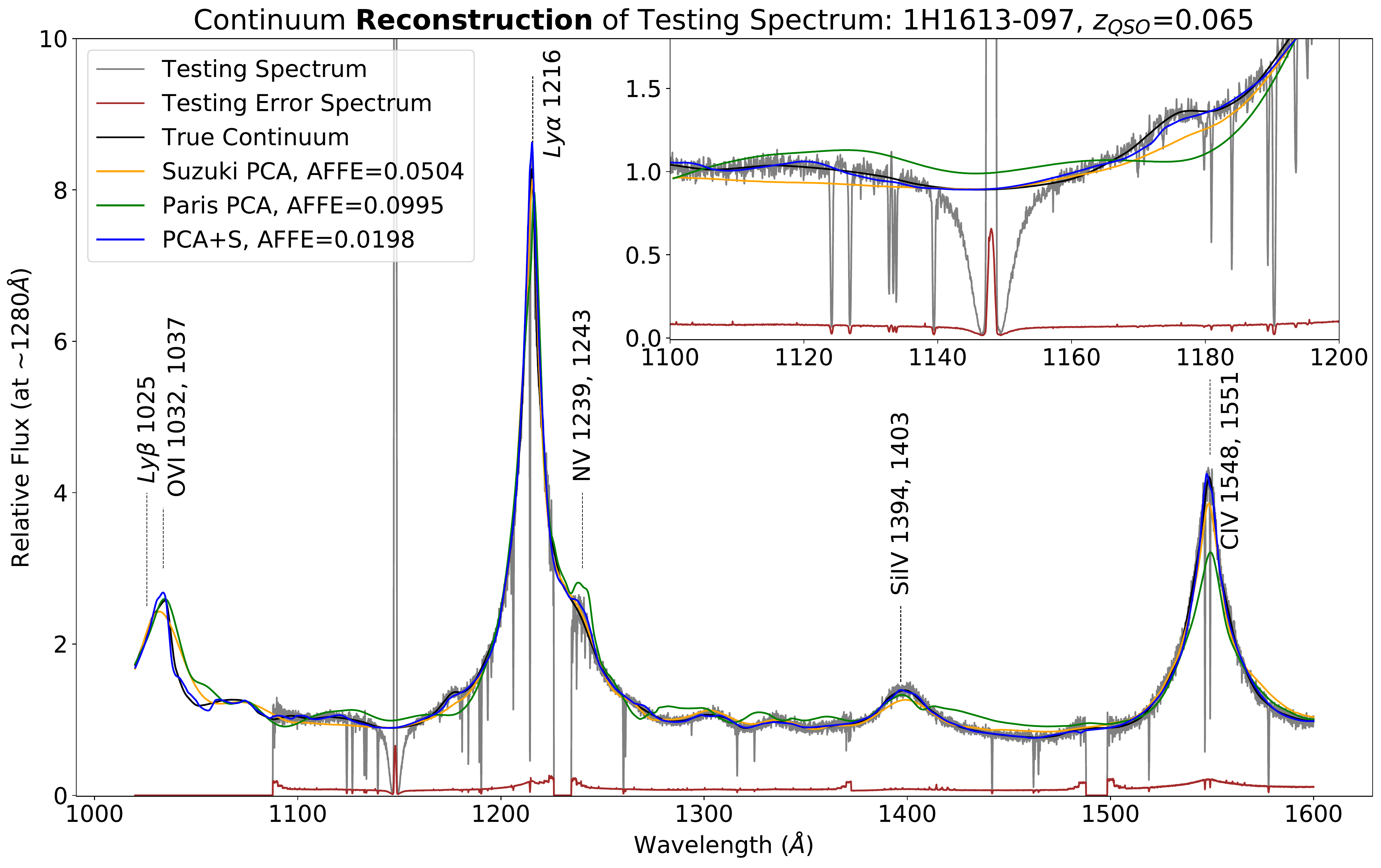}
    \caption{PCA \textit{reconstruction} model comparison. The background testing spectrum and its corresponding ground truth continuum are shown in grey and black, respectively. The reconstructed continuum by \citet{Suzuki_2005} is shown in orange and the reconstructed continuum by \citet{Paris_2011} in green. The reconstructed continua from all PCA models covering the whole wavelength of interest ([1020, 1600]\text{\AA}), and the inset in the upper right corner shows a zoomed view on the wavelength range [1100, 1200]\text{\AA} of the same quasar continuum reconstruction. Our PCA reconstructed continuum with Standardization-Normalization is shown in blue. Other works fail to reconstruct the quasar continuum and have higher absolute fractional flux error (AFFE) (shown in legend) due to the lack of standardization process. Only the strongest emission features are marked in this figure. We recommend readers to consult \citet{Shull_2012_linelist} for a more complete line-list.}
    \label{fig:pca_testing}
\end{figure*}

We gauge the performance of all these models with the testing quasar spectra, which are not used by any of the models to construct the PCA. This use of the blind testing sample was not used in neither \citet{Suzuki_2005} nor \citet{Paris_2011}, and the PCA model performance was tested on the same sample with which the original PCA models were constructed. We avoid this circularity to have a fair comparison between different models. Figure \ref{fig:pca_testing} shows a testing quasar spectrum, which is not used in the construction of PCA. Only our PCA model with the standardization process closely resembles the quasar continuum. Other PCA prediction models manage to predict the $\lya$ emission peak but cannot accurately predict the continuum in the regions with no emission lines. This demonstrates the reason why the standardization process is necessary to properly scale down the dominant features like $\lya$ and CIV.

We quantify the PCA performance in reconstructing the quasar continua with the absolute fractional flux error (AFFE) as defined in equation \ref{eqn:affe} for the full testing sample and summarize the results in Table~\ref{tab:affe_pca} for PCA reconstruction models. It is evident that the PCA model with the standardization process obtains the lowest AFFE and has two times lower AFFE as compared to \citet{Suzuki_2005} and four times lower than \citet{Paris_2011} models. This is mainly because we implement the standardization process. The standardization process is a necessary step as PCA methods are sensitive to dominant features in the data set and can easily \textit{overfit} the data if the dominant features are not properly scaled.

\begin{table}
	\centering
	PCA \textit{Reconstruction} Model
    \begin{tabular}{lcccr} 
		\hline
		Model Name & \multicolumn{2}{c}{Training Set} & \multicolumn{2}{c}{Testing Set}\\
		 & Median & Mean & Median & Mean\\
		\hline
		Suzuki PCA$^{(a)}$ & N/A & N/A & 0.0593 & 0.0685\\
		Paris PCA$^{(a)}$ & N/A & N/A & 0.1180 & 0.1130\\
	    PCA (This work) & 0.0668 & 0.0869 & 0.0663 & 0.0700\\
		PCA+S$^{(b)}$ (This work) & 0.0323 & 0.0352 & 0.0345 & 0.0364\\
		\hline
	\end{tabular}
	\caption{Absolute Fractional Flux Error (AFFE) on reconstructed continua of $z\sim 0.2$ HST-COS training and testing 1-D spectra.
	$^{(a)}$AFFEs are not available for \citet{Suzuki_2005} and \citet{Paris_2011} as the training set is used to construct these PCA models. The performances of these PCA-based models are evaluated only on the testing quasar spectra since they all have not used the testing data. $^{(b)}$ PCA+S represents our PCA model after applying the standardization process. The AFFE on PCA+S is calculated after the inverse transformation of our standardization process.}
	\label{tab:affe_pca}
\end{table}

We further use these PCA models to \textit{predict} the quasar continua blueward of $\lya$ emission. In this case, the PCA model only utilizes the quasar spectra in the range $1216 \text{\AA} \leq \lambda_{\text{rest}} \leq 1600 \text{\AA}$ and predicts the continuum in the range of $1020 \text{\AA} \leq \lambda_{\text{rest}} \leq 1216 \text{\AA}$. This method can be used to predict the quasar continuum in the $\lya$ forest region at high redshift. We again only use the testing data set to make an independent assessment of all PCA prediction models. We compute the AFFE as the evaluation metric for the full testing sample. Table \ref{tab:affe_pca_predict} shows that our PCA prediction model with the standardization process outperforms all other PCA-based models and predicts the underlying continua more accurately and with less bias on the testing quasar spectra. In the continuum prediction scenario, compared to the reconstruction case, our PCA with the standardization process surpasses other PCA models by lowering AFFE by approximately 25\%--50\%. In all cases, we demonstrate that the standardization process is a necessary pre-processing step before model training. 

It should be noted that the \cite{Paris_2011} PCA model is not optimized to predict the quasar continua at $z \sim 0$ as they used $z \sim 3$ quasar spectra to construct their PCA model. At low redshifts more low-luminosity quasars can be detected and that observational bias coupled with the quasar luminosity-emission line equivalent width anti-correlation (Baldwin effect, \citealt{Baldwin1977}) implies that the observed population of $z \sim 0$ quasars will have higher $\lya$ and CIV emission equivalent widths than the $z \sim 3$ \cite{Paris_2011} sample. This might be one of the reasons why the \cite{Paris_2011} PCA model is the least accurate among the four PCA models studied (Table \ref{tab:affe_pca_predict}).

\begin{table}
	\centering
	PCA \textit{Prediction} Model
	\begin{tabular}{lcccr} 
		\hline
		Model Name & \multicolumn{2}{c}{Training Set} & \multicolumn{2}{c}{Testing Set}\\
		 & Median & Mean & Median & Mean\\
		\hline
		Suzuki's PCA & N/A & N/A & 0.0861 & 0.0951\\
		Paris's PCA & N/A & N/A & 0.1210 & 0.1290\\
		PCA+S (This work) & 0.0474 & 0.0590 & 0.0670 & 0.0628\\
		\hline
	\end{tabular}
	\caption{Absolute Fractional Flux Error of \textit{predicted} continua of $z< 1$ HST-COS training and testing quasar Spectra.
	AFFEs of training set are not available for \citet{Suzuki_2005} and \citet{Paris_2011} as the training set is used to construct these PCA models. The performances of PCA-based models are evaluated only on the testing quasar spectra as the testing sample is blind to the PCA construction process. PCA+S represents our PCA model after the data standardization process. Clearly PCA+S model gives the best performance. }
	\label{tab:affe_pca_predict}
	
\end{table}

\subsubsection{iQNet deep learning performance}
\begin{figure*}
    \centering
        \includegraphics[width=0.8\textwidth]{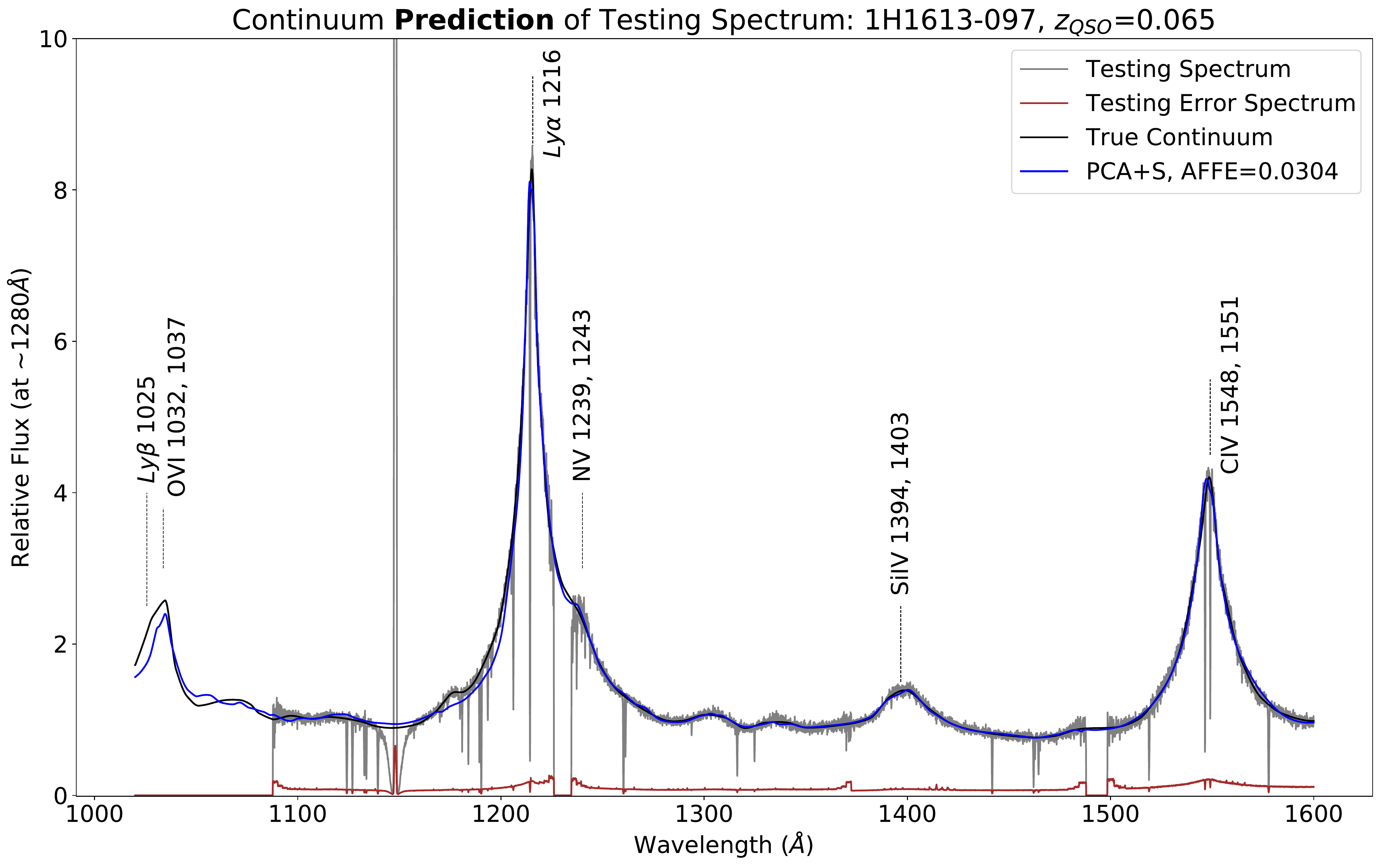}
        \includegraphics[width=0.8\textwidth]{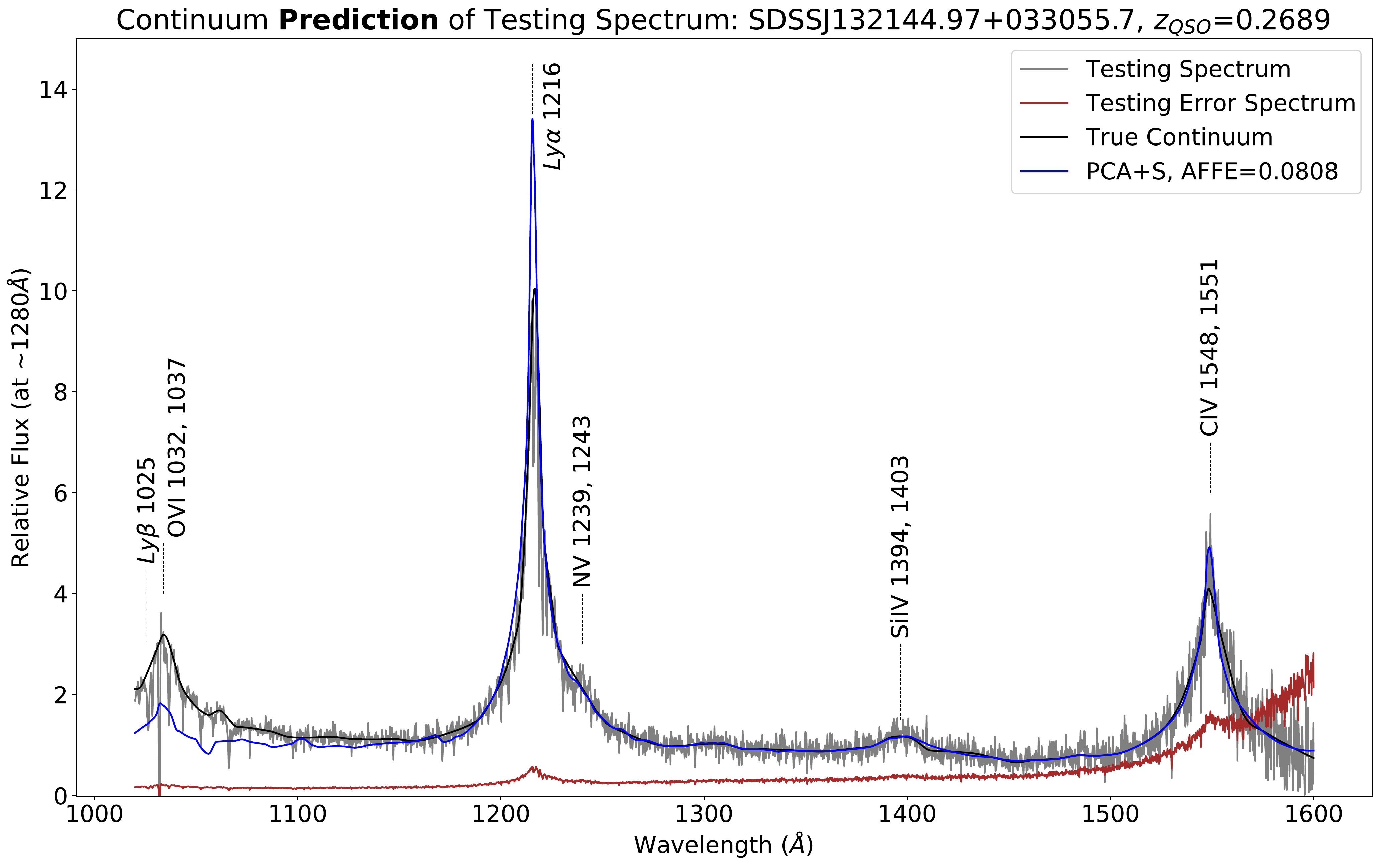}
    	
    \caption{PCA \textit{predictions} of quasar continuum on two $z<1$ HST-COS testing quasar spectra. The HST-COS testing spectra and the corresponding ground truth continua are shown in grey and blue, respectively. The predicted continua of our PCA-based prediction model with standardization is shown in blue. Top panel shows an example of a good prediction of the quasar continuum and the bottom panel show a bad prediction of the quasar continuum highlighting the shortcomings of all PCA based continuum prediction methods.}
    \label{fig:pca_testing_set}
\end{figure*}

Even though our PCA-based model outperforms the two PCA-based models studied, the transformation matrix in our PCA prediction model is fixed and cannot be optimized to reduce the AFFE further without increasing the number of principal components. However, increasing the number of principal components will also increase the size of the transformation matrix (thereby increasing computational time) without significant gain in accuracy (see Section \ref{subsec: pca}). This limitation of these PCA prediction models inspired us to apply a deep learning approach for intrinsic quasar continuum prediction. Figure \ref{fig:pca_testing_set} shows that the PCA \textit{prediction} model cannot always  predict the testing continuum correctly because of the limited linear superposition of our PCA prediction models. Therefore, our traditional PCA-based \textit{prediction} models cannot be robustly generalized for unseen quasar continua.

Another issue with PCA-based models is that we need to fit a continuum on the red-side of $\lya$ first before applying PCA prediction models. PCA-based models cannot work directly with the quasar spectra but need human intervention to fit a spline curve and mask out absorption/emission lines before predicting the quasar continua. Different curve-fitting techniques may result in additional uncertainty in each spectrum that is ready to be fed into the model for the prediction of the whole continuum requiring researchers' inspection. 

To reduce the amount of human inspection and to use the full information matrix that exists in each spectral pixel in predicting the quasar continuum, we apply the iQNet model directly on the quasar spectra and build a pipeline to automate the entire process. In the iQNet prediction model, the network takes a raw 1-D quasar spectrum within $1216 \text{\AA} < \lambda_{\text{rest}} \leq 1600 \text{\AA}$ as input, and predicts the quasar continuum for the wavelength range $1020 \text{\AA} < \lambda_{\text{rest}} \leq 1600 \text{\AA}$ as the output.

\begin{figure*}
	\includegraphics[width=0.8\textwidth]{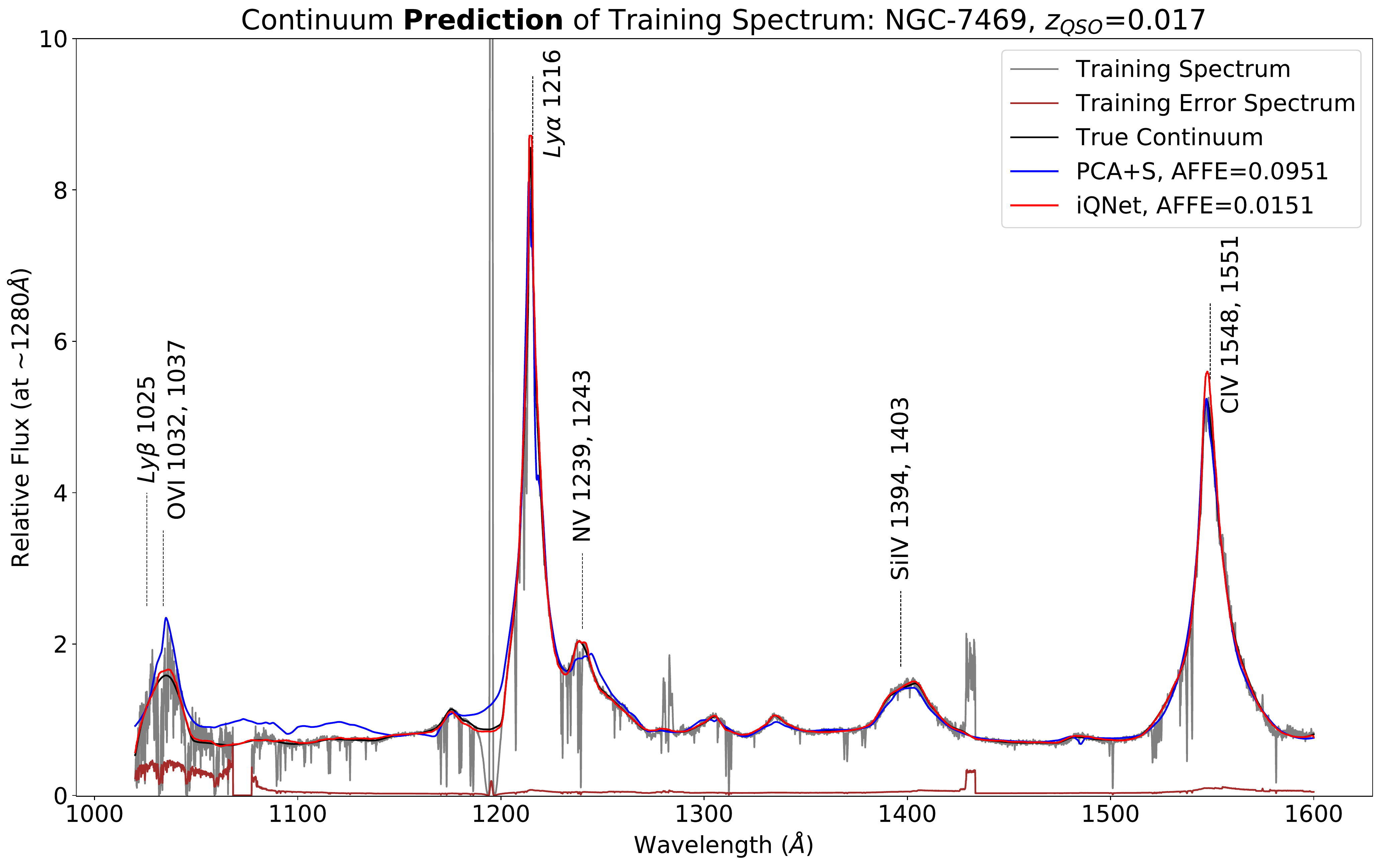}
    \caption{Example of an HST-COS training spectrum and its corresponding predicted continua with PCA prediction model and iQNet network. AFFEs are shown at the top. The training spectrum and its corresponding ground-truth continuum are shown in grey and black, respectively. The PCA Predicted Continuum in blue cannot predict the quasar continuum correctly blueward of $\lya$ emission due to its limited extrapolation ability. The predicted continuum from our neural network in red almost completely overlaps the ground truth continuum.}
    \label{fig:nn_training}
\end{figure*}

\begin{figure*}
    \includegraphics[height=0.31\textheight]{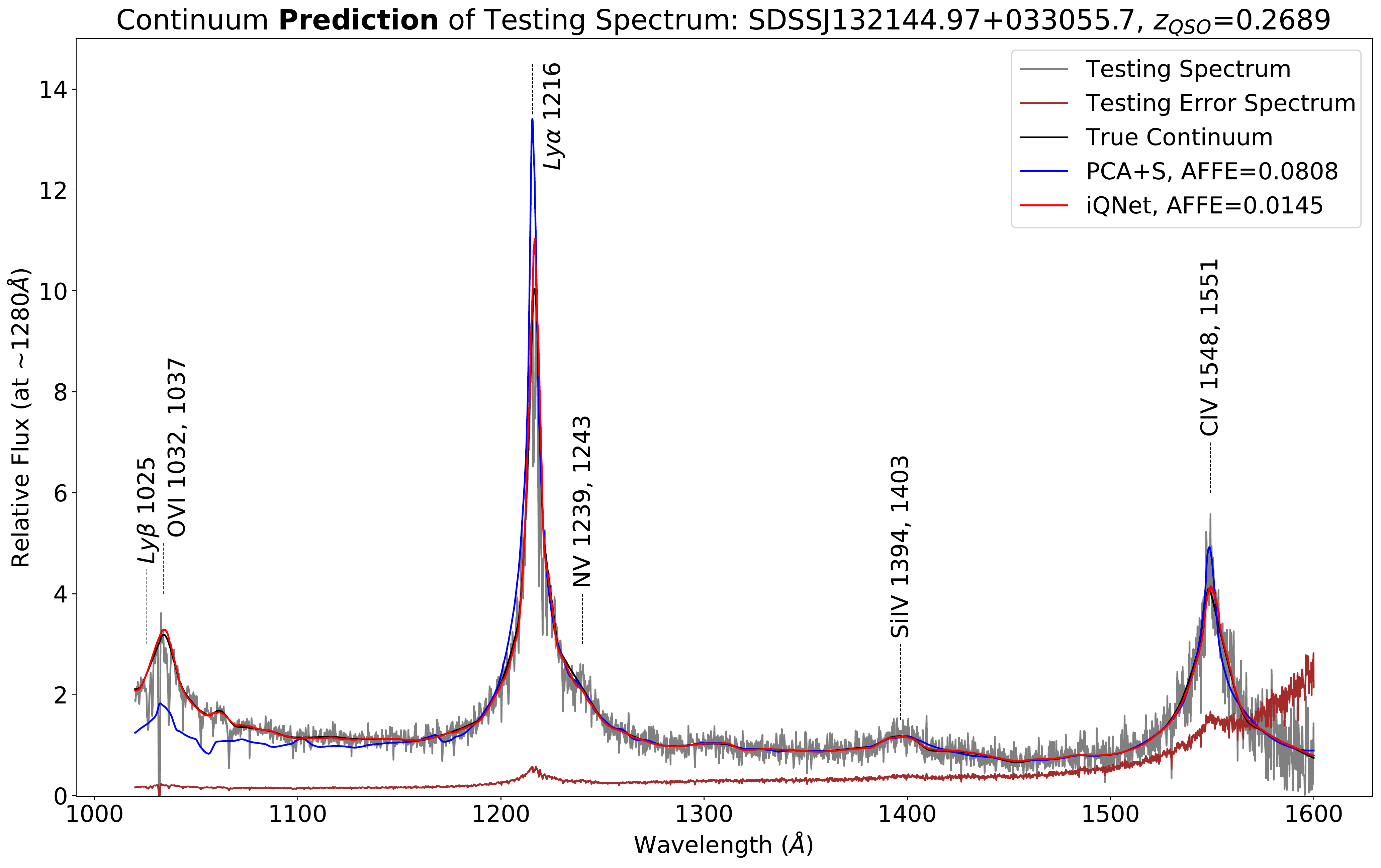}
    \includegraphics[height=0.31\textheight]{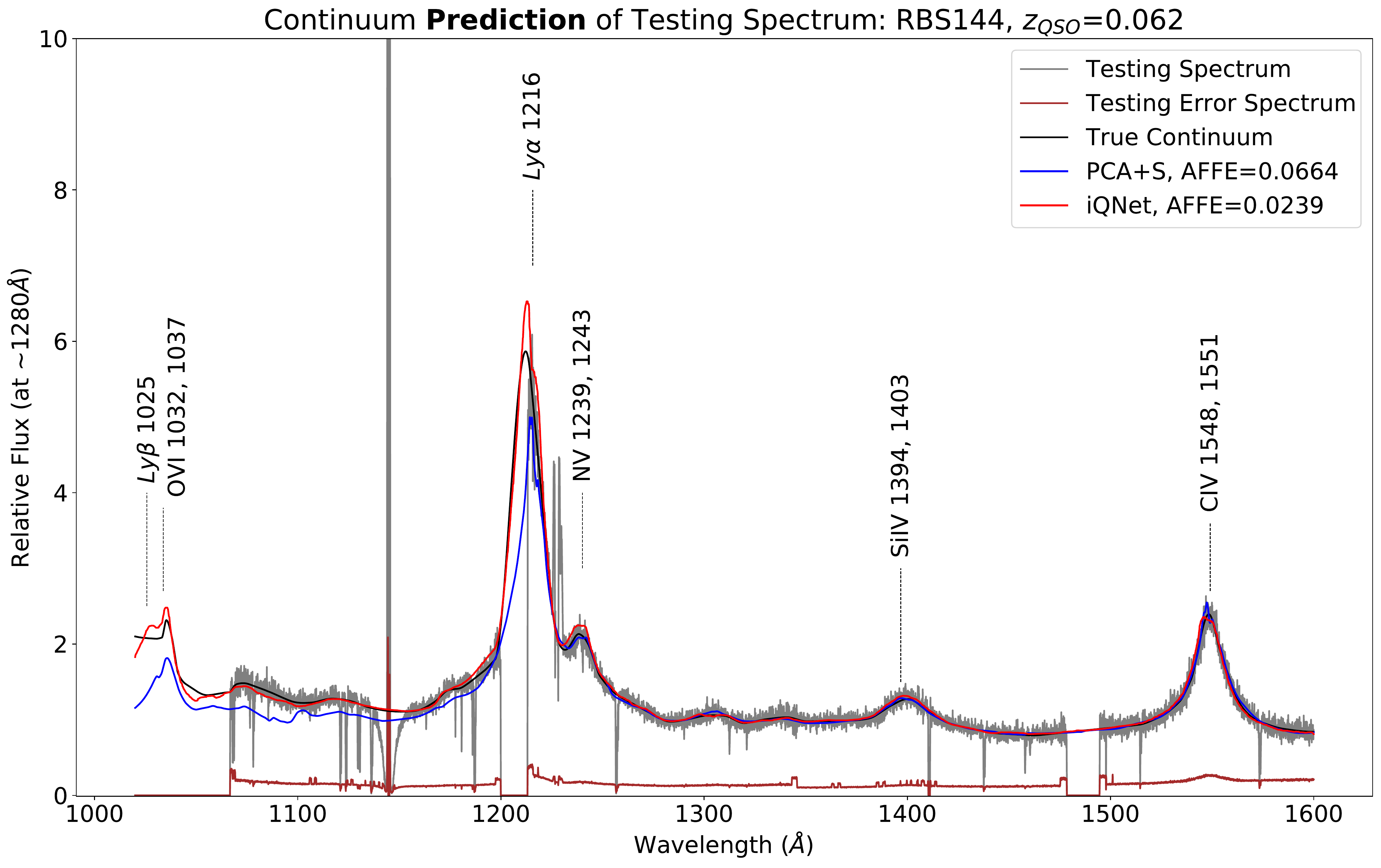}
    \includegraphics[height=0.31\textheight]{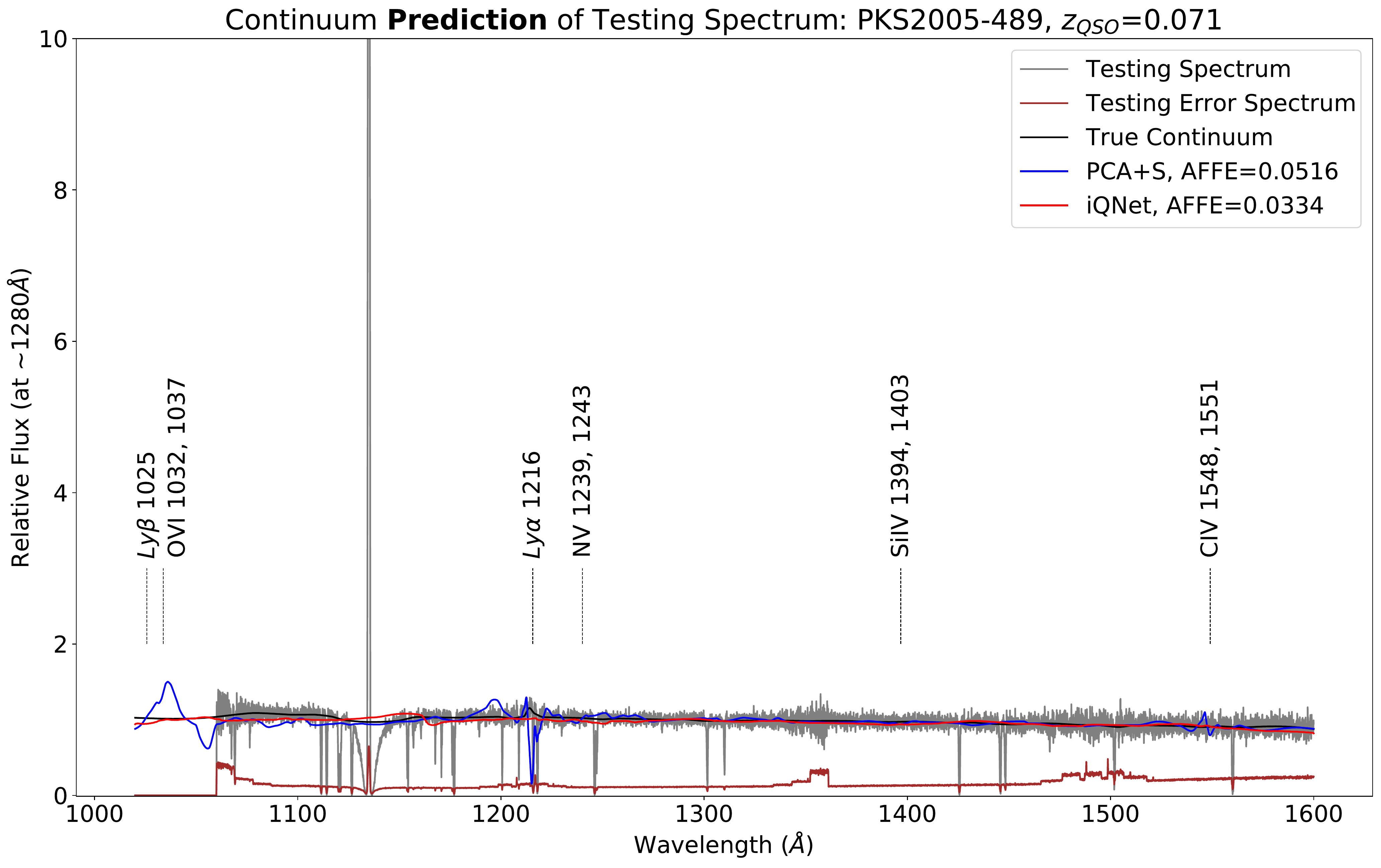}
    \caption{Examples of quasar continua prediction with PCA and iQNet network on three HST-COS testing quasar spectra. Top and middle panels show a quasar spectrum with strong $\lya$ and CIV emission features and the bottom panel shows a quasar spectrum with no intrinsic emission line features. The testing spectra and the corresponding ground truth continua are shown in grey and black, respectively. The PCA Predicted Continuum (blue line)  cannot predict the continua blueward of Ly-$\alpha$ emission correctly owing to its limited extrapolation ability. The predicted continua from iQNet (red line) almost overlap the ground truth continua.}
    \label{fig:nn_testing}
\end{figure*}

Table \ref{tab:affe_nn_pca} presents the AFFE of both training and testing quasar spectra using iQNet. The median AFFE of predicted training continua is around 2.24\%, which indicates that neural network can replicate our hand-fit continua with approximately 2\% error, whereas our PCA-based methods cannot achieve such low AFFE.  More importantly, the iQNet has enhanced the performance on predicting the testing spectra and is able to generate a quasar continuum with only approximately 4\% to 5\% AFFE given a quasar spectrum that our model has not seen or trained. The additional advantage of the proposed iQNet model is that it is able to easily handle quasar spectra with missing data. Figures \ref{fig:nn_training} and \ref{fig:nn_testing} demonstrate that the iQNet manages to predict a continuum close to the true continuum even though there are missing spectra within the wavelength range from $1020\text{\AA}$ to $1600\text{\AA}$.

Figure \ref{fig:nn_training} shows an example of the training spectrum comparing our PCA model and iQNet model. The iQNet is able to capture the emission features, denoise the spectra, and generate a smooth continuum. This results in the smallest AFFE between the predicted continuum and the ground truth continuum among all the models studied in this work. The main advantage of applying a deep learning model instead of the traditional PCA model is that the traditional PCA models are linear transformations, whereas the neural networks are non-linear transformations. Therefore, the neural networks have better generalization ability than traditional PCA models. By minimizing the binary cross-entropy loss function, the neural network model learns and updates its weights to generate outputs as close to the ground truth as possible.

Evaluating a model performance on a training data set is not sufficient because the model has seen the training data and there is a chance that the same model may over-fit the training data set in order to obtain a low error or high accuracy. Therefore, we keep a testing set of quasar spectra that the model has never used. Figure \ref{fig:nn_testing} demonstrates three different quasar spectra in the testing set. In all cases, the results generated from iQNet outperforms the PCA model. The iQNet prediction generates moderately higher $\lya$ emission than the ground truth, but the whole continuum traces the ground truth continuum successfully, whereas the PCA prediction fails to predict the continuum on the blue side of $\lya$. Even for a spectrum without strong emission features (e.g. $\lya$ and CIV) the iQNet model is able to successfully generate a continuum accurately (Figure \ref{fig:nn_testing}, bottom panel). The PCA-based prediction model, on the other hand, fails to predict the continuum at regions where there should have been no emission lines shown in the quasar spectrum. The fluctuation in the PCA prediction shows that the PCA extracts emissions as important and representative features, and the superposition of different principal components cannot fit a flat continuum at the location of the $\lya$ emission line.

\begin{table}
	\centering
	\caption{Absolute Fractional Flux Error of Predicted Continuum on HST Training and Testing Quasar Spectra.
	}
	\label{tab:affe_nn_pca}
	\begin{tabular}{lcccr} % four columns, alignment for each
		\hline
		Model Name & \multicolumn{2}{c}{Training Set} & \multicolumn{2}{c}{Testing Set}\\
		 & Median & Mean & Median & Mean\\
		\hline
		PCA+S (This work) & 0.0474 & 0.0590 & 0.0670 & 0.0628\\
		iQNet (This work) & 0.0224 & 0.0315 & 0.0417 & 0.0514\\
		\hline
	\end{tabular}
\end{table}

\subsection{Predicting Quasar Continuum on $2 < z\leq 5$ SDSS Spectra}
In this section, we present results of quasar continuum predictions when we apply the PCA+S prediction model and the deep learning iQNet model to 3196 $2 < z  \leq 5$ quasar spectra from SDSS DR16 Release  \citep{ahumada2019sixteenth}. The fully trained deep learning method, iQNet, takes 3.379 seconds to successfully predict the 3196 selected SDSS 1D spectra (about 945 predictions per second). We ran the network on Google Colab without GPU or TPU enabled. This is three times faster than the PCA+S model on the same architecture. The PCA+S needs 10.19 seconds to predict the 1d spectra, which is equivalent to making about 313 predictions per second.

\begin{figure*}
        \includegraphics[width=0.49\textwidth]{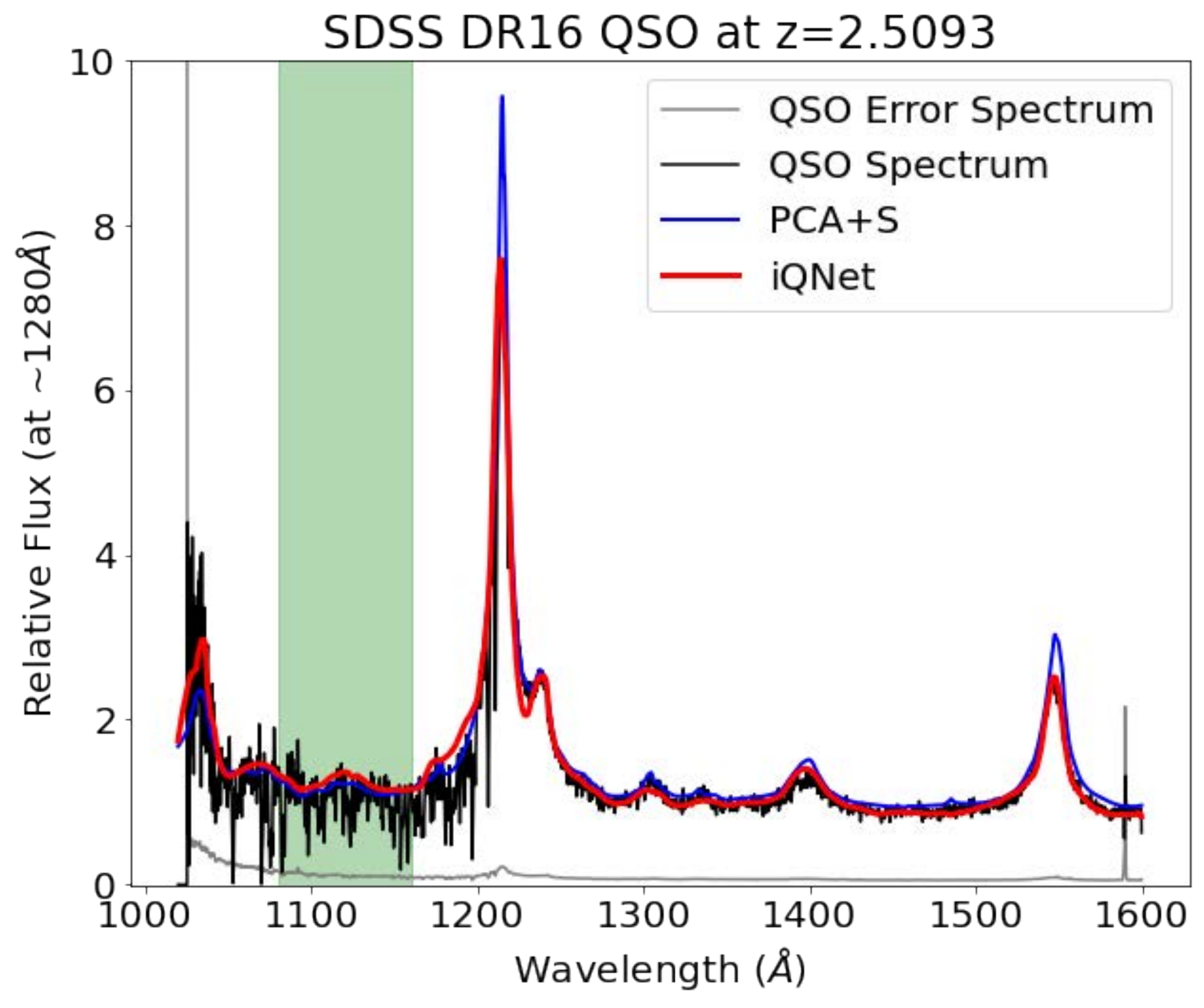}
        \includegraphics[width=0.49\textwidth]{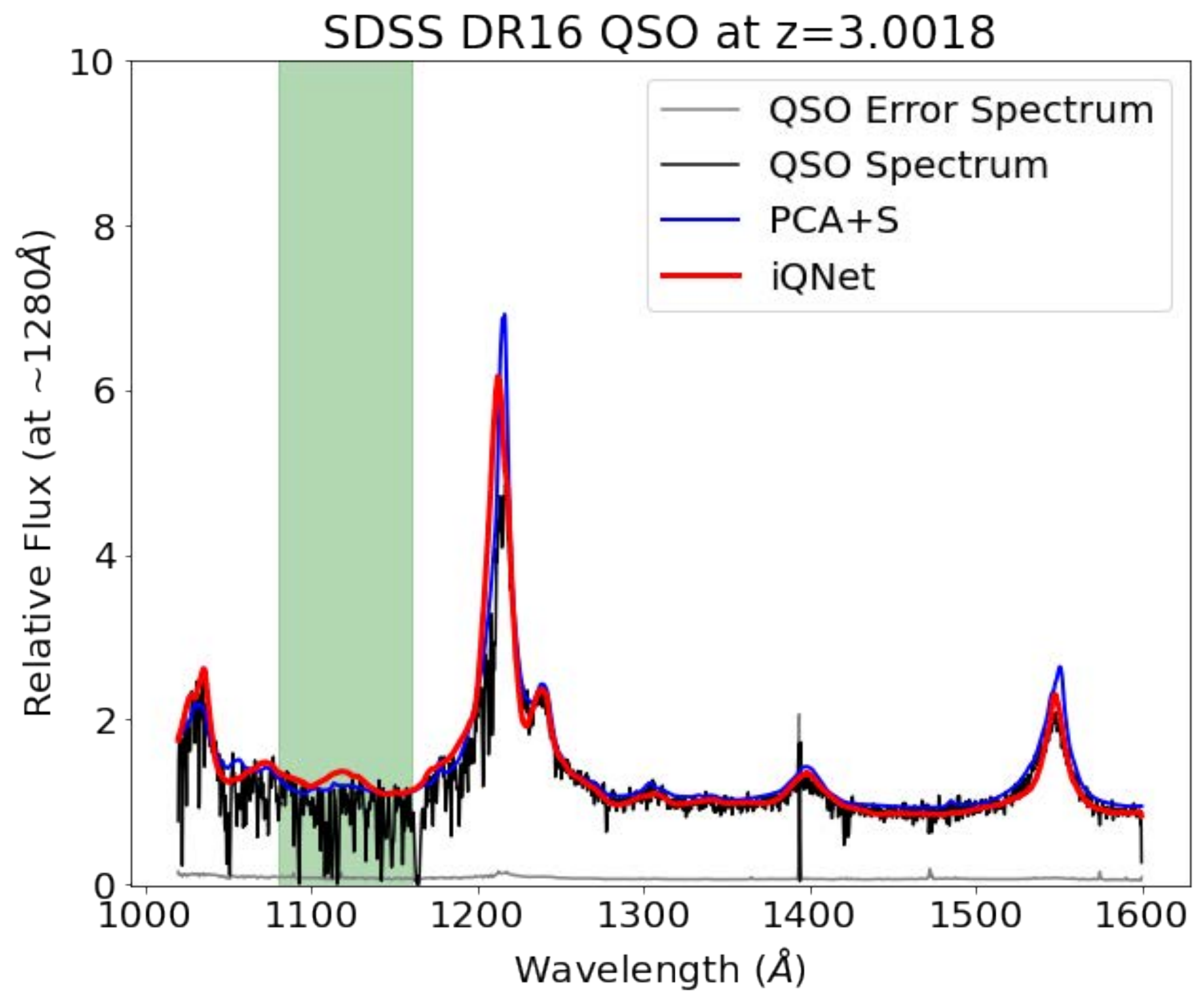}
        \includegraphics[width=0.49\textwidth]{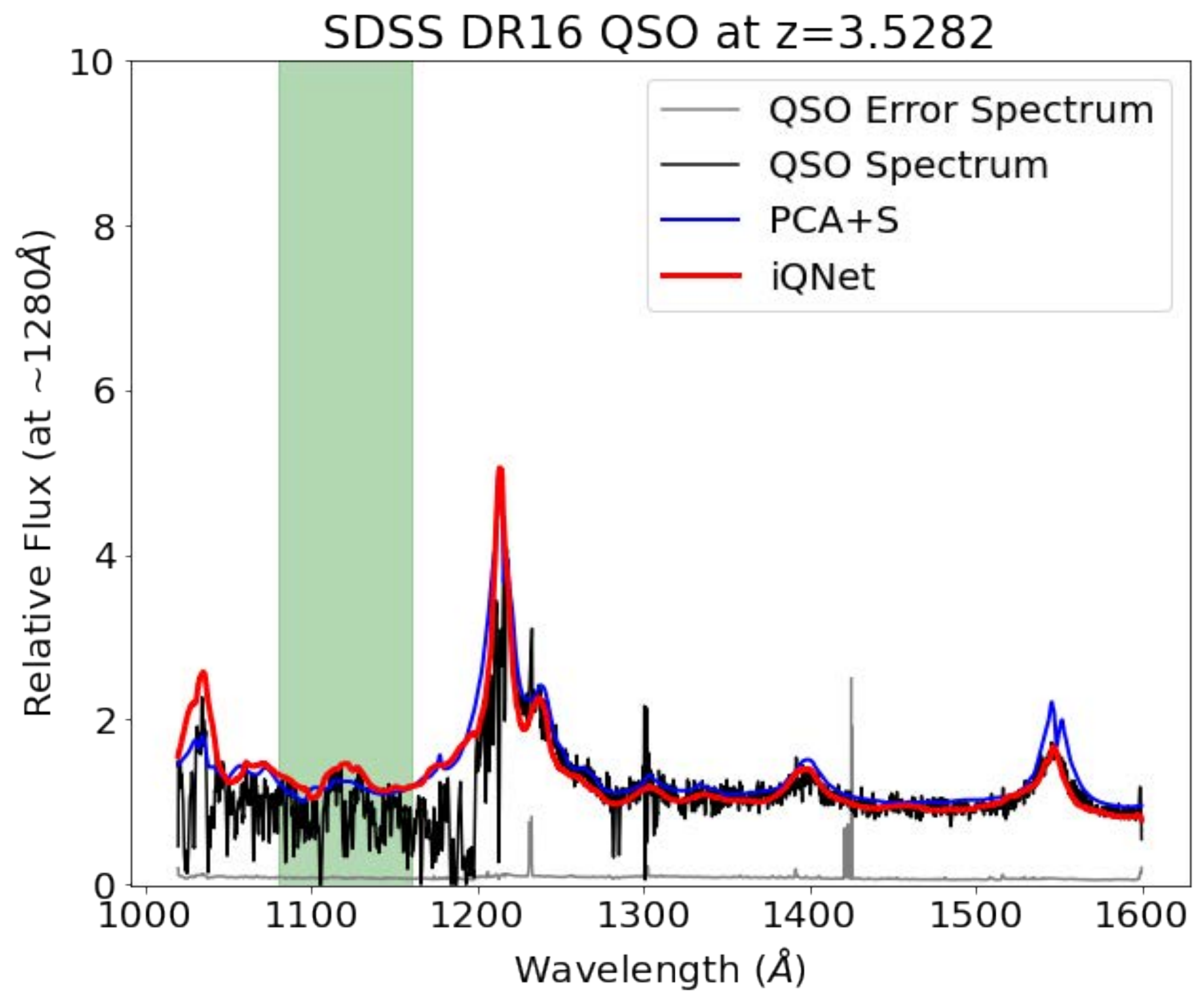}
        \includegraphics[width=0.49\textwidth]{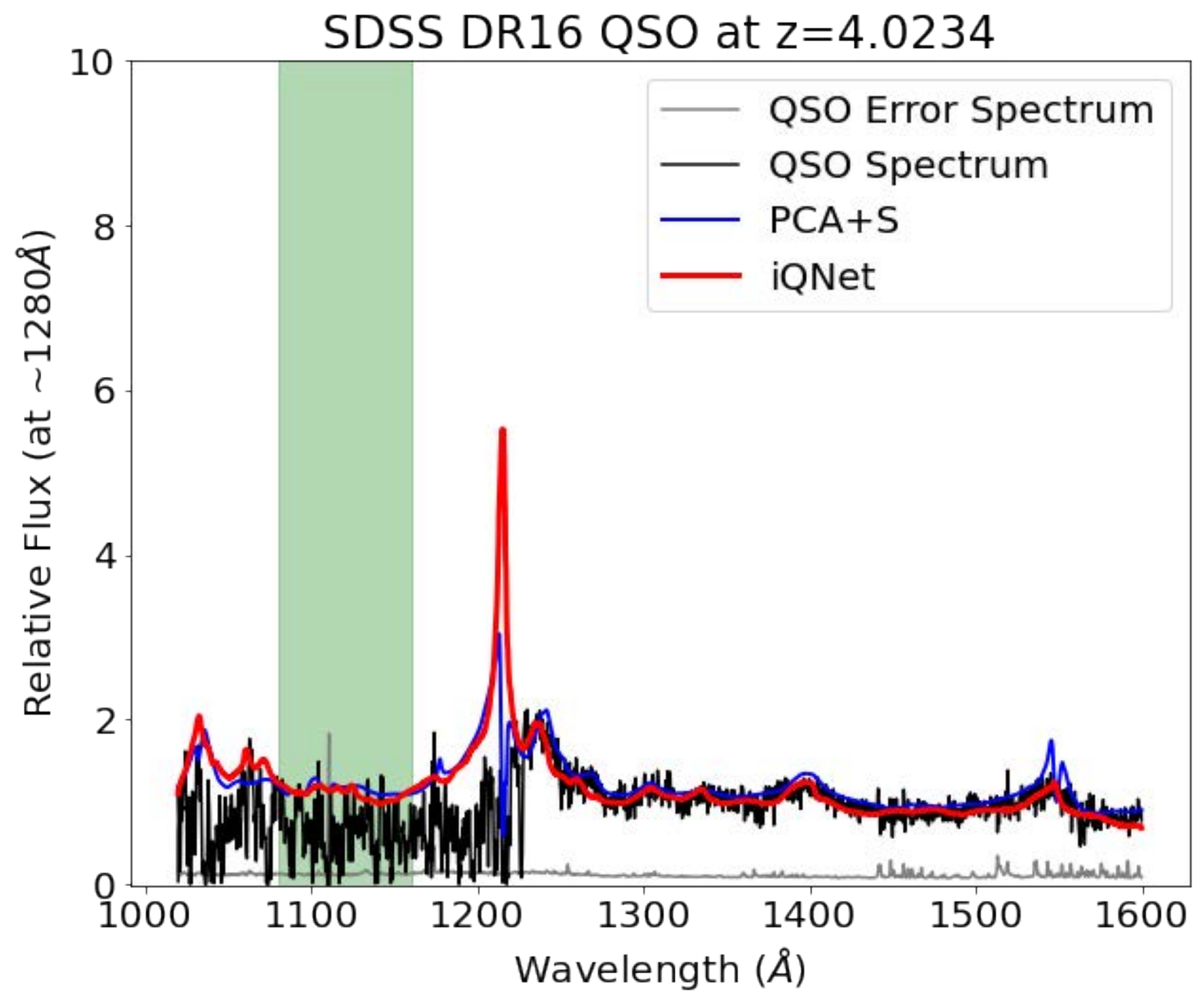}
        \includegraphics[width=0.49\textwidth]{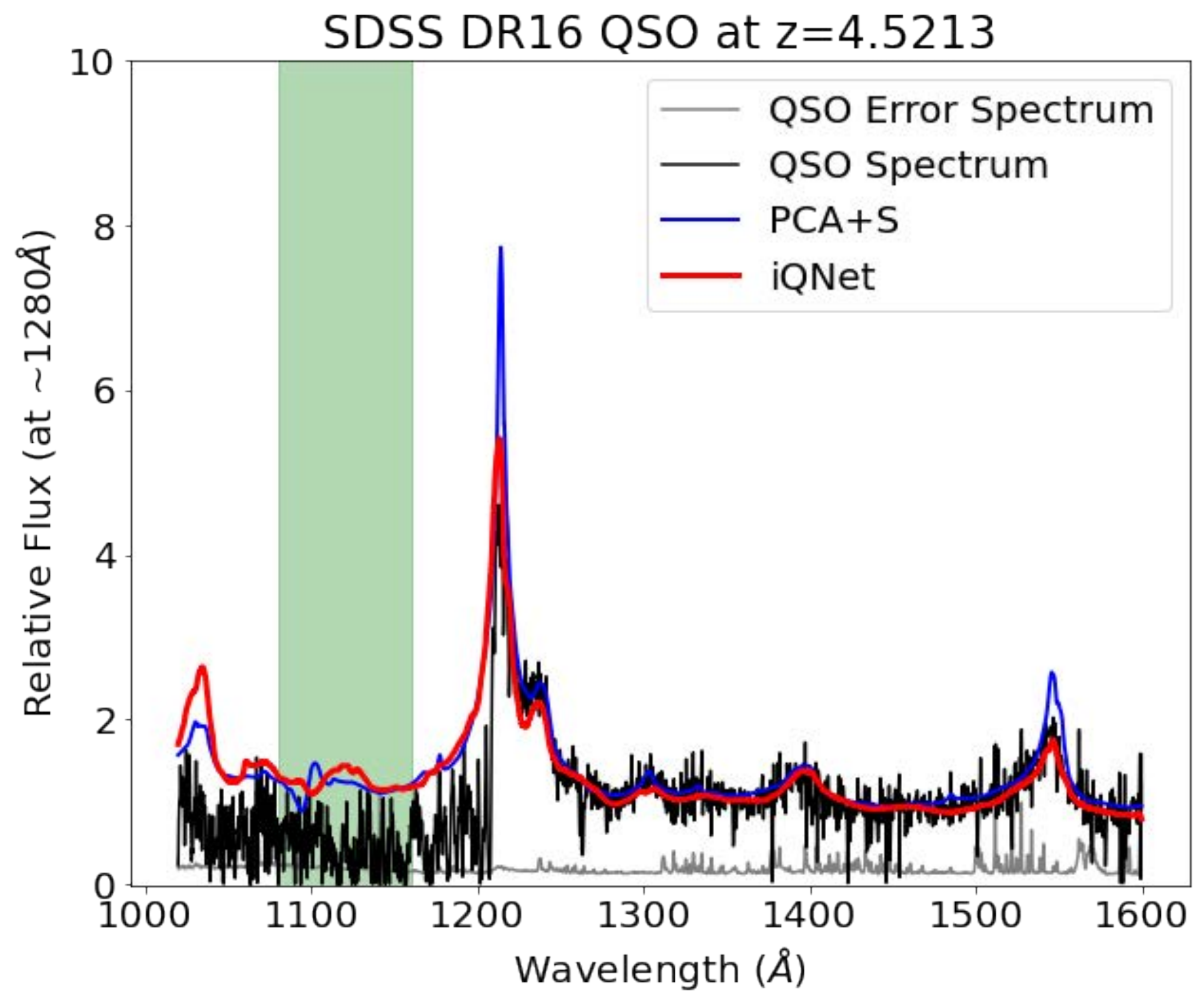}
        \includegraphics[width=0.49\textwidth]{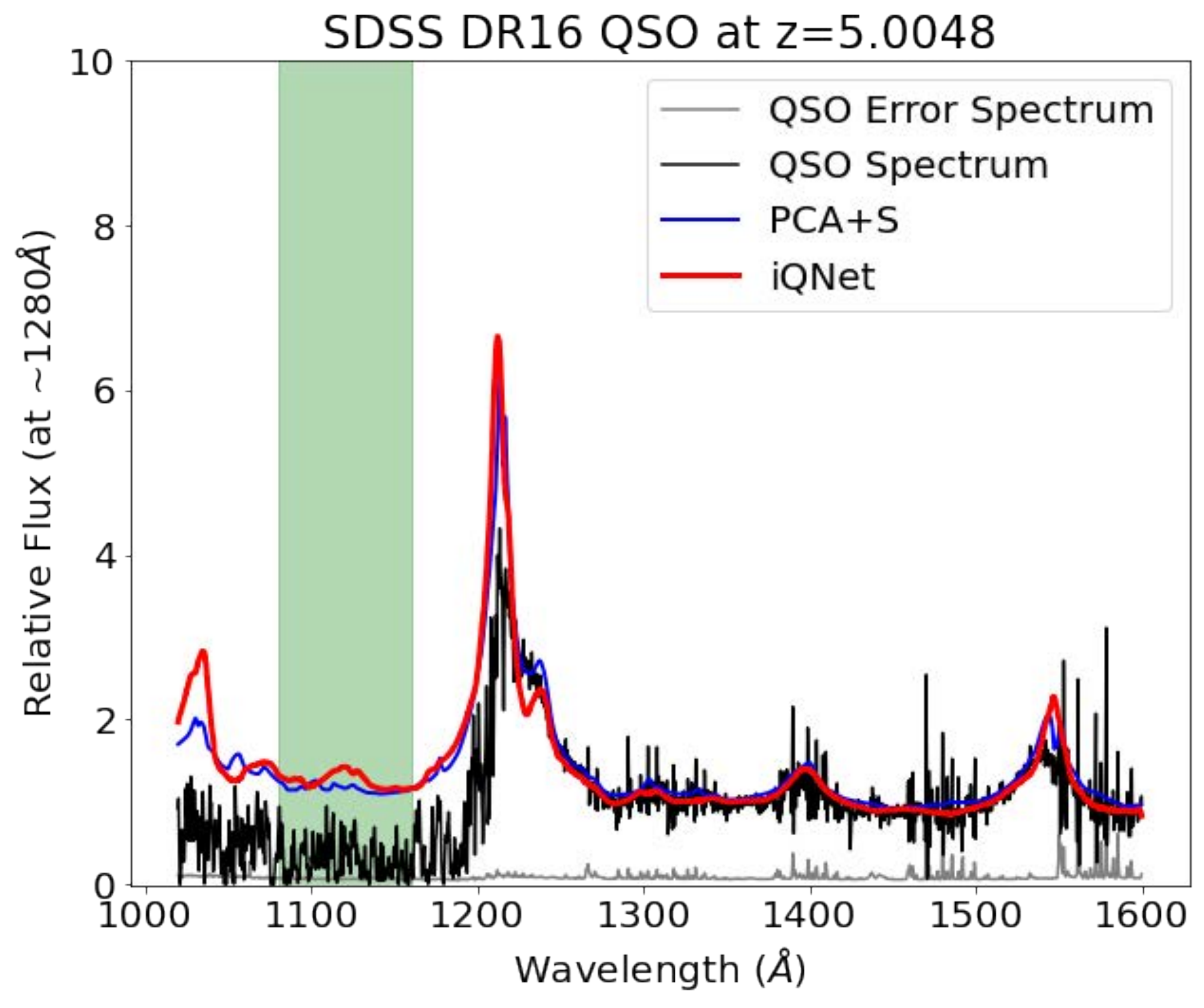}
    \caption{1-D SDSS DR16 spectra of $2 < z \leq 5$ quasars (black lines) along with their corresponding error spectra (gray lines) are presented in the quasar rest-frame. The PCA+S model fit (blue line) and deep neural network iQNet (red lines) predictions are also shown. For each case,  iQNet outperforms the PCA+S model in predicting the overall 1-D quasar spectra.  The mean transmitted flux is calculated within the green region shown in each panel.}
    \label{fig:sdss_varz}
\end{figure*}

Figure \ref{fig:sdss_varz} presents six SDSS quasar spectra and their corresponding predicted continua based on PCA+S prediction model (blue line) and iQNet model (red line) within  $2 <z \lesssim 5$. The PCA+S prediction model consistently overestimates the $\lya$ and CIV emission peaks in all these spectra whereas the iQNet model successfully predicts those emission features. In addition, the iQNet model is also better at adapting and predicting the full continuum even if there are no strong emission features redward of $\lya$ emission in the spectrum (second row in Figure \ref{fig:sdss_varz}).

We note that both the PCA+S and iQNet models are  built with $z \sim 0.2$ HST/COS quasar spectra. However, it is well known that the observed quasar emission lines are changed at higher ($z >2$) redshifts owing to observational selection effects on quasar luminosity and its activity \citep{Baldwin_1977}. This results in broader and shallower emission peaks for lines such as $\lya$, CIV, NV. This might be one reason why the PCA+S model is over-predicting the strength of $\lya$ and CIV emission lines at $z >2$. Interestingly, although the iQNet model is not trained explicitly to account for this trend, the network is flexible enough (by design) to account for any variability in emission  line strengths. As can be seen from Figure \ref{fig:gemm_spec}, different emission line ratios are present in the training set, which already allows the network to train for such variations. Moreover, the addition of a fifth quasar class with a constant relative flux representing a BL-Lac like AGN spectra (Figure \ref{fig:nn_testing}, bottom panel) helps the network training process and allows the network to update the weight matrix to generate quasar continua with weak/broad emission features. This is the main reason why iQNet can successfully predict weak and broad emission features at high-$z$ even though it was not trained with $z \sim$ 2 quasar spectra. This is one of the inherent strengths of using a non-linear machine learning approach for predicting quasar continua. As the performance of the iQNet network is superior to the traditional PCA based methods, we will only use the iQNet predicted continua for the next analysis.

\subsection{Evolution of Mean Transmitted Flux of the IGM}
We use the iQNet predicted quasar continuum to estimate the mean transmitted flux ($\langle F(z_{\lya})\rangle$) in the $\lya$ forest region $2 < z \lesssim 5$. 
Following the steps in Section \ref{subsec:mean_flux}. We calculate $\langle F(z_{\lya})\rangle$ in the rest-frame wavelength range [1080, 1160]\text{\AA} to account for contribution only from the $\lya$ forest region.

\begin{figure}
    \centering
    \includegraphics[width=\columnwidth]{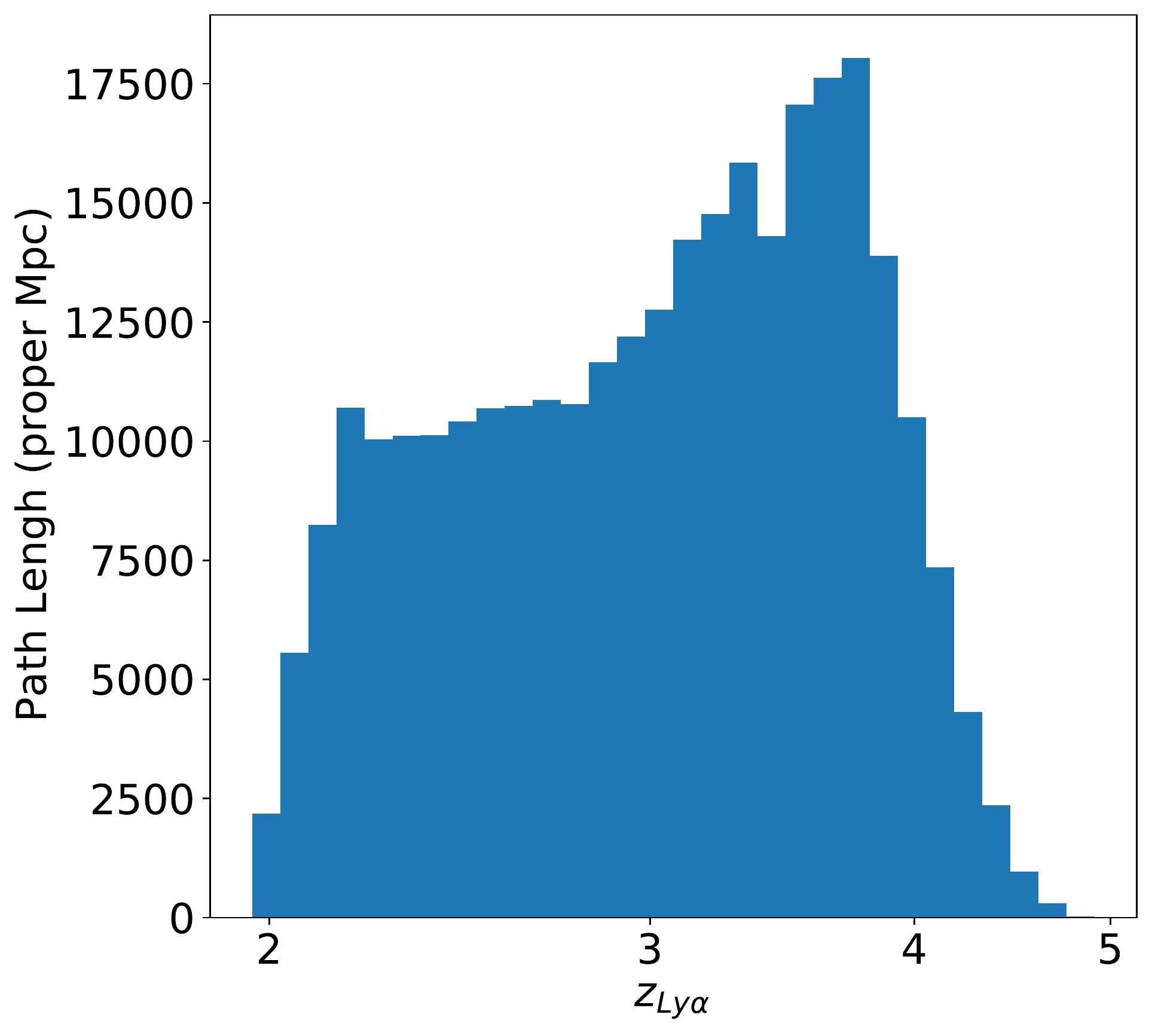}
    \caption{Proper path length distribution of $\lya$ absorption, with $\delta z_{\lya}=0.1$ interval, of the $\sim$ 3200 quasar spectra used from the SDSS survey. The  spectra gives a near uniform coverage of the $\lya$ forest up to $z \sim  5$.}
    \label{fig:hist_lyaz}
\end{figure}

The sample of SDSS spectra presents a large statistical sampling of the $\lya$ forest region. Figure \ref{fig:hist_lyaz} illustrates the corresponding path length distribution of all the quasar 1-D spectral pixels in the $\lya$ forest. Our sample covers the full redshift range $1.96< z_{\lya} <4.92$ with a mean $\langle z_{\lya}\rangle =3.12$ and standard deviation of 0.618.  

\begin{figure*}
    \centering
    \includegraphics[width=\textwidth]{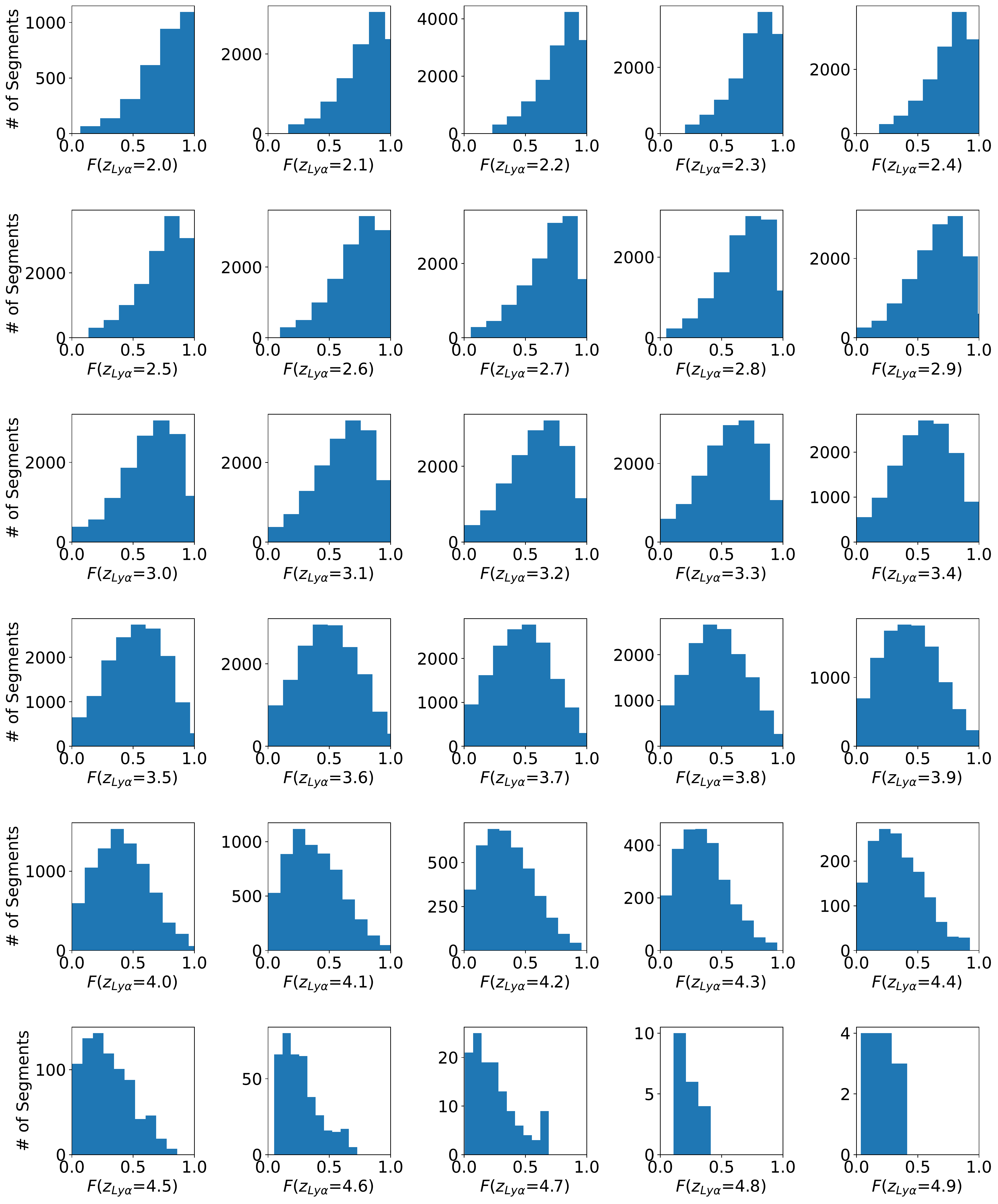}
    \caption{Distribution of transmitted flux $F(z_{\lya})$ in 3-Mpc segment bins for different redshift bins. Each of the panels represents a redshift bin with width of $\Delta z=0.1$. The mean transmitted flux in each bin continuously decreases with redshift. The $\langle F(z_{\lya})\rangle$ and the associated uncertainties are computed by bootstrapping the transmitted flux distribution in each redshift bin. }
    \label{fig:all_histo_lyaz}
\end{figure*}

\begin{figure*}
    \includegraphics[width=0.49\textwidth]{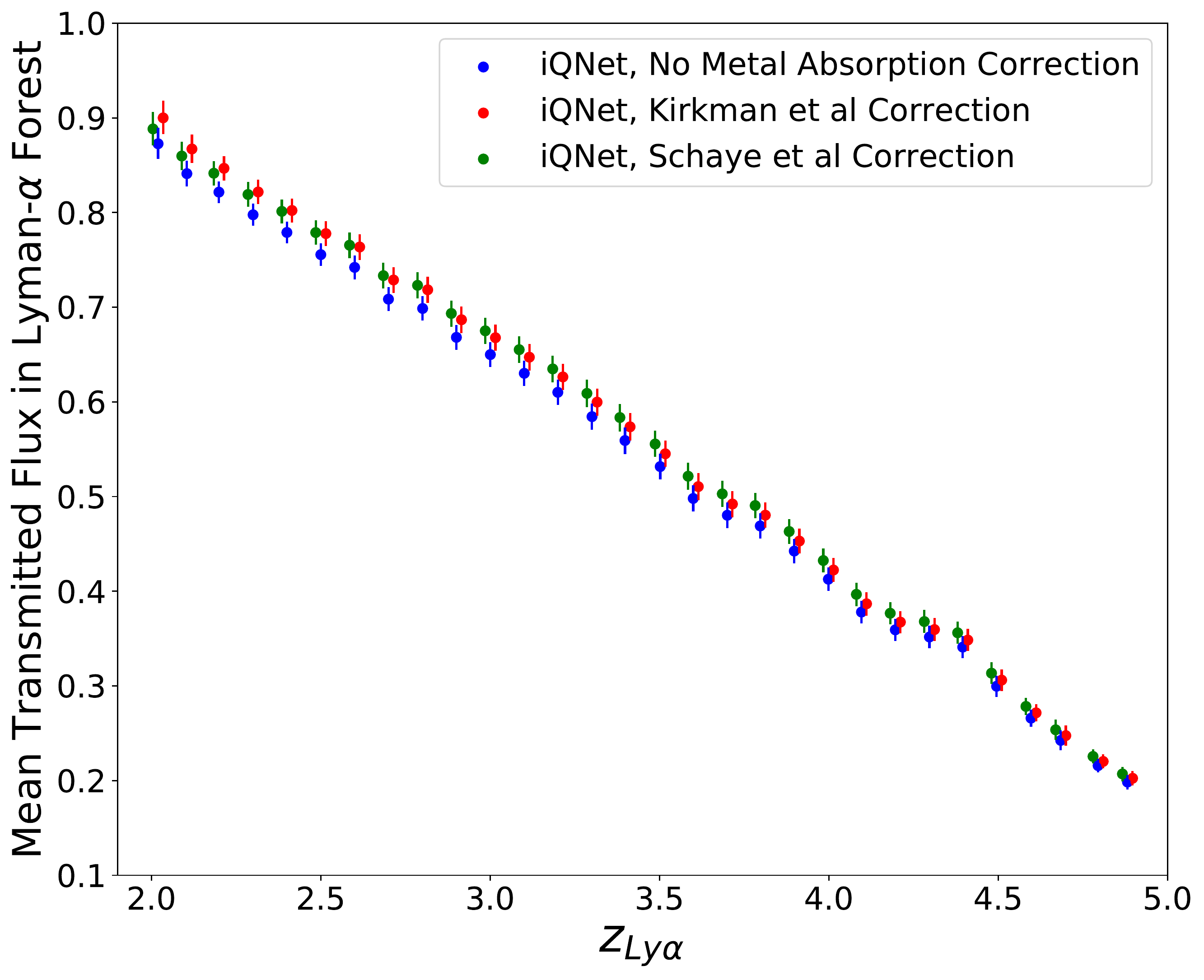}
    \includegraphics[width=0.49\textwidth]{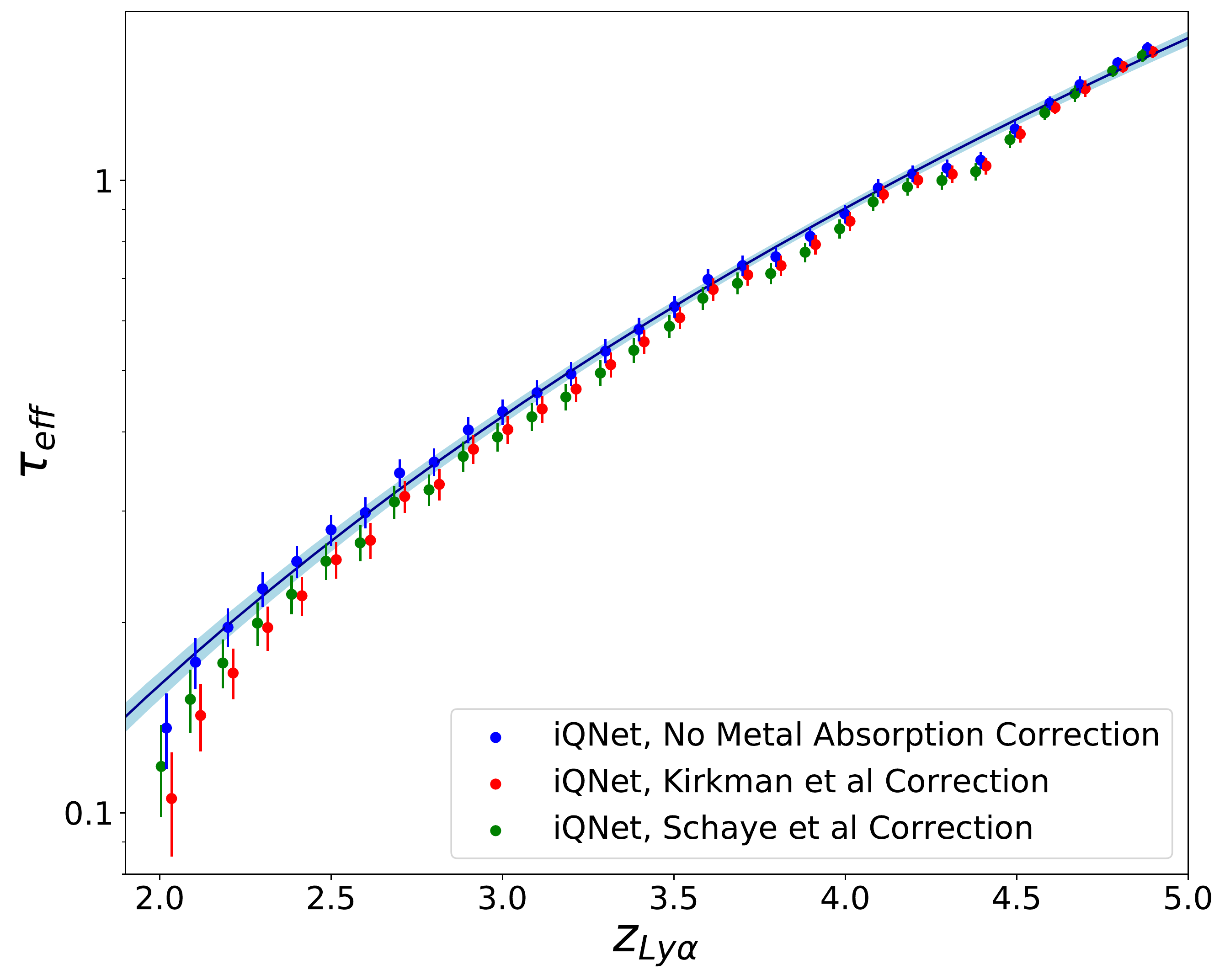}
    \caption{Comparison of Metal Absorption Correction methods. Left panel shows the mean transmitted flux in $\lya$ forest with iQNet continuum predictions (blue points) and the metal line absorption corrections by \citet{Kirkman_2005} (red points) and \citet{Schaye_2003} (green points). Right panel demonstrates the corresponding effective optical depth to the iQNet predictions (blue points) with metal line absorption corrections by \citet{Kirkman_2005} (red points) and \citet{Schaye_2003} (green points). For better visualization, we uniformly shift $z_{\lya}$ by 0.015 to the left for the results corrected by \citet{Kirkman_2005} method and 0.015 to the right for those by \citet{Schaye_2003} method, respectively. The blue curve indicates the fitting function in Equation \ref{eq:log_tau}}
    \label{fig:scatter_ours}
\end{figure*}

We compute the transmitted flux values and the corresponding effective optical depth values in the $\lya$ forest region following the steps discussed in Section \ref{subsec:mean_flux}.
Figure \ref{fig:all_histo_lyaz} shows distribution of $F(z_{\lya})$ in each redshift bin of interest. We can clearly see that the peak of the distribution shifts from $\langle F(z_{\lya})\rangle> \sim 0.9$ at $z \sim 2$ to $\langle F(z_{\lya})\rangle \sim 0.2$ at $z \sim 5$. These measurements are quantified and presented in Figure \ref{fig:scatter_ours} as blue circles. The left panel shows the evolution of $\langle F(z_{\lya})\rangle$ with redshift and the right panel shows the evolution of $\tau_{\text{eff}}$. The uncertainties are 16th and 84th percentiles of the bootstrapped mean distribution, computed from in each panel of Figure \ref{fig:all_histo_lyaz}. We see a clear trend of smoothly declining $\langle F(z_{\lya})\rangle$ with redshift.

These measurements are robust and more direct measurement of $\tau_{\text{eff}}$ compared to literature results, as we are making no assumptions regarding the ionizing history of the Universe, and the most uncertain element of this measurement, the quasar continua, are estimated at $z \sim 0.2$. However, the secondary effects of contamination from the intervening line of sight metal absorption lines remain. These metal absorption line systems can arise from foreground CGM or IGM gas \citep{Schaye_2003, Steidel_2010, Bordoloi2011, Zhu_2013,Cooper_2019}. Our sigma clipping approach will account for some metal line contamination but it is not sufficient to get rid of all the contamination. To correctly measure $\tau_{\text{eff}}$, we need to remove the impact of these metal absorption line systems in the $\lya$ forest, even though this in itself is a small effect (6-8\%, \citealt{Faucher_Gigu_re_2008}). We apply the methods discussed in \citet{Schaye_2003} and \citet{Kirkman_2005}, and Figure \ref{fig:scatter_ours} shows how the metal line corrections affect the mean transmitted flux as well as the effective optical depth. In Figure \ref{fig:scatter_ours}, the red and green points show how each metal absorption line correction prescription impacts the evolution of $\tau_{\text{eff}}$ with redshift. Most impact is seen at $z < 3.5$, where $\tau_{\text{eff}}$ changes by 9.88\% for \citet{Schaye_2003} metal line correction prescription and 9.61\% for the  \citet{Kirkman_2005} metal line correction prescription. These values as well as the raw  $\langle F(z_{\lya})\rangle$ without metal line correction are tabulated in Table \ref{tab:f_tau_log}.

\begin{table*}
\renewcommand{\arraystretch}{1.5}
	\centering
	\caption{Effective optical depth values with and without metal line correction prescriptions. The first column, $z_{\lya}$, describes the mean $\lya$ redshift where the effective optical depth, $\tau_{\text{eff}}$, computed. The second column shows the raw measurements of the effective optical depth without metal absorption line corrections. The third and the fourth columns show the optical depth with metal line correction prescriptions by \citet{Schaye_2003} and \citet{Kirkman_2005}, respectively.}
	\label{tab:f_tau_log}
	\begin{tabular}{lcrr} 
	    \hline
	    \multicolumn{4}{c}{Effective Optical Depth, $\tau_{\text{eff}}$}\\
	    \hline
		$z_{\lya}$ & Raw Measurements & \multicolumn{2}{c}{Metal Absorption Line Correction from} \\
		  & & Schaye et al & Kirkman et al \\
		\hline
        \input{tau_table}
		\hline
	\end{tabular}
\end{table*}

We quantify the evolution of  $\tau_{\text{eff}}$ by fitting a power law of a functional model
\begin{equation}
    \log\tau_{\text{eff}}=b + m\log(1+z_{\lya})
    \label{eq:log_tau}
\end{equation}
to the three estimates of $\tau_{\text{eff}}$. The best fit model parameters are also shown in Table \ref{tab:model_param}. Our fits show a smooth evolution of $\tau_{\text{eff}}$ with redshift and we find no evidence of a "bump" at $z=3.2$, in agreement with \citet{Becker_2013} and \citet{Paris_2011}. We find that the slopes of the effective optical depth start to increase beyond redshift $z=3.2$, which indicates that the evolution of effective optical depth is nonlinear. This is broadly consistent with other literature results, which we will discuss in the next section.

\begin{table}
\renewcommand{\arraystretch}{1.5}
	\centering
	\caption{Model parameters of fitting power-law curves of three models in Equation \ref{eq:log_tau}. The fitting curves among all models are plotted in Figure \ref{fig:comparison_schaye_kirkman}.}
	\label{tab:model_param}
	\begin{tabular}{lcccc} % four columns, alignment for each
		\hline
		Model Name & $b$ & $m$ \\
		\hline
		Raw Measurements & $-2.419_{-0.020}^{+0.021}$ & $3.398_{-0.031}^{+0.029}$ \\
		Schaye Correction & $-2.565_{-0.024}^{+0.024}$ & $3.573_{-0.033}^{+0.034}$\\
		Kirman Correction & $-2.565_{-0.024}^{+0.023}$ & $3.586_{-0.033}^{+0.033}$\\
		\hline
	\end{tabular}
\end{table}

\subsection{Comparison with Literature}
\label{subsec:comparison_literature}

In this section, we present the $\tau_{\text{eff}}$ measurements of previous studies from the literature, which characterized the evolution of  $\tau_{\text{eff}}$ with redshift. We compare our $\tau_{\text{eff}}$ measurements with those reported in \citet{Faucher_Gigu_re_2008}, \citet{Becker_2013},  and \citet{Paris_2011} respectively.

\begin{figure*}
    \includegraphics[width=0.49\textwidth]{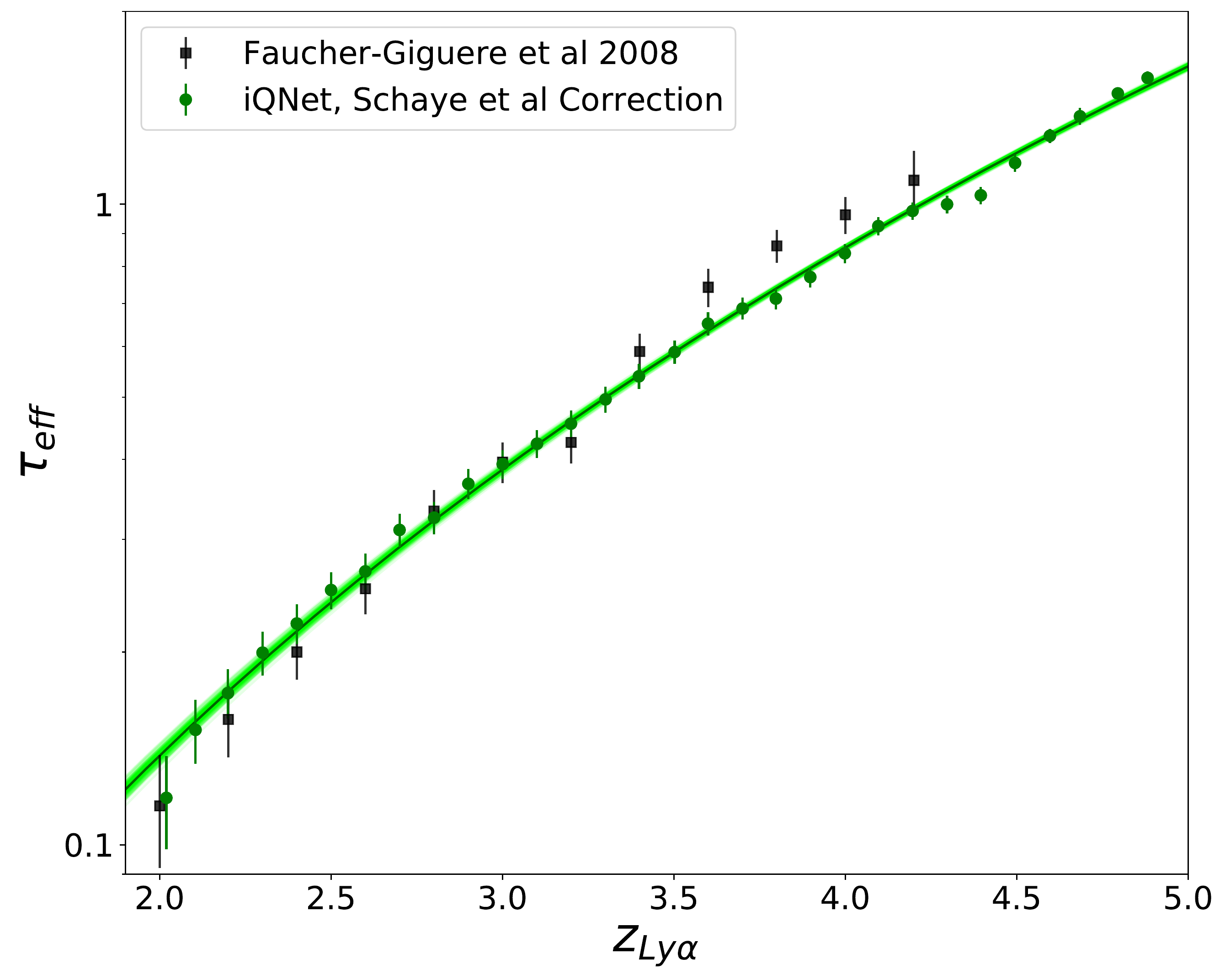}
    \includegraphics[width=0.49\textwidth]{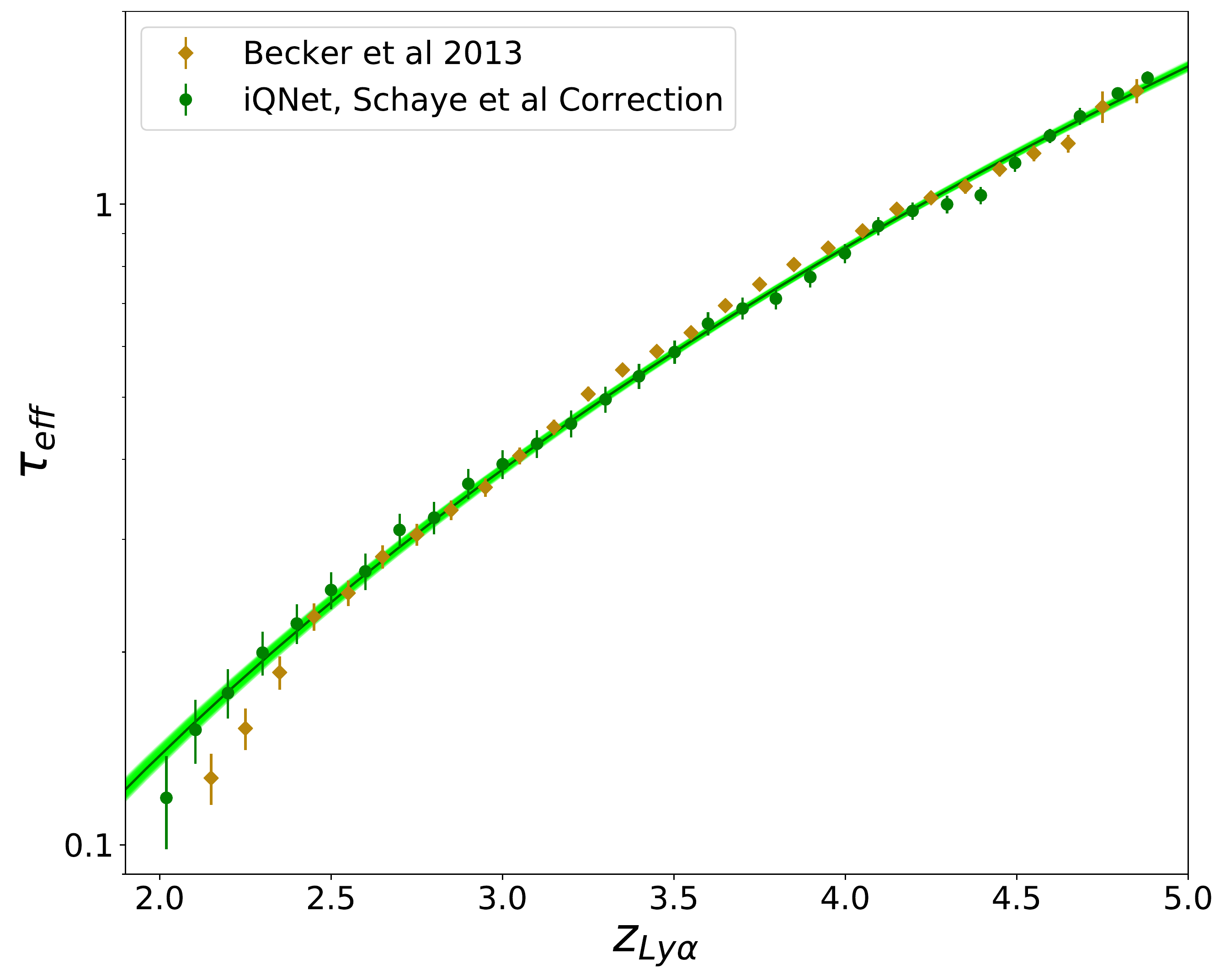}
    \includegraphics[width=0.49\textwidth]{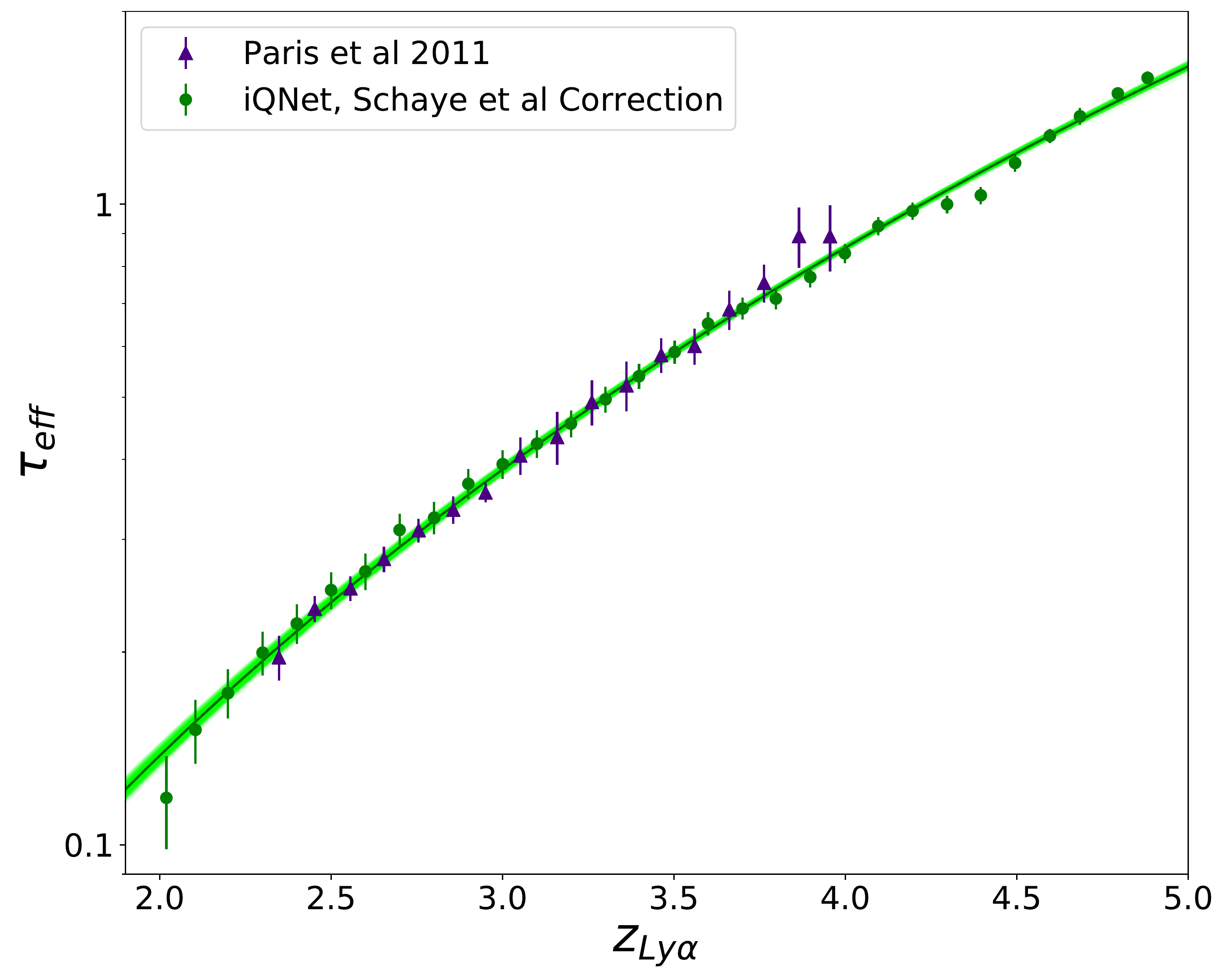} 
    \caption{Comparison of our effective optical depth values, $\tau_{\text{eff}}$, with other published results. Top left panel shows our results along with the results from \citet{Faucher_Gigu_re_2008} employing the \citet{Schaye_2003} metal-line contamination correction. Top right panel shows our results and \citet{Becker_2013}. Bottom right shows the comparison between our measurements and \citet{Paris_2011}. The green fitting curves represent the fitting function in Equation \ref{eq:log_tau} with the best-fit parameters listed in Table \ref{tab:model_param}.}
    \label{fig:comparison_schaye_kirkman}
\end{figure*}

Figure \ref{fig:comparison_schaye_kirkman}, top left panel shows $\tau_{\text{eff}}$ evolution as a function of redshift from \citet{Faucher_Gigu_re_2008} as black squares. \citet{Faucher_Gigu_re_2008} directly measured the effective optical depth in the redshift range $2\leq z\leq 4.2$ using 86 quasar spectra from moderate resolution Keck/ESI, and high-resolution Keck/HIRES, and magellan/MIKE spectrographs. They used the peak flux in the $\lya$ forest regions as their continuum estimates and performed spline fitting to obtain the quasar continuum in the $\lya$ forest region. They further corrected their continuum estimates to correct for underestimating the true quasar continuum using mock spectra from theoretical models. They also accounted for intervening metal absorption lines from the IGM and corrected for these biases. Our $\tau_{\text{eff}}$ measurements are shown as green circles in Figure \ref{fig:comparison_schaye_kirkman} and the best fit $\tau_{\text{eff}}$ profile is shown as the green band. These estimates are broadly consistent with \cite{Faucher_Gigu_re_2008} over the range $2<z_{\lya}\leq 3.5$. The difference beyond $z_{\lya}>3.5$ is primarily owing to the fact that their continuum bias correction at higher redshift is too large. Whereas, in this work, we are directly estimating the quasar continuum from $z \sim 0$ quasars, and do not require any correction to continuum estimates. In addition, we do not find the "dip" feature identified by \citet{Faucher_Gigu_re_2008} at $z \sim 3.2$ . Our findings are consistent with those of \citet{Paris_2011} and \citet{Becker_2013} that this feature is not observed for more precise measurements of $\tau_{\text{eff}}$.

We proceed to compare our $\tau_{\text{eff}}$ measurements with results from \citet{Paris_2011} and \citet{Becker_2013}. Since our $\tau_{\text{eff}}$ measurements are only from SDSS DR16 quasar spectra, we restrict our comparison to these two studies. We refer the readers to \cite{Becker_2013} for a detailed comparison of other studies of $\tau_{\text{eff}}$ evolution   (e.g. \citealt{Bernadi_2003, McDonald_2005,Dall'Aglio_2009}).

Figure \ref{fig:comparison_schaye_kirkman}, top right panel shows that the results from \citet{Becker_2013} (gold diamonds). These are the most consistent results with our $\tau_{\text{eff}}$ estimates over the whole redshift range $2.5\leq z_{\lya}<5$.  \citet{Becker_2013} applied the method of composite quasar spectra on SDSS DR7 to measure the mean transmitted flux in the $\lya$ forest over $2<z<5$. They combined 6065 quasar spectra into 26 composites with mean redshift at $2.25\leq z_{\text{composites}}\leq 5.08$, and then corrected their measurements with data from \citet{Faucher_Gigu_re_2008} at $z\leq 2.5$, however they did not apply metal absorption line correction in their work, because the flux ratio measurement in their composite spectra is an equivalent method to reduce the metal line contamination in the $\lya$ forest region. Indeed this is a valid approach, as our $\tau_{\text{eff}}$ estimates after metal line correction are more or less consistent with that from \citet{Becker_2013}. Our $\tau_{\text{eff}}$ measurements slightly differ from these measurements over $2\leq z< 2.5$. This is mainly owing to the fact that \citet{Becker_2013} assumes that the true quasar continuum without contamination from $\lya$ forest can be measured in this redshift range, whereas we are estimating the quasar continuum with $z \sim 0.2$ quasar spectra. Therefore, the \cite{Becker_2013} composite spectra may have some $\lya$ forest contamination at $z_{\lya}\sim 2$ and under-estimate the true continuum blueward of $\lya$ transition. Otherwise, \citet{Becker_2013}  $\tau_{\text{eff}}$ estimates are very close to our results after correcting for the metal absorption line contamination.

Lastly, we compare our results with those of \citet{Paris_2011} to compare the neural-network-based model results with that of a PCA-based model. \citet{Paris_2011} measured the mean transmitted flux by applying the PCA-prediction model on the quasar spectra in the SDSS DR7 to obtain the true quasar continua. They assumed that the true  quasar continua blueward of the quasar $\lya$ emission can be found by fitting a spline curve to the flux peaks in the $\lya$ forest region at $z\sim 3$. \citet{Paris_2011} did not perform any metal absorption line correction or absolute scaling when computing the mean transmitted flux. We show the $\tau_{\text{eff}}$ estimates from \citet{Paris_2011} as purple triangles in Figure \ref{fig:comparison_schaye_kirkman}, bottom panel. Their mean estimates of $\tau_{\text{eff}}$ over the range $3.7<z_{\lya}<4.0$ are slightly higher than our measurements. These differences could be because of how the true continuum model was constructed. While we are using low redshift quasar spectra,  \citet{Paris_2011} fitted the peaks of $\lya$ forest region at higher redshift $z_{\text{QSO}}$ where there are some $\lya$ forest present. This method of fitting the peaks may be overestimating the true quasar continuum at $z_{\text{QSO}}$.

Figure \ref{fig:comparison_schaye_kirkman} also shows the best fit power law evolution of $\tau_{\text{eff}}$ as a function of redshift (green lines).  Overall, the  evolution of $\tau_{\text{eff}}$ with redshift from literature are broadly consistent with our measurements. Our measurements suggest a smooth evolution of $\tau_{\text{eff}}$  with redshift suggesting a more or less a smooth evolution of ionization and thermal state of the IGM. All subtle differences between these studies and our current measurements are mainly coming from the fact that we are estimating the most accurate quasar continuum blueward of $\lya$ transition uncontaminated by $\lya$ forest absorption.

\section{Conclusions}
\label{sec:conclusions}
In this paper, we introduce a novel deep learning approach, intelligent quasar continuum neural network (iQNet), to predict the quasar continuum in the rest frame wavelength range of $1020 \text{\AA} \leq \lambda_{\text{rest}} \leq 1600 \text{\AA}$. We train the iQNet model with high-resolution quasar spectra at low redshift $z_{\text{QSO}} \sim 0.2$ from the Hubble Spectroscopic Legacy Archive (HSLA). Our flexible model can predict the quasar continua at any $2<z_{\text{QSO}}\leq 5$ for any arbitrary survey. We test the iQNet network performance with a testing sample of quasar spectra from HSLA and apply it to predict quasar continua of 3196 SDSS DR16 quasars at  $2<z_{\text{QSO}}\leq 5$. Our main findings are summarized below:

\begin{itemize}
    \item Before we train a network or perform any PCA analysis, a standardization process of the data is necessary. This standardization process helps reduce model training bias and over-fitting \citep{gewers2018principal}. In typical quasar spectra, the $\lya$ and CIV emission lines are dominant features within the rest-frame wavelength range over $1020\text{\AA}\leq \lambda_{\text{rest}}\leq 1600\text{\AA}$. The PCA models investigated in this work will be biased towards those features if these two emission features are not properly scaled. This will result in a biased prediction of the quasar continua. 
    
    \item We construct a PCA prediction model with a standardization process (PCA+S) on the quasar spectra and find that our model gives a mean AFFE (Equation \ref{eqn:affe}) of 0.0590 on training data and 0.0628 on testing data. Even though our PCA+S prediction model outperforms all other published PCA models, the generalizability of PCA+S is limited. The PCA prediction model cannot predict a constant-flux continuum, as we discuss in Figure \ref{fig:nn_testing}, whereas the iQNet model can easily handle this scenario.
    
    \item We can characterize all quasar spectra into 4 distinct classes, using a GMM and the decomposed principal components. These classes are representative of all the quasars used for this work with the additional 5th class representing a BL-Lac like AGN spectra. The main differences among those 5 classes of quasar spectra are the emission strengths of the $\lya$ and CIV features. We also find a smaller difference in the slope of the continua blueward of $\lya$ emission.
    
    \item We use the PCA+GMM classes to create 11947 synthetic quasar spectra (12000 training spectra in total including 53 HSLA real quasar spectra). We inject strong ISM and HI absorption lines at random redshifts and change the S/N of the spectra to create a diverse range of training samples. We train The iQNet model on these 12000 training spectra with a binary cross-entropy loss function for over 40 minutes on the Google Colab platform with GPU enabled. We also add a standardization process and its corresponding inverse transformation process before the input layer of iQNet and after the output layer, respectively, to properly scale quasar continua after iQNet model prediction.
    
    \item We find that the iQNet model can achieve a median absolute fractional flux error 2.24\% in the training data set at $z\sim 0.2$ on the entire wavelength range of interest ([1020, 1600]\text{\AA}), which is approximately twice lower than our PCA prediction model with the standardization process (PCA+S), and the iQNet median AFFE on the predicted continua blueward of $\lya$ emission ([1020, 1216]\text{\AA}) is approximately three times lower than the PCA+S prediction model and four times lower than PCA prediction model introduced in this work. Comparing with AFFE reports on training sets from other literature (AFFE $\approx 6\%$ in \citealt{Paris_2011}, and AFFE $\approx 9\%$ in \citealt{Suzuki_2005}), our iQNet model outperforms these PCA prediction models based on the performance to predict the training data. In the blind testing set, the iQNet model outperforms these PCA prediction models and achieves an AFFE 0.0417, which is approximately half the AFFE from \citet{Suzuki_2005} PCA prediction model and only one-third of AFFE from \citet{Paris_2011} PCA. We find an AFFE 4.17\% on the testing data set at $z \sim 0.2$. The iQNet median AFFE on the predicted continua blueward of $\lya$ emission ([1020, 1216]\text{\AA}) in the testing set is approximately four times lower than the PCA+S prediction model and five times lower than PCA prediction model introduced in this work.

    \item We run the iQNet model on a standard Google Colab CPU architecture and it can predict about 945 high-resolution quasar continua per second, without utilizing GPU or TPU. This is almost three times faster than the PCA prediction models that were run on the same hardware architecture. In addition, the usage of iQNet model does not require a fitted continuum of the redward of $\lya$ emission of a quasar spectrum, but only needs a quasar spectrum with the proper standardization applied. This further reduces the processing time to generate a quasar continuum compared among other quasar continuum estimation techniques.
    
    \item We apply the iQNet model to predict the rest-frame quasar continua of 3196 $2<z_{\text{QSO}}\leq 5$ quasars from the SDSS survey. We use these continua predictions to estimate the mean transmitted flux ($\langle F\rangle$) and the corresponding effective optical depth ($\tau_{\text{eff}}$) in the $\lya$ forest region. We find that $\tau_{\text{eff}}$ evolves smoothly with redshift and can be characterized by a power-law evolution as $\tau_{\text{eff}}=0.00273^{+0.000142}_{-0.000141}(1+z_{\lya})^{3.571_{-0.031}^{+0.032}}$.
    
    \item Our $\langle F\rangle$ estimates are broadly consistent with those reported in the literature although these works used different methods to measure $\langle F\rangle$  \citep{Faucher_Gigu_re_2008, Paris_2011,Becker_2013}. Our approach presents an alternative method of measurement of $<F>$ compared to other methods, as we are directly predicting the quasar continua (the most uncertain element of these measurements) from $z \sim 0.2$ observations, where the impact of $\lya$ forest is the least. 
    This confirms that the iQNet model achieves accurate predictions of quasar continua at higher redshift $2<z_{\text{QSO}}\leq 5$.
\end{itemize}

Moreover, the architecture of our iQNet reveals a new approach to generate a quasar continuum, predict the continuum blueward of $\lya$ emission at high redshift, and study the $\lya$ forest. The number of neurons in each layer and the number of hidden layers in the neural network can be adapted to various quasar spectra and different instruments. Future studies may need to find another criterion, similar to the absolute fractional flux error, to evaluate the goodness of continuum fit and optimize the neural network models. Our iQNet proves that even the simplest vanilla neural network with fully-connected layers is able to generate quasar continua within an AFFE of ~1\% on a training set and that of ~4\% on a testing data set, far outperforming the studied PCA based methods. Thus, the architectures of the convolutional variational autoencoder(ConvVAE) and the generative adversarial network(GAN) may be the next model structures that researchers would want to apply to quasar spectra because those neural networks have better nonlinear representation and generalization ability than neural networks with only fully-connected neurons. We will explore these options in future work as well as extend our work to $z >6$ quasars.

\section*{Acknowledgements}

This research has made use of the HSLA database, developed and maintained at STScI, Baltimore, USA, and Astropy,\footnote{http://www.astropy.org} a community-developed core Python package for Astronomy \citep{astropy:2013, astropy:2018}. We are extremely grateful to J. X. Prochaska, J. Chisholm, and J. Burchett  for their careful reading of the draft and constructive comments and suggestions, that helped improve the paper. We also acknowledge the computing resources provided on Henry2, a high-performance computing cluster operated by North Carolina State University.

\section*{Data Availability}
The data underlying this article are available in HST Spectroscopic Legacy Archive (HSLA) and Sloan Digital Sky Survey Data Release 16. The models and manual-fitting HSLA QSO continua underlying this article are available in the iQNet Repository, at https://github.ncsu.edu/bordoloiastro/iQNet.

%%%%%%%%%%%%%%%%%%%%%%%%%%%%%%%%%%%%%%%%%%%%%%%%%%



\bibliographystyle{mnras}
\bibliography{qcnn_ref}% if your bibtex file is called example.bib

%\begin{thebibliography}{99}
%\end{thebibliography}

%%%%%%%%%%%%%%%%%%%%%%%%%%%%%%%%%%%%%%%%%%%%%%%%%%
\appendix

\section{Model bias and uncertainty}
\label{appendix: bias_uncer}
\begin{figure*}
    \includegraphics[width=\textwidth]{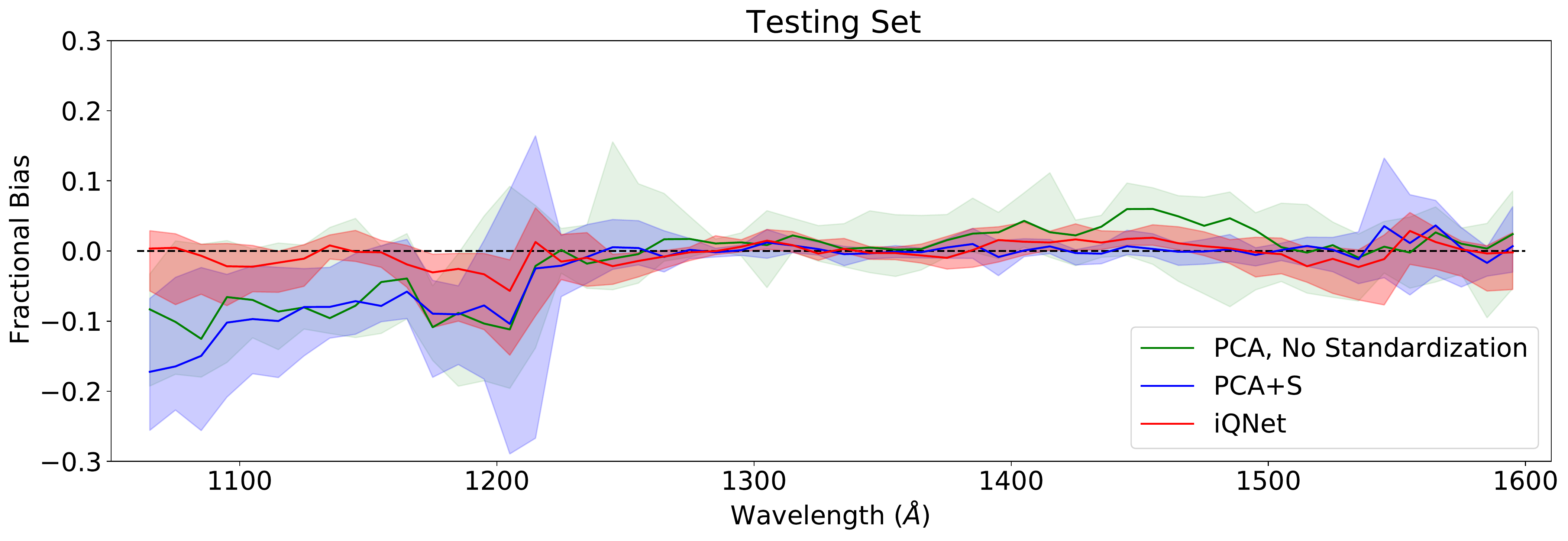}
    \includegraphics[width=\textwidth]{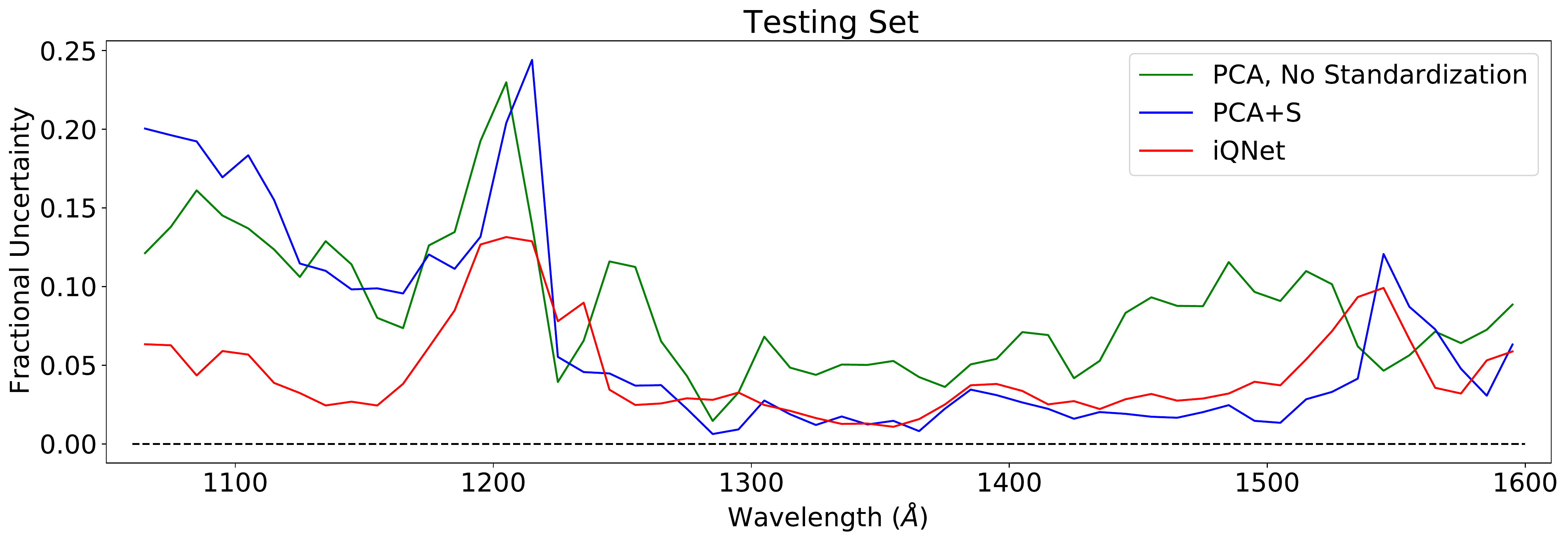}
    \caption{Fractional bias (top panel) and uncertainty (bottom panel) on the testing set among different models as a function of wavelength. For better visualization, the fractional bias and uncertainty are binned into 10\text{\AA} wavelength bins with the median values in each bin. The shaded areas correspond to the 16th and 84th percentiles of the computed bias distribution per pixel for each model. The green color represents the results using PCA prediction model without standardization, the blue color is the PCA prediction with standardization (PCA+S), and the red is for iQNet model.}
    \label{fig:bias_uncer_allmodels}
\end{figure*}

In this section we present the fractional bias and the fractional uncertainty as a function of rest-frame wavelength in the testing sample for all three models discussed in this paper, PCA prediction model without standardization, PCA prediction model with standardization (PCA+S), and the iQNet deep learning model. Figure \ref{fig:bias_uncer_allmodels} presents the fractional bias (top panel) and fractional uncertainty (bottom panel) in the PCA model without standardization (green line), PCA with proper standardization (PCA+S,  blue line), and the  deep learning iQNet model (red line), respectively. The shaded regions in Figure \ref{fig:bias_uncer_allmodels} correspond to the 16th and 84th percentiles of the bias distribution per pixel in the testing sample. While both fractional biases for PCA (green) and PCA+S (blue) blueward of $\lya$ emission are comparably the same, the fractional bias for PCA is higher at [1280, 1500]\text{\AA} than that for PCA+S. This confirms that the standardization process we introduced in Section \ref{subsec:standardization} enhances the performance of the PCA prediction model by scaling down the dominant $\lya$ and CIV emission features. The fractional uncertainty for the PCA model in the same wavelength region is also larger than that in the PCA+S model. The PCA prediction model performs slightly better than the PCA+S model near the $\lya$ and CIV emission regions. This may be because the PCA prediction model without standardization over-fits those two features and ignore non-obvious features in the quasar spectra. In comparison to both these PCA prediction models, the iQNet model shows significantly lower fractional bias and lower fractional uncertainty in the wavelength range blueward of $\lya$ emission, while reaching the same performance of the PCA+S model on the redward of $\lya$.

This is further quantified in Table \ref{tab:affe_blue_red}, which shows the AFFE estimates  among the PCA prediction models and the iQNet model.  Here we compute the AFFE in two wavelength regions (1) [1020, 1216]\text{\AA}, the region blueward of $\lya$ emission and (2)  [1216, 1600]\text{\AA}, the region redward of $\lya$ emission in both the training and testing samples.
 In all cases, the deep learning iQNet model outperforms the two PCA prediction models significantly. In particular, while predicting the quasar continua blueward of $\lya$ emission in the testing sample, the iQNet median AFFE is $\sim$ 4 times lower than that of the PCA+S model and $\sim$ 5 times lower than the PCA model. In case of the two PCA models we see that the standardization process helps reduce continuum prediction errors redward of $\lya$ emission. The PCA+S model shows a median AFFE $\sim$ 2 times lower than the PCA model in that wavelength range. This suggests that the standardization process helps reduce the continuum prediction error on the quasar continuum redward of $\lya$ emission, while keeping the error on the blueward continua approximately the same. These values are consistent with Figure \ref{fig:bias_uncer_allmodels} that shows that the iQNet model outperforms the two PCA prediction models by achieving the lowest fractional bias and uncertainty.

\begin{table*}
	\centering
	\caption{Absolute Fractional Flux Error of predicted continua in the training and testing Quasar spectra, blueward and redward of $\lya$ emission, respectively. PCA prediction model without standardization (PCA), PCA prediction model with standardization process (PCA+S) and the deep learning iQNet model are shown. The mean and median AFFE's for each case are reported. }
	\label{tab:affe_blue_red}
	\begin{tabular}{lcccccccc}
		\hline
		 & 
		\multicolumn{4}{c}{Blueward ([1020, 1216]\text{\AA})} & \multicolumn{4}{c}{Redward ([1216, 1600]\text{\AA})} \\
		 & \multicolumn{2}{c}{Training Set} & \multicolumn{2}{c}{Testing Set} & \multicolumn{2}{c}{Training Set} & \multicolumn{2}{c}{Testing Set}\\
		 Model Name & Median & Mean & Median & Mean & Median & Mean & Median & Mean\\ 
		\hline
		PCA & 0.123 & 0.156 & 0.141 & 0.148 & 0.0499 & 0.0560 & 0.0446 & 0.0496\\
		PCA+S & 0.0954 & 0.107 & 0.111 & 0.116 & 0.0207 & 0.0222 & 0.0230 & 0.0250 \\
		iQNet & 0.0288 & 0.0408 & 0.0291 & 0.0483 & 0.0097 & 0.0179 & 0.0163 & 0.0269\\
		\hline
	\end{tabular}
\end{table*}

\begin{figure*}
    \includegraphics[width=\textwidth]{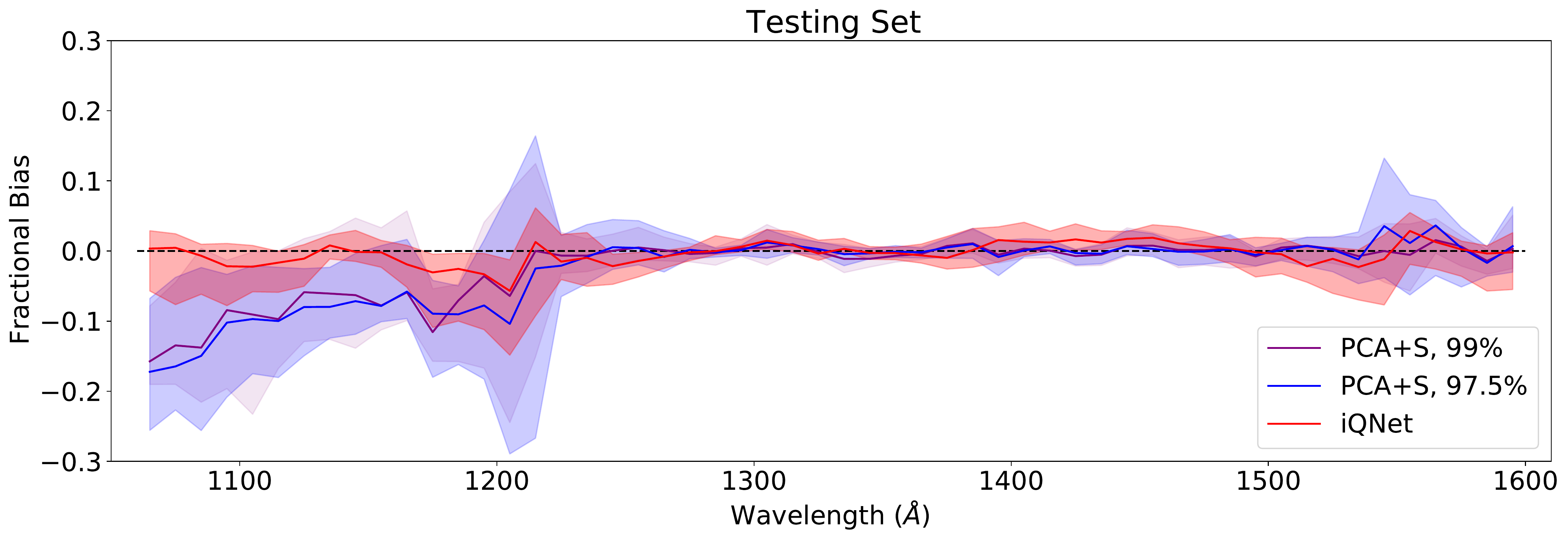}
    \includegraphics[width=\textwidth]{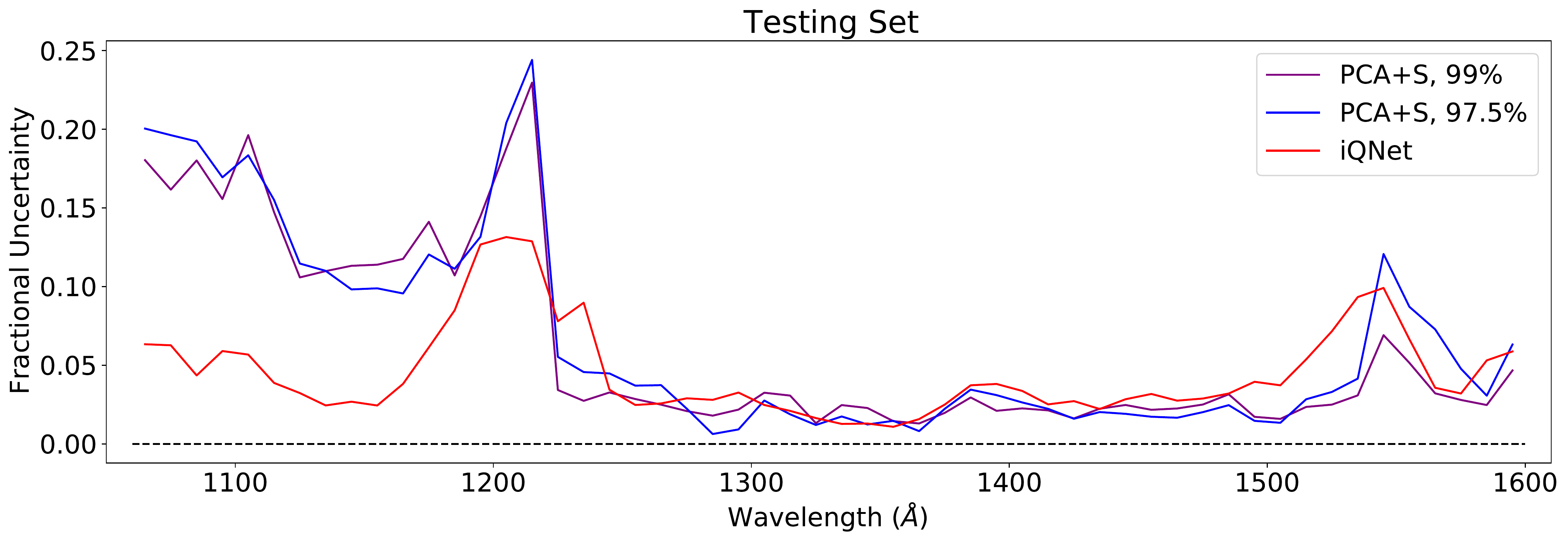}
    \caption{Fractional bias (top panel) and uncertainty (bottom panel) on the testing set among different models as a function of wavelength. For better visualization, the fractional bias and uncertainty are binned into 10\text{\AA} wavelength bins with the median values in each bin. The shaded areas correspond to the 16th and 84th percentiles of the computed bias distribution per pixel for each model. The violet color represents the results using PCA prediction model with standardization and explained variance of 99\% (PCA+S, 99\%), the blue color is the PCA prediction with standardization and explained variance of 97.5\% (PCA+S, 97.5\%), and the red is for iQNet model.}
    \label{fig:bias_uncer_allmodels_pca99}
\end{figure*}

We also present the fractional bias and the fractional uncertainty as a function of rest-frame wavelength in the testing sample for two PCA prediction models, with standardization and explained variances of 97.5\% and 99\%, respectively. We compare their performances with that from the iQNet model. Figure \ref{fig:bias_uncer_allmodels_pca99} shows the fractional bias (top panel) and fractional uncertainty (bottom panel) of these models. The two PCA models considered, the PCA+S model with 97.5\% explained variance (blue line) and the PCA+S model with 99\% explained variance (violet line) show similar performance in fractional bias. However, the second model exhibit slightly improved uncertainty around the CIV emission peak. The iQNet model (red line) exhibits better performance than both the PCA models. The shaded regions correspond to the 16th and 84th percentiles of the bias distribution per pixel in the testing sample.

\begin{figure*}
    \includegraphics[width=6in, height=3in]{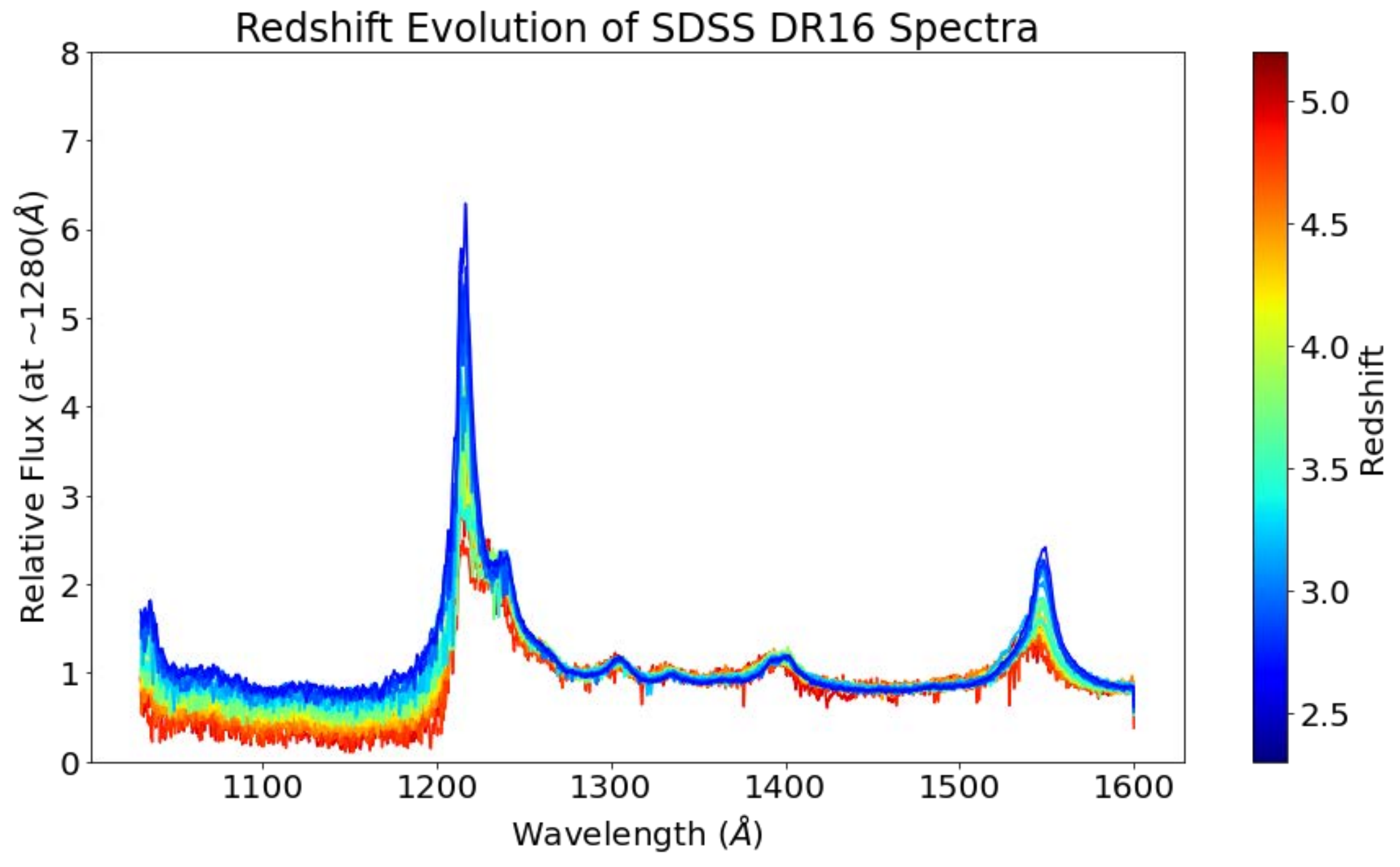}
    \includegraphics[width=6in, height=3in]{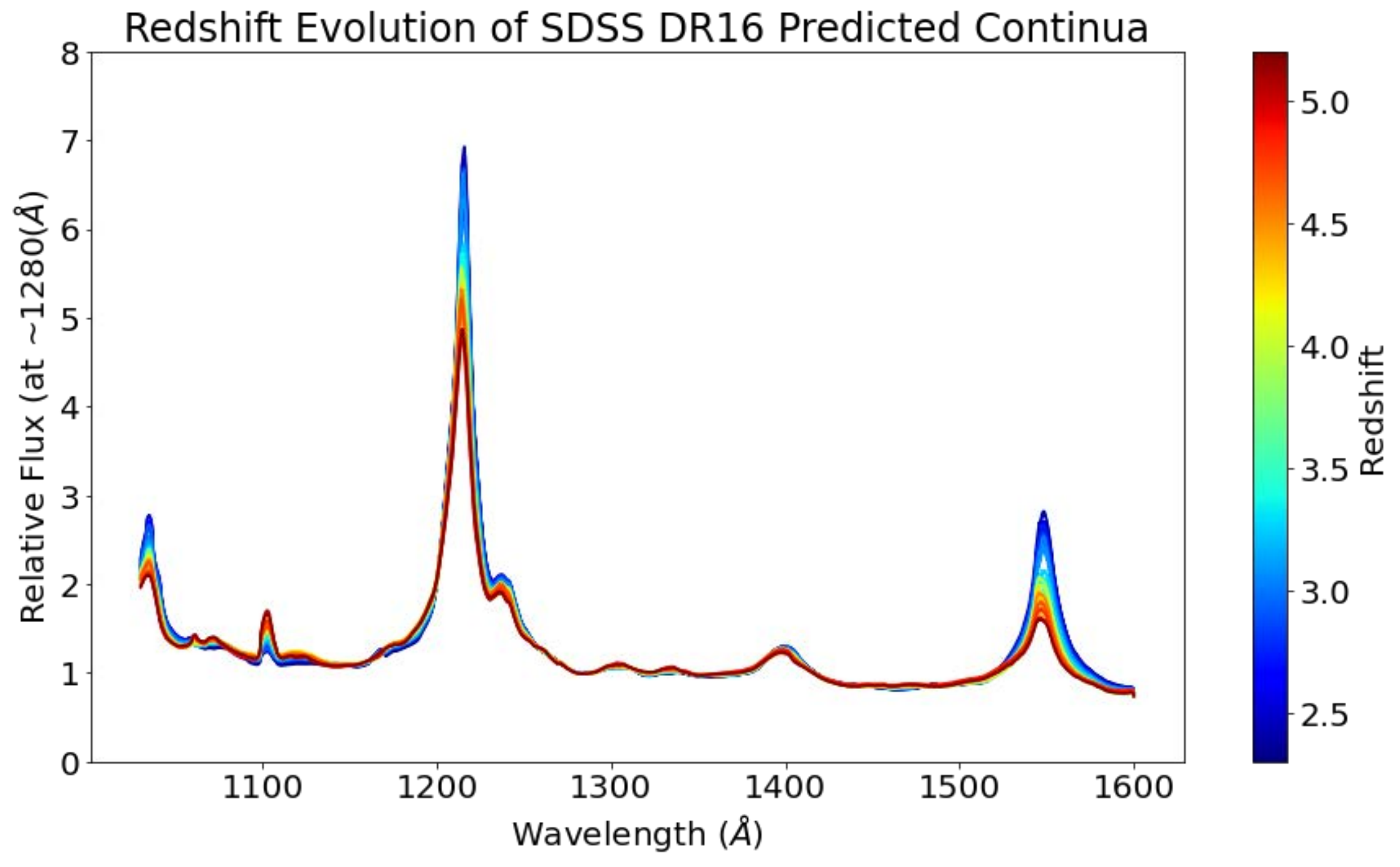}
    \caption{\textit{Top Panel:} Redshift evolution of co-added mean quasar spectra from SDSS DR16 quasar sample is shown. The spectra are co-added within $\delta z$ = 0.1 redshift bins. \textit{Bottom Panel:} The co-added mean   quasar continua predicted by iQNet are shown for the same objects. In both panels, the individual lines are color-coded as a function of redshift. The average line strengths of $\lya$ and CIV decrease with redshift consistent with observations, which demonstrates the Baldwin effect.}
    \label{fig:baldwin}
\end{figure*}

We also investigate whether the iQNet model can capture any observed redshift evolution of the intrinsic quasar spectra (e.g. the Baldwin effect \citep{Baldwin_1977}). Figure \ref{fig:baldwin}, top panel shows the co-added SDSS DR16 quasar spectra in different redshift bins. The lines are color coded to reflect the mean redshift of the quasars within each bin. The Baldwin effect is clearly visible in the data, and the emission line strengths vary with redshift. For each individual $z>2$ SDSS quasar spectra, iQNet model predicts the intrinsic quasar continuum. Figure \ref{fig:baldwin}, bottom panel shows the mean iQNet predicted quasar continua in the same redshift bins ($\Delta z = 0.1$). It can be seen that the iQNet can reasonably reproduce the observed evolution in emission line strengths. We note that, although the emission line strengths vary with redshift, the blueward continuum over the $\lya$ forest region remains more or less unchanged. This is one of the reason why we can use $z<1$ training data to predict the continuum in the $\lya$ forest region.

Although our iQNet deep learning model is not trained explicitly to account for this Baldwin effect, the network is flexible enough to account for any variability in $\lya$ and CIV line strengths. Figure \ref{fig:gemm_spec} shows that different $\lya$ to CIV line ratios are present in the training set, which already allows the network to train for different line ratios. Notice that even though the mean spectra of the four GMM quasar classes do not show weak-broad-line features, we add a fifth quasar class with a constant relative flux representing a BL-Lac like AGN spectra, which is not well represented in our training spectra (Figure \ref{fig:nn_testing} bottom panel). This helps the network training process and allows the network to update the weight matrix to generate quasar continua with weak/broad emission features. This is the main reason why the network can handle the weak and broad emission features at high-z.

\begin{figure*}
    \includegraphics[width=0.6\textwidth]{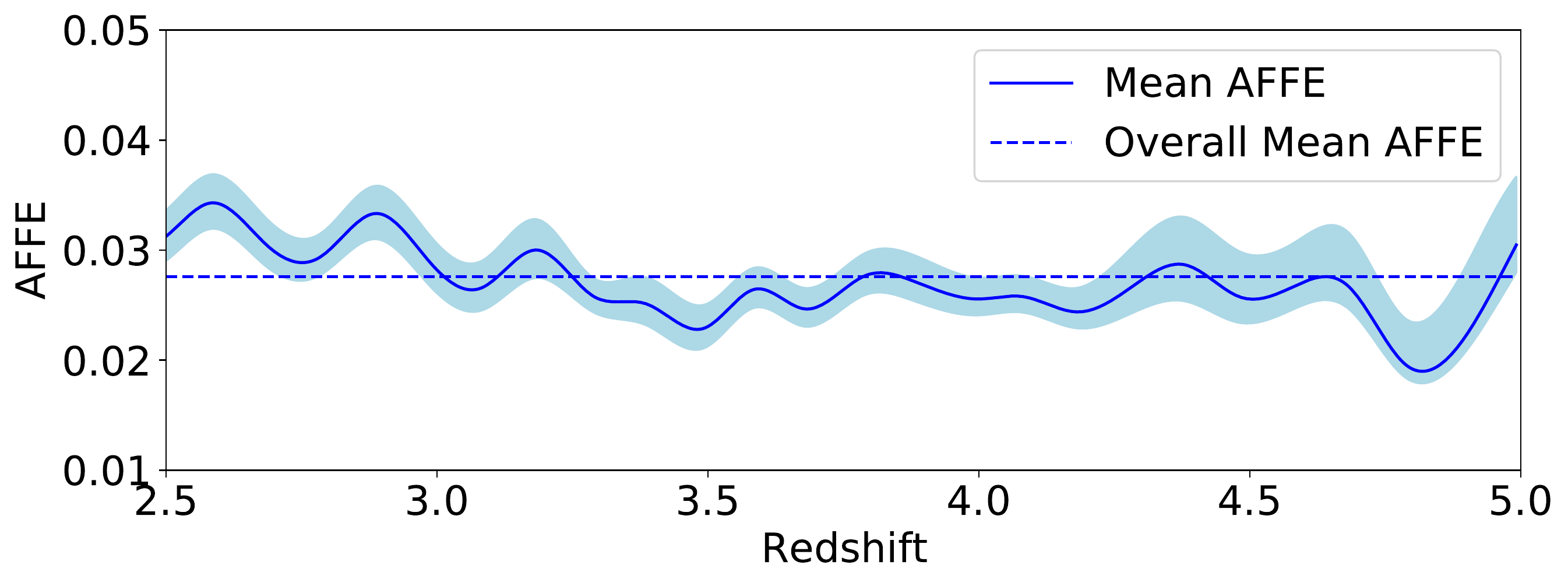}
    \caption{Redshift evolution of AFFE in co-added SDSS DR16 quasar spectra redward of $\lya$ emission. The AFFE is computed in $\Delta z=0.1$ redshift bins is shown as the solid blue line. The shaded region shows the 16th and 84th percentiles in AFFE estimates in each redshift bin. The AFFE of the red part of the SDSS spectra do not show any evolution with redshift and is consistent with the mean AFFE (dashed line) for the full sample.}
    \label{fig:affe_redshift}
\end{figure*}

We also investigate if the AFFE changes as we predict the continuum of different redshift quasars. This can only be tested for the part of the quasar continuum redward of $\lya$ emission, as it is not contaminated by the $\lya$ forest. Figure \ref{fig:affe_redshift} shows the redshift evolution of AFFE redward of $\lya$ emission for the SDSS DR16 quasar sample. For computational simplicity, we compute the AFFE on the stacked SDSS quasar spectra presented in Figure \ref{fig:baldwin}. We estimate the bootstrapped mean spectra in each redshift bin with 1000 repetitions. The 16th and 84th percentiles in bootstrapped mean distribution (per pixel) gives the error spectrum for each co-add. We then compute the mean AFFE redward of $\lya$ emission (solid blue line) in each redshift bin. The uncertainties in co-adding are propagated and shown as the shaded region in Figure \ref{fig:affe_redshift}. We clearly see that there is no redshift evolution in the mean AFFE redward of $\lya$ emission in this data-set. This demonstrates the flexibility of iQNet to predict the $z>2$ QSO continua. Some careful considerations while creating the training set help our model generate a more generalized continua consistent with observed spectra at z>2.

% Don't change these lines
\bsp	% typesetting comment
\label{lastpage}
\typeout{get arXiv to do 4 passes: Label(s) may have changed. Rerun}
\end{document}

%% file: tau_table.tex
2.019 & $0.136_{-0.018}^{+0.019}$ & $0.118_{-0.019}^{+0.020}$ & $0.105_{-0.019}^{+0.020}$ \\2.104 & $0.173_{-0.016}^{+0.016}$ & $0.151_{-0.017}^{+0.017}$ & $0.143_{-0.017}^{+0.017}$ \\2.199 & $0.197_{-0.014}^{+0.014}$ & $0.173_{-0.015}^{+0.015}$ & $0.166_{-0.015}^{+0.015}$ \\2.300 & $0.226_{-0.014}^{+0.015}$ & $0.200_{-0.016}^{+0.016}$ & $0.196_{-0.016}^{+0.016}$ \\2.400 & $0.250_{-0.014}^{+0.014}$ & $0.222_{-0.016}^{+0.016}$ & $0.220_{-0.016}^{+0.016}$ \\2.500 & $0.280_{-0.016}^{+0.016}$ & $0.250_{-0.017}^{+0.017}$ & $0.251_{-0.017}^{+0.017}$ \\2.600 & $0.298_{-0.017}^{+0.017}$ & $0.267_{-0.018}^{+0.018}$ & $0.270_{-0.018}^{+0.018}$ \\2.700 & $0.345_{-0.018}^{+0.018}$ & $0.310_{-0.018}^{+0.018}$ & $0.316_{-0.018}^{+0.018}$ \\2.801 & $0.359_{-0.018}^{+0.018}$ & $0.324_{-0.019}^{+0.019}$ & $0.331_{-0.019}^{+0.019}$ \\2.900 & $0.403_{-0.020}^{+0.019}$ & $0.366_{-0.020}^{+0.020}$ & $0.376_{-0.020}^{+0.020}$ \\3.001 & $0.431_{-0.020}^{+0.020}$ & $0.393_{-0.020}^{+0.020}$ & $0.404_{-0.020}^{+0.020}$ \\3.101 & $0.462_{-0.021}^{+0.021}$ & $0.423_{-0.021}^{+0.021}$ & $0.435_{-0.021}^{+0.021}$ \\3.200 & $0.494_{-0.022}^{+0.022}$ & $0.454_{-0.022}^{+0.022}$ & $0.468_{-0.022}^{+0.022}$ \\3.301 & $0.537_{-0.024}^{+0.023}$ & $0.496_{-0.023}^{+0.023}$ & $0.511_{-0.023}^{+0.023}$ \\3.398 & $0.581_{-0.025}^{+0.025}$ & $0.539_{-0.025}^{+0.024}$ & $0.556_{-0.025}^{+0.024}$ \\3.502 & $0.632_{-0.025}^{+0.025}$ & $0.588_{-0.024}^{+0.025}$ & $0.607_{-0.024}^{+0.025}$ \\3.599 & $0.697_{-0.028}^{+0.027}$ & $0.651_{-0.027}^{+0.027}$ & $0.672_{-0.027}^{+0.027}$ \\3.700 & $0.733_{-0.028}^{+0.027}$ & $0.688_{-0.027}^{+0.027}$ & $0.709_{-0.027}^{+0.027}$ \\3.797 & $0.757_{-0.027}^{+0.028}$ & $0.712_{-0.027}^{+0.027}$ & $0.733_{-0.027}^{+0.027}$ \\3.898 & $0.816_{-0.028}^{+0.029}$ & $0.770_{-0.028}^{+0.028}$ & $0.792_{-0.028}^{+0.028}$ \\3.999 & $0.885_{-0.029}^{+0.030}$ & $0.838_{-0.029}^{+0.029}$ & $0.862_{-0.029}^{+0.029}$ \\4.096 & $0.973_{-0.032}^{+0.031}$ & $0.925_{-0.031}^{+0.030}$ & $0.950_{-0.031}^{+0.030}$ \\4.196 & $1.024_{-0.032}^{+0.031}$ & $0.976_{-0.031}^{+0.030}$ & $1.001_{-0.031}^{+0.030}$ \\4.297 & $1.045_{-0.033}^{+0.033}$ & $1.000_{-0.032}^{+0.033}$ & $1.023_{-0.032}^{+0.033}$ \\4.395 & $1.076_{-0.033}^{+0.034}$ & $1.033_{-0.032}^{+0.033}$ & $1.054_{-0.032}^{+0.033}$ \\4.495 & $1.205_{-0.037}^{+0.037}$ & $1.160_{-0.036}^{+0.036}$ & $1.184_{-0.036}^{+0.036}$ \\4.597 & $1.325_{-0.033}^{+0.033}$ & $1.279_{-0.032}^{+0.032}$ & $1.304_{-0.032}^{+0.032}$ \\4.684 & $1.417_{-0.043}^{+0.043}$ & $1.372_{-0.042}^{+0.042}$ & $1.396_{-0.042}^{+0.042}$ \\4.795 & $1.534_{-0.033}^{+0.033}$ & $1.490_{-0.032}^{+0.033}$ & $1.513_{-0.032}^{+0.033}$ \\4.882 & $1.617_{-0.038}^{+0.037}$ & $1.575_{-0.037}^{+0.036}$ & $1.598_{-0.037}^{+0.036}$ \\